%% file: vowinckel_cohesive_sediment.tex
\documentclass{jfm}
\usepackage{graphicx}
\usepackage{epstopdf, epsfig}
\usepackage{amsmath}
\usepackage[usenames,dvipsnames]{xcolor} 
\usepackage{textcomp}
\usepackage{gensymb}
\usepackage{soul}

\newcommand{\rone}{\textcolor{black}}
\newcommand{\rtwo}{\textcolor{black}}

\DeclareMathOperator\erf{erf}

\shorttitle{Settling of cohesive sediment}
\shortauthor{B. Vowinckel, J. Withers, E. Meiburg, P. Luzzatto-Fegiz}

\title{Settling of cohesive sediment: particle-resolved simulations}

\author{B. Vowinckel\aff{1}
  \corresp{\email{vowinckel@engineering.ucsb.edu}},
  J. Withers\aff{1,}\aff{2},
  Paolo Luzzatto-Fegiz\aff{1},\\
  \and E. Meiburg\aff{1}}

\affiliation{\aff{1} Department of Mechanical Engineering, University of California, Santa Barbara, CA, USA
             \aff{2} The University of Queensland, School of Mechanical and Mining Engineering, Brisbane, Queensland, Australia}

\begin{document}

\maketitle
\begin{abstract}
We develop a physical and computational model for performing fully coupled, grain-resolved Direct Numerical Simulations of cohesive sediment, based on the Immersed Boundary Method. The model distributes the cohesive forces over a thin shell surrounding each particle, thereby allowing for the spatial and temporal resolution of the cohesive forces during particle-particle interactions. The influence of the cohesive forces is captured by a single dimensionless parameter in the form of a cohesion number, which represents the ratio of cohesive and gravitational forces acting on a particle. We test and validate the cohesive force model for binary particle interactions in the Drafting-Kissing-Tumbling (DKT) configuration. Cohesive sediment grains can remain attached to each other during the tumbling phase following the initial collision, thereby giving rise to the formation of flocs. The DKT simulations demonstrate that cohesive particle pairs settle in a preferred orientation, with particles of very different sizes preferentially aligning themselves in the vertical direction, so that the smaller particle is drafted in the wake of the larger one. This preferred orientation of cohesive particle pairs is found to remain influential for systems of higher complexity. To this end, we perform large simulations of 1,261 polydisperse settling particles starting from rest. These simulations reproduce several earlier experimental observations by other authors, such as the accelerated settling of sand and silt particles due to particle bonding, the stratification of cohesive sediment deposits, and the consolidation process of the deposit. They identify three characteristic phases of the polydisperse settling process, viz. (i) initial stir-up phase with limited flocculation; (ii) enhanced settling phase characterized by increased flocculation; and (iii) consolidation phase. The simulations demonstrate that cohesive forces accelerate the overall settling process primarily because smaller grains attach to larger ones and settle in their wakes. For the present cohesion number values, we observe that settling can be accelerated by up to 29\%. We propose \rtwo{physically based parametrization of classical} hindered settling functions introduced by earlier authors, in order to account for cohesive forces. An investigation of the energy budget shows that, even though the work of the collision forces is much smaller than that of the hydrodynamic drag forces, it can substantially modify the relevant energy conversion processes. 
\center
\Large
\end{abstract}


\section{Introduction}\label{sec:introduction}
\input{01_introduction.tex}
\section{Computational method}\label{sec:method}
\input{02_computational_method.tex}

\section{Cohesive forces in particle-resolving simulations}\label{sec:cohesion}
\input{03_cohesive_force_model.tex}
\section{Binary interaction}\label{sec:binary_interaction}
\input{04_drafting_kissing_tumbling.tex}
\section{Settling of a large ensemble}\label{sec:settling}
\input{05_settling.tex}

\section{Conclusions}\label{sec:conclusions}
 \input{99_conclusions.tex} 
\section*{Acknowledgements}
This research is supported by the National Science Foundation (NSF) through grant CBET-1638156.  BV gratefully acknowledges the Feodor-Lynen scholarship provided by the Alexander von Humboldt Foundation, Germany. The authors thank J. Israelachvili and R. Seto for stimulating discussions on the interaction between colloids. Computational resources for this work used the Extreme Science and Engineering Discovery Environment (XSEDE), which was supported by the National Science Foundation, USA, Grant No. TG-CTS150053.

\begin{appendix}
\section{Comparison to natural systems} \label{app:comparison}
 \input{App_comparison.tex}  
\section{Non-dimensional particle equation of motion} \label{app:eom}
 \input{App_particle_eom_dimesionless.tex} 
\section{Definition of the characteristic velocity} \label{app:char_vel} 
 \input{App_char_velocity.tex} 
\end{appendix} 

\bibliographystyle{jfm}

\end{document}

%% file: 01_introduction.tex
The term ‘cohesive sediment’ commonly refers to particles with diameters smaller than 63$\mu$m \citep{grabowski2011}. At this size, cohesive van-der-Waals (vdW) forces can dominate over gravitational forces and trigger particle aggregation or flocculation. These cohesive forces result from correlations in the fluctuating polarizations of nearby particles, and they play an important role in such environments as rivers \citep{seminara2010}, lakes and estuaries \citep{deswart2009}, fisheries and coastal ecosystems, and benthic habitats near the seafloor \citep{rhoads1974}. \rtwo{Due to the modified particle-particle interaction, the dynamics of cohesive sediment is significantly more complex than for its non-cohesive counterpart.} 

Even though the concept of vdW forces dates back to the late 19{\em th} century, the physical mechanisms responsible for these forces were not explained until the development of the theory of quantum mechanics, as reviewed by \cite{visser1989}. First scaling laws were presented by \rone{\cite{hamaker1937}} for the idealized situation of spherical particles. This author suggested that cohesive forces on an individual particle under dry conditions scale as $F_\text{coh} \propto A_H R_p /(6 \zeta_n^2)$, where the Hamaker constant $A_H$ accounts for particle properties such as mineralogy and surface coating, $R_p$ denotes the particle radius, and $\zeta_n$ is the gap size between two approaching particles. These forces can lead to the formation of flocs through the \rtwo{binding of individual particles}, thereby resulting in much larger aggregates. Since gravitational and hydrodynamic forces scale as $F_g \propto R_p^3$ and $F_h \propto R_p^2$, respectively, flocculated aggregates typically settle more rapidly than the Stokes settling velocity of the individual particles \citep[e.g][]{mehta1989,zinchenko2014}. Several investigations have addressed the impact of the ambient fluid properties such as salinity \citep[e.g.][]{aberle2004,sutherland2015} on flocculation. In these studies, it was found that the magnitude of the cohesive forces, along with the related flocculation behavior, can depend strongly on the salinity. However, reliable predictive tools for the sedimentation and erosion characteristics of cohesive sediment have not yet been developed  \citep{debnath2010}. For example, it remains unclear how sediment composition \citep{aberle2004}, salinity \citep{sutherland2015}, the grain size distribution \citep{teslaa2015}, or a combination of these \citep{huang2017} affect the settling rate of fine-grained sediments.

\rone{This lack of predictive tools can be attributed to difficulties in precisely measuring cohesive forces on the grain scale in natural systems, such as silt settling in water. In principle, the Hamaker constant can be derived from the Lifshitz theory depending on the properties of the particles and the ambient fluid. For example, \cite{visser1972} reported values up to $A_H = 1.8 \cdot 10^{-18} \, \text{J}$  for ionic crystals in water, \cite{bergstrom1997} found $A_H=1\cdot10^{20}$ for quartz in water, while \cite{lick2004} measured a value of $A_H = 6.4 \cdot 10^{-23} \, \text{J}$ for silicate particles in water, which spans a range of five orders of magnitude. However, the actual vdW forces must then  be derived by integrating the electromagnetic fluctuations at microscopic scales over the volume of the particles. This can be done for engineered colloids of spherical shapes but is less trivial for natural silica materials with complex shapes. The issue of parameterizing cohesive forces has, hence,  been the subject of an ongoing debate in the literature \citep[e.g.][]{israelachvili1992, ho2002, leong2003,lick2004, liang2007, righetti2007, kosinski2010, breuer2015}.}

Our incomplete understanding of how cohesive forces depend on the experimental conditions prevents us from deriving universal scaling relationships for the settling rates of flocculated sediments. Numerical investigations have attempted to tackle this issue by employing point-particle approaches in conjunction with a hard-sphere model to account for particle-particle interactions \citep[e.g.][]{ho2002, kosinski2010, breuer2015, sun2018}. The hard-sphere model resolves collisions instantaneously by changing the particle velocity according to a restitution coefficient for inelastic collisions. This approach has well known deficiencies when dealing with denser systems such as flocculated sediment, since the empirical relationship for computing the hydrodynamic drag of a particle is typically based on undisturbed flow conditions \citep{loth2000}. A more realistic approach that has gained popularity for noncohesive sediment in recent years involves fully coupled particle-resolving Direct Numerical Simulations (DNS) \citep{balachandar2010}. In particular, computational tools have been developed that are able to capture the dynamics of very dense, polydisperse systems with a minimal number of tunable parameters \citep{biegert2017}. It is hence desirable to extend this computational approach to include cohesive forces. For example, \cite{gu2016} proposed a cohesive force model that scales inversely with gap size and is capped at a critical value, but only at the cost of changing the stiffness of the spring-dashpot system for the underlying soft-sphere model. 

In the present study, we extend the particle-resolved DNS framework developed by \cite{biegert2017} to include cohesive forces \rone{for macroscopic particles, i.e. within the range of $2\mu\text{m} \leq D_p \leq 63 \mu\text{m}$, where $D_p$ is the particle diameter.} Subsequently, we will employ this computational approach in order to study the influence of cohesive forces on binary particle-particle interactions, such as the classical Drafting-Kissing-Tumbling scenario of two settling particles \citep{fortes1987}. This case will provide insight into the physical mechanisms by which interacting, cohesive particles arrange themselves into steady-state settling configurations. The knowledge gained from this simple test case is then applied to the more complex situation of a large ensemble of polydisperse sedimenting particles, for which we will compare numerical observations with experimental studies investigating fine sand \citep{lick2004} and silt \citep{teslaa2015}, respectively, albeit on a much smaller spatial scale.

The paper is structured along the following lines. We briefly state the governing equations of motion for the fluid and the particles in \S \ref{sec:method}, where we also summarize the numerical approach underlying the fully-coupled, particle-resolving DNS simulations, including the collision model for cohesionless grains. A novel computational model for cohesive forces is introduced and validated in \S \ref{sec:cohesion}. By distributing the cohesive forces over a thin shell surrounding each particle, this model allows for their spatial and temporal resolution during particle-particle interactions, while preserving certain integral properties. It thus enables us to analyze the influence of the cohesive forces on the processes by which kinetic and potential energies are converted into each other, in terms of a dimensionless cohesion number. The implications of cohesive forces on binary particle interactions are discussed in \S \ref{sec:binary_interaction}. By focusing on the well-known Drafting-Kissing-Tumbling problem, we identify preferred quasisteady geometric configurations in which cohesive particle pairs tend to settle, as a function of the particle size ratio. A polydisperse ensemble of 1,261 settling particles is analyzed in \S \ref{sec:settling}, for different values of the cohesion number. We find that cohesive forces accelerate the settling process, primarily because smaller grains attach to larger ones, which speeds up their downward motion. For the parameter values of the present study, we find that settling is accelerated by up to 29\%. Based on the simulation results, we propose a parameterization of the hindered settling function for cohesive sediment. In addition, we carry out a detailed investigation of the energy budget, in order to obtain quantitative information on the work performed by the hydrodynamic and collision forces. We observe that the preferred settling configurations of isolated particle pairs remain influential even within the large ensemble. Finally, \S \ref{sec:conclusions} summarizes the main findings of the investigation.

%% file: 02_computational_method.tex
\subsection{Fully coupled grain-resolving simulations}
We solve the unsteady Navier-Stokes equations for an incompressible Newtonian fluid, given by

\begin{equation} \label{eq:navier_stokes}
\frac{\partial{\textbf{u}}}{\partial{t}}+\nabla\cdot(\textbf{u}\textbf{u}) = -\frac{1}{\rho_f}\:\nabla p + \nu_f \nabla^2 \textbf{u} + \textbf{f}_\textit{IBM} \hspace{0.5cm},
\end{equation}
along with the continuity equation
\begin{equation}\label{eq:continuity}
\nabla\cdot\textbf{u}=0 \qquad ,     \hspace{0.5cm}
\end{equation}
on a uniform rectangular grid with grid cell size $\Delta x = \Delta y = \Delta z = h$. Here, $\textbf{u}=(u,v,w)^{T}$ designates the fluid velocity vector in Cartesian components, $p$ denotes the pressure, $\nu_f$ is the kinematic viscosity, $t$ the time, and $\textbf{f}_\textit{IBM}$ represents an artificial volume force introduced by the Immersed Boundary Method \cite[IBM;][]{uhlmann2005,kempe2012a}. This volume force, which acts on the right-hand side of \eqref{eq:navier_stokes} in the vicinity of the inter-phase boundaries, connects the motion of the particles to the fluid phase. We integrate equations \eqref{eq:navier_stokes} and \eqref{eq:continuity} by a third order low-storage Runge-Kutta (RK) scheme and a finite differencing approach in time and space, respectively. The pressure is treated with a direct solver based on Fast Fourier Transforms.

Note that gravity has been omitted from the equation of motion for the fluid \eqref{eq:navier_stokes}, because the contribution from hydrostatic pressure is not relevant to the problems presented in the following. Gravity is, however, explicitly accounted for in the equations of motion for the particles. Within the framework of the IBM, we calculate the motion of each individual spherical particle by solving an ordinary differential equation for its translational velocity $\textbf{u}_p=(u_p,v_p,w_p)^{T}$
       %
       \begin{equation}\label{eq:eom_trans}
        m_p\: \frac{\text{d}\textbf{u}_p}{\text{d} t} = \underbrace{\oint_{\Gamma_p} \boldsymbol{\tau} \cdot \textbf{n}\: {\text{d}A}}_{=\textbf{F}_{h,p}} +
        \underbrace{V_p\:( \rho_p-\rho_f )\: \textbf{g}}_{=\textbf{F}_{g,p}} + \textbf{F}_{c,p} \qquad ,
       \end{equation}
        and its angular velocity $\boldsymbol{\omega}_p=(\omega_{p,x},\omega_{p,y},\omega_{p,z})^{T}$
       \begin{equation}\label{eq:eom_rot}
       I_p \:\frac{ \text{d}\boldsymbol{\omega}_p}{\text{d} t} = \underbrace{\oint_{\Gamma_p} \textbf{r}\times(\boldsymbol{\tau}\cdot\textbf{n})\:{\text{d}A}}_{=\textbf{T}_{h,p}} + \textbf{T}_{c,p} \hspace{0.5cm}.
      \end{equation}
Here, $m_p$ denotes the particle mass, $\Gamma_p$ the fluid-particle interface, $\boldsymbol{\tau}$ the hydrodynamic stress tensor, $\rho_p$ the particle density, $V_p$ the particle volume, $g$ the gravitational acceleration, $I_p=8\pi\rho_p R_p^{5}/15$ the moment of inertia, and $R_p$ the particle radius. Furthermore, the vector $\textbf{n}$ represents the outward-pointing normal on the interface $\Gamma_p$, $\textbf{r} = \textbf{x} - \textbf{x}_p$ is the position vector of the surface point with respect to the center of mass $\textbf{x}_p$ of a particle, and $ \textbf{F}_{c,p}$ and $\textbf{T}_{c,p}$ indicate the force and torque due to particle collisions, respectively. For the sake of brevity, we denote the hydrodynamic force and torque as $\textbf{F}_{h,p}$ and $\textbf{T}_{h,p}$, respectively, and the gravitational force as $\textbf{F}_{g,p}$. We use a RK-scheme that subdivides the three-step procedure of the fluid into a total of 15 substeps per fluid time step to integrate the particles' equations of motion \eqref{eq:eom_trans} and \eqref{eq:eom_rot} in time. It was shown by \cite{biegert2017} that this is a necessity to resolve short-range effects of lubrication forces in time. 

The fluid-particle interaction was validated in \cite{biegert2017} by comparing our simulation results to experimental data for a settling sphere in a large container \citep{mordant2000}, and for a particle settling above a wall \citep{tencate2002}, yielding excellent agreement.

\subsection{Cohesionless particle-particle interaction}\label{sec:collision}

The computational approach for modeling cohesionless particle-particle interactions is described in detail in \cite{biegert2017}, and validation results are provided for normal and oblique binary collisions, as well as for the collective motion of a sediment bed sheared by a Poiseuille flow. In order to keep this paper self-contained, we provide a brief summary in the following. 

The particle-particle interaction comprises short-range hydrodynamic effects due to lubrication forces $\textbf{F}_{l}$, as well as forces acting in the normal and tangential directions for direct particle contact, denoted as $\textbf{F}_{n}$ and $\textbf{F}_{t}$, respectively. The resulting collision force on particle $p$ is the sum off all these effects
\begin{equation}\label{eq:particle_forces}
	\textbf{F}_{c,p} = \sum_{q,\:q \neq p}^{N_p} \left( \textbf{F}_{l,pq} + \textbf{F}_{n,pq} + \textbf{F}_{t,pq}\right) + \textbf{F}_{l,pw} + \textbf{F}_{n,pw} + \textbf{F}_{t,pw}\hspace{0.5cm},
\end{equation}
where the subscripts $pq$ and $pw$ indicate interactions with particle $q$ or a wall, respectively. In what follows, we present the algebraic expressions for particle-particle interaction only. Analogous formulations for particle-wall interactions can be found in \cite{biegert2017}.
Consistent with the findings of \cite{cox1967}, we model the unresolved component of the lubrication forces in our simulations as
\begin{equation}\label{eq:lubrication}
  \textbf{F}_{l,pq} = 
  \begin{cases}
  - \frac{6 \pi \rho_f \nu_f R_\text{eff}^2}{\max(\zeta_n,\zeta_\text{min})} \textbf{g}_n  & 0 < \zeta_n \leq 2h \\
    0 & \text{otherwise}
  \end{cases}
\end{equation} 
where $\textbf{g}_n = \textbf{u}_p - \textbf{u}_q$ is the relative velocity of the two colliding particles. To prevent the unresolved lubrication forces from diverging to infinity with decreasing gap size, the force is limited by $\zeta_\text{min}$, which can be interpreted as a surface roughness of the particles. The value of $\zeta_\text{min} = 3\cdot10^{-3}R_m$ was calibrated in \cite{biegert2017} for particle-wall collisions to match the rebound trajectories of the experiments by \cite{gondret2002}. The mean radius becomes $R_m = R_p$ and $R_m = \frac{1}{2}(R_p + R_q)$ for particle-wall and particle-particle interactions, respectively. The effective radius $R_\text{eff}$ is defined as $R_\text{eff} = R_p \, R_q /(R_p + R_q)$, where we set $R_q = \infty$ for particle-wall collisions. We also note that \eqref{eq:lubrication} only accounts for the part that cannot be resolved with the IBM as $\zeta_n$ becomes smaller than $2h$. \rone{We have conducted detailed tests repeating the test case of \cite{tencate2002} for the particle sizes of interest here. Our results show that the particles rapidly decelerate as soon as they come as close as $\zeta_n = 2D_p$ illustrating that nearly all of the work required to squeeze the fluid out of the gap is fully resolved in our simulations.} 
	
Direct particle contact is accounted for by a normal and a tangential component of the collision force. The repulsive normal component is represented by a nonlinear spring-dashpot model for the normal direction
	\begin{equation} \label{eq:acm_force}
	\textbf{F}_{n,pq} = -k_n |\zeta_n|^{3/2} \textbf{n} - d_n \textbf{g}_{n,cp} \qquad ,
	\end{equation}
where $\textbf{g}_{n,cp}$ denotes the normal component of the relative velocity at the surface contact point \citep{kempe2012b}. Furthermore,  $k_n$ and $d_n$ represent stiffness and damping coefficients that are adaptively calibrated for every collision as described by \cite{biegert2017}, in order to yield a prescribed restitution coefficient $e_\text{dry} = - u_\text{out} / u_\text{in}$. Here, $u_\text{out}$ and $u_\text{in}$ indicate the normal components of the relative particle speed immediately after and right before the particle impact, respectively. The forces in the tangential direction are modeled by a linear spring-dashpot model capped by the Coulomb friction law as
\begin{equation} \label{eq:lin_tan}
\textbf{F}_{t,pq} = \min \left(-k_t  \boldsymbol{\zeta}_t - d_t \textbf{g}_{t,cp} , ||\mu \textbf{F}_n|| \textbf{t} \right)  \qquad ,
\end{equation}
where $\mu$ represents the coefficient of friction between the two surfaces and $\boldsymbol{\zeta}_t$ is the tangential displacement integrated over the time interval for which the two particles are in contact. The tangential stiffness and damping coefficients $k_t$ and $d_t$ are adapted to account for zero-slip rolling or sliding according to the Coulomb friction law \citep{thornton2013}. For all of the simulations to be presented in the following, we have chosen $e_\text{dry}=0.97$ and $\mu = 0.15$, which is a common parameterization for silicate materials \citep[e.g.][]{biegert2017, joseph2001, joseph2004, vowinckel2014, vowinckel2017a, vowinckel2017b}.

%% file: 03_cohesive_force_model.tex
\subsection{Physical background}\label{sec:cohesion_physics}

To derive a cohesive force model suitable for the framework of the IBM, we start with the classical theory by Derjaguin-Landau-Verwey-Overbeek \citep[DLVO,][]{derjaguin1941,verwey1948}. This theory was derived for colloids and is based on the assumption that there are two dominant short-range forces that can be interpreted as opposing potentials surrounding particles with grain sizes in the micro- to nanometer range. On one hand, there exists a repulsive force when equally charged surfaces are in close proximity. On the other hand, as one particle causes correlations in the fluctuating polarization of a nearby particle surface an attractive force is generated. The former effect is usually called the repulsive `double-layer' (DL) force, while the latter effect is commonly referred to as van-der-Waals (vdW) force. These forces become important for gap sizes \rone{$\zeta_0 < \zeta_n < \zeta_\infty$, where $\zeta_0$ defines the microscopic size of surface asperities and 
$\zeta_\infty$ is the distance for which these forces decay to zero \cite{israelachvili1992}}. The repulsive DL force and the attractive vdW force due to polarization scale as $F_\text{rep} \propto e^{-\zeta_n}$ and  $F_\text{att} \propto \zeta_n^{-2}$, respectively. Note that the quadratic scaling of $F_\text{att}$ applies to radii much larger than $\zeta_0$ \citep{kosinski2010,breuer2015}, while linear scaling of vdW forces has been reported for cylinders of smaller size \citep{israelachvili1992}. A qualitative sketch of the DLVO theory is given in figure \ref{fig:DLVO}a. \rone{We elaborate further on the shape of this figure in Appendix \ref{app:comparison}}. The superposition of the two potentials yields a net force as a function of the gap size $\zeta_n$. \rone{Depending on properties of the particles surface charge and  the salinity of the ambient fluid, this net force} exhibits several distinct characteristics. \rone{Hence, figure \ref{fig:DLVO}a displays the characteristics of silica particles with small to medium surface charge \citep{wu2017} and rather low salinity}: (i) for larger gap sizes, there exists an outer range where attractive forces dominate over repulsive forces. (ii) Within this outer range the net attractive force displays a maximum. (iii) In the inner range with a maximum, a force barrier dominated by the repulsive double-layer force can be found; and (iv) the attractive forces diverge to infinity for gap sizes smaller than $\zeta_0$. The latter condition is equivalent to merging two separate objects into one, although evidence suggests that this condition is never reached for rough surfaces, since asperities on the particle surfaces prevent them from coming into such close contact \citep{parsons2014}. 

\setlength{\unitlength}{1cm}
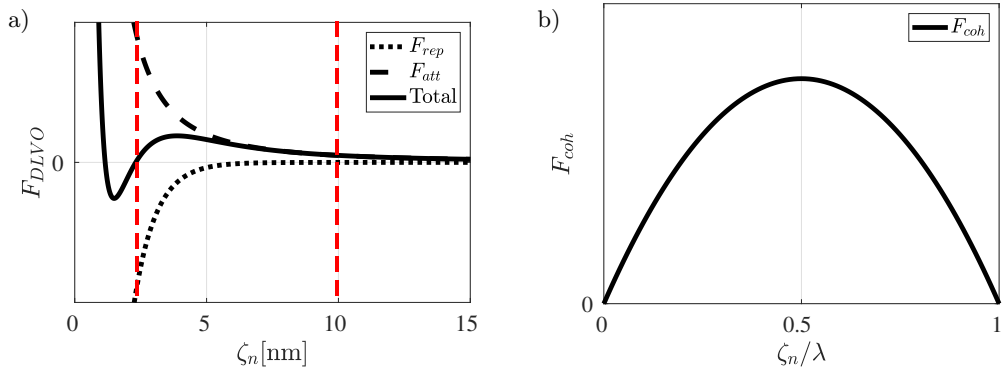
\begin{figure}
\begin{picture}(7,5)
  \put(0.0 ,   0  ){\includegraphics[width=0.5\textwidth]{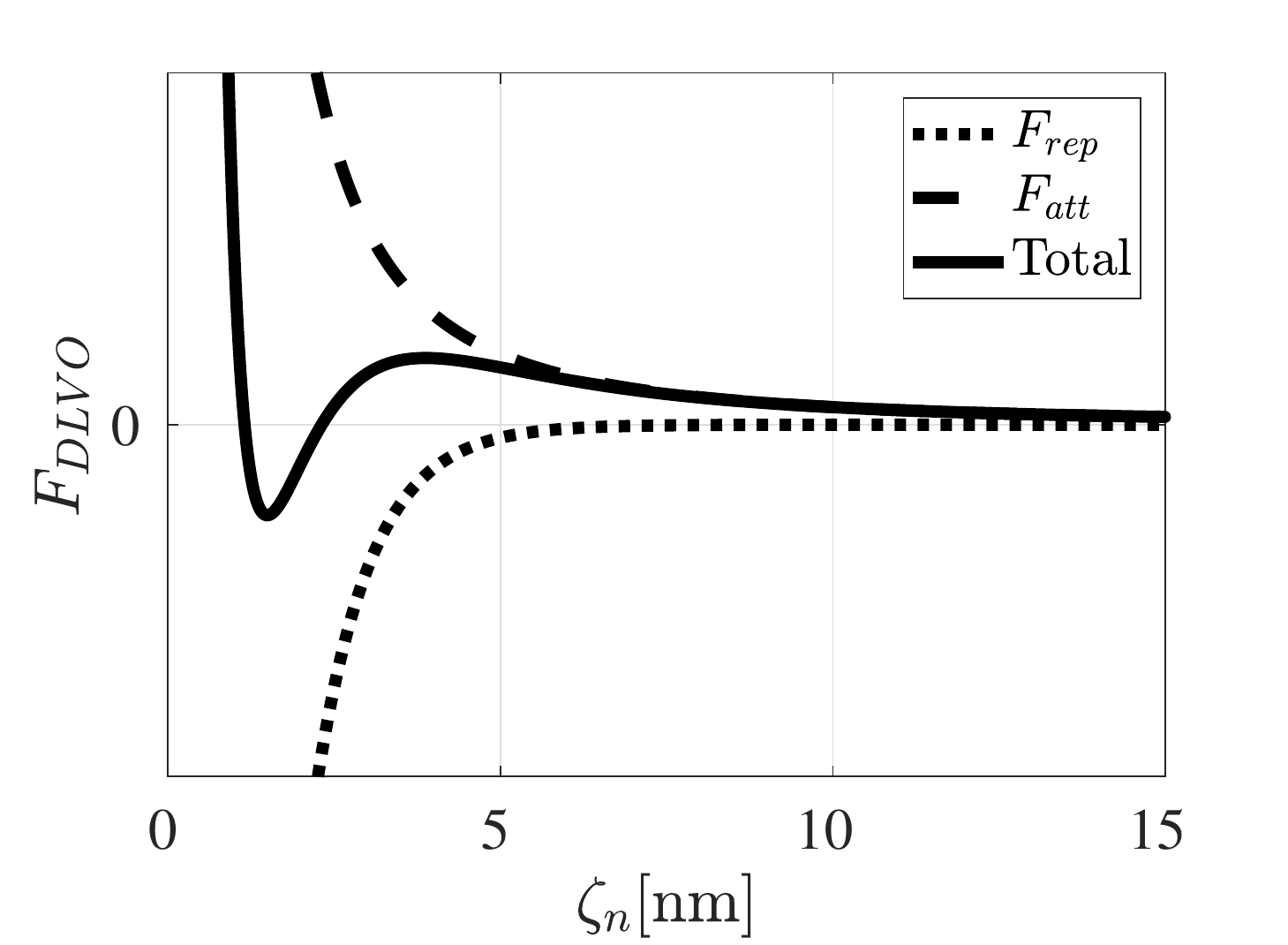}}
  \put(7.0 ,   0  ){\includegraphics[width=0.5\textwidth]{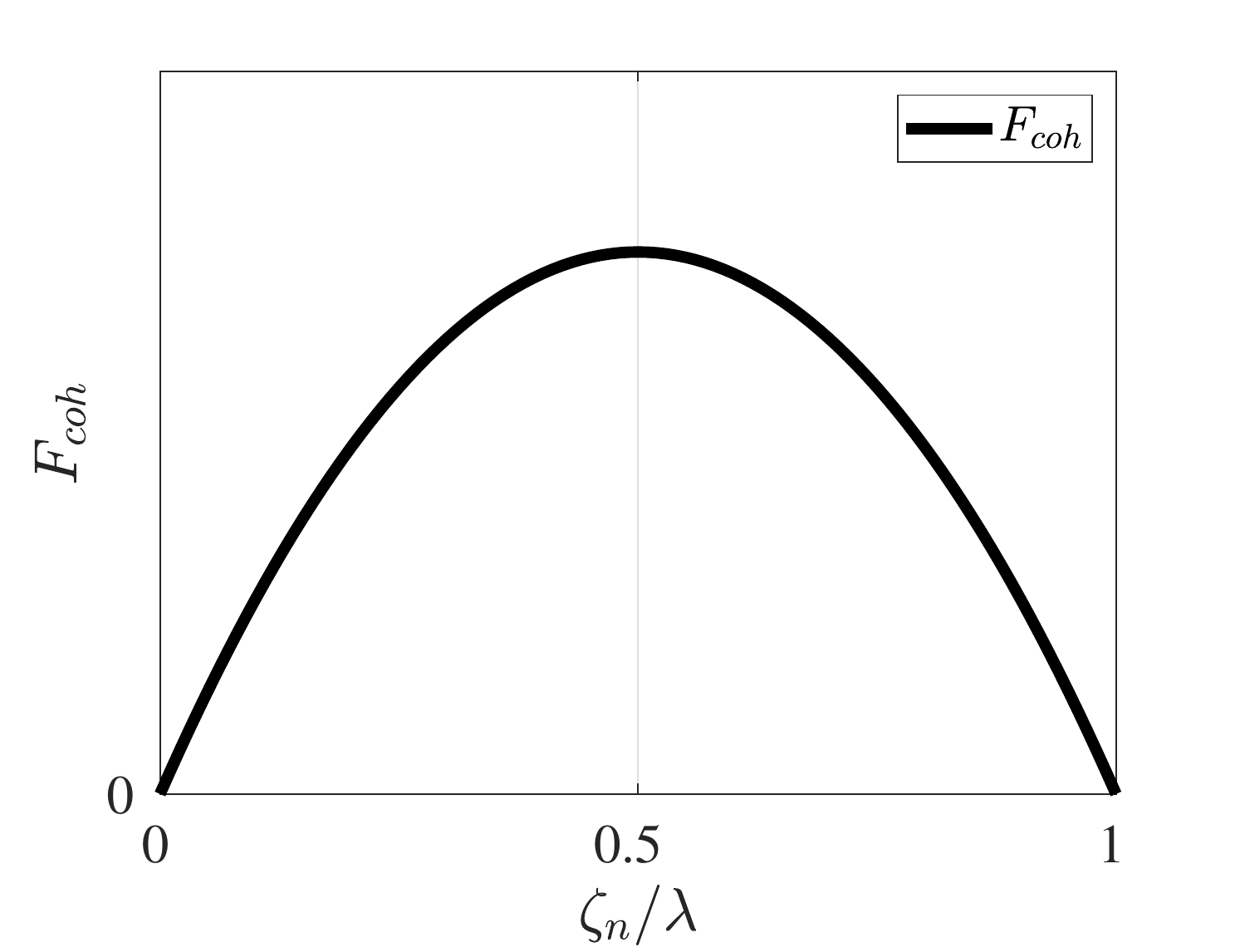}}  
 %
    \linethickness{0.5mm}
    \multiput(1.7 ,0.97)(0,0.31){12}{\color{red}{\line(0,1){0.2}}}
    \multiput(4.35,0.97)(0,0.31){12}{\color{red}{\line(0,1){0.2}}}    
    \put( 0,  4.5)     {a) }
    \put( 7.0,4.5)     {b) }    
\end{picture}
        \caption{Schematic of short range forces vs. gap size. (a) force profiles according to the DLVO theory. (b) the model ansatz given by \eqref{eq:ansatz} The red vertical lines in (a) indicate the range considered by the model ansatz shown in (b). }
    \label{fig:DLVO}
\end{figure}

The DLVO theory holds only for gap sizes in the nanometer range. Since we cannot resolve this length scale in simulations involving hundreds of micrometer-size particles, we instead employ a simplified algebraic expression for the vdW forces that reproduces its integral properties. The most common expression for vdW forces scales linearly with the particle diameter, as reviewed by \cite{visser1989} and \cite{israelachvili1992}
 \begin{equation}\label{eq:classic_VdW}
  \textbf{F}_\text{vdW} = \frac{A_H R_\text{eff}}{6 \zeta_0^2} \textbf{n} \qquad.
 \end{equation}
Due to its simplicity and ease of implementation and interpretation, this scaling assumption has been popular in Discrete Element Methods (DEM) and point-particle approaches, as well as for experimental analysis \cite[e.g.][]{pandit2005,ye2004,breuer2015,righetti2007}. However, such methods do not resolve the gap size and cohesive forces effectively act only when particles come into contact. This approach is hence equivalent to lowering the restitution coefficient of the inelastic collision, in line with the underlying hard-sphere collision model. The hard-sphere model modifies the velocity right after the impact as $u_\text{out} = -e \, u_\text{in}$, where $e$ is the restitution coefficient of the collision and $u_\text{in}$ denotes the normal component of the impact velocity. Hence, accounting for cohesive forces using the hard-sphere model involves manipulating $u_\text{out}$ and introducing thresholds for particle escape, but it prevents quantifying the intergranular stresses or work required for floc break-up. 
 
Recently, various models for cohesive collisions were tested in the framework of a DEM by \cite{thornton2017}. The authors report that piecewise cohesive force models that distinguish between approach and rebound give rise to unphysical properties. In particular, this approach leads to an overestimation of the tensile forces during the rebound process. As a consequence, flocculation may be overestimated because of unphysical sticking conditions at high impact velocities. Hence, studies treating the particles as mass points that collide according to a hard-sphere model are difficult to interpret physically, because this approach does not capture and quantify the forces associated with cohesive particle-particle interaction. Consequently, it becomes impossible to distinguish between the different contributions to particle-particle interactions, such as repulsive collision, tangential friction and lubrication forces as outlined in \S \ref{sec:collision}, as repulsive collision forces and cohesive forces are lumped into the inelastic restitution coefficient.

\cite{derksen2014} carried out particle-resolving simulations using a finite-size square-well potential to account for flocculation. This square-well potential considers two particles attached as soon as $u_\text{in}$ falls below a critical threshold and, vice-versa, to break up if the escape velocity lies above a critical threshold. For the case of floc break up, this model converts the effect of cohesion from potential to kinetic energy. This treatment represents an improvement over point-particle approaches, although it does not allow for the space-resolved computation of the cohesive forces and stresses acting on flocculated particles. Similarly, \cite{gu2016} proposed a cohesive force model that scales inversely with $\zeta_n$. We tested this approach and found it to be prone to numerical instabilities when dealing with rather stiff particles, as it introduces large attractive forces for small gap sizes that are discontinuously shut off at a minimal gap size.

In order to computationally simulate realistic cohesive sediment dynamics, we aim to resolve in space and time the following three phases of particle-particle interaction, cf. figure \ref{fig:binary_contact_scenario}: (a) particle approach/flocculation, (b) capture/steady state contact, and (c) separation in the presence of external forces. In doing so, we will obtain detailed information on the work performed by the inter-particle forces. We furthermore aim for a computational model that recovers the original DEM scheme proposed by \cite{biegert2017} for cohesionless grains. These goals will be achieved by the approach to be described in the following.
 
 \setlength{\unitlength}{1cm}
\begin{figure}
\begin{picture}(7,5)
  \put(5.5 ,   0  ){\includegraphics[width=0.18\textwidth]{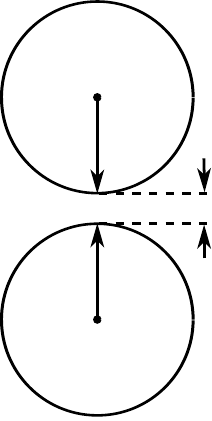}}

    \put( 6.7,  3.2 )     {$R_p$ }
    \put( 6.7,  1.5 )     {$R_q$ }    
    \put( 8.0,  2.35)     {$\zeta_n$ }     
\end{picture}
  \caption{Computational scenario for the binary interaction of two cohesive particles.}
    \label{fig:binary_contact_scenario}
\end{figure}
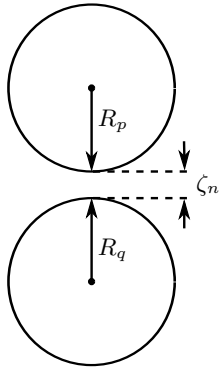    
%

\subsection{Cohesive force model}\label{sec:cohesive_force_model}
To exploit the advantages offered by the soft-sphere model of \cite{biegert2017} for grain-resolving simulations, we develop an approach that is consistent with the DLVO theory as sketched in figure \ref{fig:DLVO}a, with an attractive inter-particle force within the interval $2 \text{nm} \leq \zeta_n \leq 10 \text{nm}$ that has a local maximum at $\zeta_n \approx 4 \text{nm}$. Note that we do not wish to resolve the layer of $\zeta_n < 2\text{nm}$, but instead consider this to be part of the surface roughness. Ideally, the cohesive forces would decay to zero for $\zeta_n = 0$ as the repulsive forces are already accounted for through \eqref{eq:acm_force}. This can be accomplished by the ansatz of a parabolic spring force with the following properties: (i) it decays to zero as the gap size goes to zero, (ii) it has a maximum at a gap width orders of magnitude smaller than the particle diameter, and (iii) it decays to zero for larger gap sizes, without any discontinuous jumps, cf. figure \ref{fig:DLVO}b.  These characteristics are incorporated by the mathematically simple model
 \begin{equation}\label{eq:ansatz}
 \textbf{F}_\text{coh} = 
  \begin{cases}
  -k_\text{coh}(\zeta_n^2 - \zeta_n \lambda) \textbf{n}  & 0 < \zeta_n \leq \lambda \\
    0 & \text{otherwise} \qquad ,
  \end{cases} 
 \end{equation}
where $k_\text{coh}$ denotes the stiffness constant and $\lambda$ represents the range over which the cohesive force is smeared. This length scale can be interpreted as a Debye length, which is typically on the order of several micrometers in DEM simulations \citep{mari2014}. As will be shown in \S \ref{sec:sensitivity_lambda}, the simulation results are insensitive to the exact value of $\lambda$. A reasonable choice that will allow us to resolve the cohesive force computationally, while limiting it to a range much smaller than the particle size, is $D_{50} / \lambda = 20$, where $D_{50}$ is the median grain size of a polydisperse ensemble of particles. Particle-resolving simulations typically employ a resolution of 20 grid cells of size $h$ per diameter, so that choosing $\lambda \approx h$ and utilizing the substepping routine as proposed by \cite{biegert2017} guarantees a proper resolution of cohesive effects in space and time. This modeling approach is also consistent with the experimental observations of \cite{delenne2004}, who coated rods of $D = 8 $mm in diameter with epoxy resin to glue them together. These rods where then put under tension to determine the cohesive forces. It was found in this study that the cohesive force increases with gap size $\zeta_n$ to a maximum at $D / \zeta_n \approx 80$. For larger gap sizes, the force decreases and eventually, the rods detach at $D / \zeta_n \approx 40$. For the present study, we have chosen the latter value as a reference for the largest particles of the considered polydisperse particle mixtures.
 
We determine the stiffness $k_\text{coh}$ of the model by preserving the energy contained in the vdW forces
\begin{subequations}
  \begin{equation}
    E_\text{coh} = E_\text{vdW} \ .
  \end{equation}
According to the DLVO theory, the vdW forces \eqref{eq:classic_VdW} are defined within the interval $\zeta_n = [\zeta_0, \infty]$, where the lower boundary $\zeta_0=0.2 \text{nm}$ is taken from \cite{israelachvili1992}. On the other hand, \eqref{eq:ansatz} is defined over the interval $\zeta_n = [0,\lambda]$. Taking the endpoints of the respective intervals as the integration limits yields
 \begin{alignat}{4}\label{eq:stiffness}
  \int_0^\lambda-k_\text{coh}(\zeta_n^2-\zeta_n \lambda) \text{d} \zeta_n          &= \int_{\zeta_0}^\infty \frac{A_HR_\text{eff}}{6 \zeta_n^2} \text{d} \zeta_n \ , \\
\text{so that}  \qquad   k_\text{coh}                                                                   &= \frac{A_H R_\text{eff}}{\zeta_0 \lambda^3 } \qquad .
 \end{alignat}
\end{subequations}
\rone{Note that we have replaced $R_p$ by $R_\text{eff}$ on the right-hand side of \eqref{eq:stiffness} to account for polydisperse particle sizes.} Substituting this expression for $k_\text{coh}$ into \eqref{eq:ansatz} provides the final expression for the dimensional cohesive force model
\begin{equation}\label{eq:cohesive_forces}
\textbf{F}_\text{coh} = 
  \begin{cases}
  -\frac{A_H R_\text{eff}}{\zeta_0 \lambda^3 }(\zeta_n^2 - \zeta_n \lambda) \textbf{n}  & 0 < \zeta_n \leq \lambda \\
    0 & \text{otherwise}
  \end{cases} 
\end{equation}

This form enables us to account for cohesive forces in our current IBM-DEM framework via an additional force term in the collision model \eqref{eq:particle_forces}
\begin{equation}\label{eq:particle_forces_cohesive}
	\textbf{F}_{c,p} = \sum_{q,\:q \neq p}^{N_p} \left( \textbf{F}_{l,pq} + \textbf{F}_{n,pq} + \textbf{F}_{t,pq} + \textbf{F}_{\text{coh},pq}\right) + \textbf{F}_{l,pw} + \textbf{F}_{n,pw} + \textbf{F}_{t,pw} + \textbf{F}_{\text{coh},pw}\hspace{0.5cm}.
\end{equation}
Equation \eqref{eq:particle_forces_cohesive} will be the basis of the simulations to be discussed in \S \ref{sec:binary_interaction} and \S \ref{sec:settling}. \rone{Note that this approach treats particles individually and is not modified for the interaction of clusters. Furthermore, no changes to the computation of $\textbf{F}_{h,p}$ and $\textbf{F}_{h,g}$ are necessary, as these are a direct result of the IBM.} The main advantage of this approach lies in its ability to resolve the particle bonding process in space and time as cohesive forces act over a finite size shell surrounding each particle. At the same time, it retains the distinction between the individual inter-particle force components via \eqref{eq:particle_forces_cohesive}, which allows for an in-depth analysis of the different effects governing the particle-particle interaction. Note that \eqref{eq:eom_trans} and \eqref{eq:particle_forces_cohesive} together take a form that resembles the minimal flocculation model proposed by \cite{vicsek1995}, which was developed for self-propelled organisms. While our particles are passive, their weight $\textbf{F}_{g}$ provides a preferred direction of motion \citep{toner2005}, the collision force $\textbf{F}_{c}$ aligns their motion through the competition of repulsion and cohesion, and the hydrodynamic force $\textbf{F}_{h}$ introduces a forcing that can cause particles to flocculate or to break up. 

The dimensional form \eqref{eq:cohesive_forces} still requires the proper parameterization of the Hamaker constant $A_H$, which implies all of the difficulties mentioned in the introduction. This issue will be addressed via the rescaling to be discussed in \S \ref{sec:rescale}.

\subsection{Nondimensionalization of the cohesive force model}\label{sec:rescale}

For the purpose of conducting numerical simulations, we wish to capture the effects of cohesive forces by means of a nondimensional similarity parameter. To this end, we render the Navier-Stokes equation \eqref{eq:navier_stokes} and the particle equation of motion \eqref{eq:eom_trans} dimensionless in Appendix \ref{app:eom}. For the Navier-Stokes equation, the only dimensionless parameter to appear is the Reynolds number \citep{biegert2017peps}. For the particle equation of motion \eqref{eq:eom_trans}, the cohesive and lubrication forces combined can be viewed as a spring-dashpot system for two interacting particles. 

As derived in Appendix \ref{app:eom}, by choosing the buoyancy velocity $u_s = \sqrt{g' D_{50}}$ (Appendix \ref{app:char_vel}), the characteristic time scale $\tau_s = D_{50} / u_s$ and the characteristic mass $m_{50} = \rho_f \pi D_{50}^3/6$, the characteristic force scale for particles settling under gravity in an otherwise quiescent fluid becomes the specific weight  $m_{50} g'$. Here, $D_{50}$ is the median diameter of an ensemble of polydisperse particles, $g' = (\rho_p - \rho_f) g/ \rho_f$ denotes the reduced gravity, and $g$ represents the gravitational acceleration. To write the algebraic expression for cohesive forces \eqref{eq:cohesive_forces} in dimensionless form, we define a cohesive number as
\begin{equation}\label{eq:cohesive_number}
 \text{Co} = \frac{\text{max}(|| \textbf{F}_{\text{coh},50} ||)}{m_{50} g'} \qquad .
\end{equation}
It represents the ratio of the cohesive force maximum for particles of diameter $D_{50}$ to the characteristic gravitational force scale of the problem. A similar characteristic number was used by \cite{sun2018} in the framework of a point-particle approach. A complete derivation for the origin of $\text{Co}$ within the present numerical framework is provided in appendix \ref{app:eom}.

By design, \eqref{eq:cohesive_forces} has its maximum at $\zeta_n = \frac{1}{2} \lambda$, so that we immediately obtain for $R_\text{eff} = D_{50} / 2$
\begin{equation}\label{eq:cohesive_forces_1}
\text{max}(|| \textbf{F}_{\text{coh},50} ||) = -\frac{A_H}{\zeta_0} \frac{D_{50}}{2 \lambda^3} \left(\frac{\lambda^2}{4} - \frac{\lambda^2}{2}\right)
= \frac{A_H}{\zeta_0} \frac{D_{50}}{8 \lambda}\qquad . 
\end{equation}
After specifying $\text{max}(|| \textbf{F}_{\text{coh},50} ||)$ for a given problem, we combine \eqref{eq:cohesive_forces_1} with \eqref{eq:cohesive_forces} to obtain the cohesive force as
\begin{equation}\label{eq:cohesive_forces_dimensional}
 \textbf{F}_\text{coh} = - \frac{8 \text{max}(|| \textbf{F}_{\text{coh},50} ||)}{D_{50}} \frac{R_\text{eff}}{\lambda^2} (\zeta_n^2 - \lambda\zeta_n) \textbf{n}\qquad .
\end{equation}
Due to the smearing of the cohesive effects over the range $\lambda$, the model now scales with $\textbf{F}_\text{coh}\propto \lambda^{-2}$ rather than $\textbf{F}_\text{coh}\propto \zeta_0^{-2}$. It is important to note, however, that this quantity does not reflect a tunable parameter but a constant property chosen in accordance with the length scales of the physical problem. 

It is then convenient to define dimensionless quantities (denoted by tilde) as
\begin{subequations}\label{eq:nondim}
\begin{equation}\label{eq:nondim_Fcoh}
  \textbf{F}_{\text{coh}} =  m_{50} g' \,  \tilde{\textbf{F}}_{\text{coh}}                                       \qquad ,
\end{equation}
\begin{equation}\label{eq:nondim_R}
  R_\text{eff} = D_{50} \tilde{R}_\text{eff}                                         \qquad ,
\end{equation}
\begin{equation}\label{eq:nondim_lambda}
  \lambda = D_{50} \tilde{\lambda}             \qquad ,
\end{equation}
\begin{equation}\label{eq:nondim_zeta}
  \zeta_n = D_{50} \tilde{\zeta}_n                             \qquad ,
\end{equation}
\end{subequations}
in order to obtain the set of dimensionless equations that will be solved numerically (Appendix \ref{app:eom}). By combining \eqref{eq:cohesive_forces_dimensional} with \eqref{eq:nondim} and normalizing with the gravitational scale $m_{50} g'$ we obtain
\begin{equation}\label{eq:cohesive_forces_dimensionless}
   \tilde{\textbf{F}}_\text{coh} =
  \begin{cases}
  - \text{Co} \, \frac{8 \, \tilde{R}_\text{eff}}{\tilde{\lambda}^2}(\tilde{\zeta}_n^2 - \tilde{\zeta}_n \tilde{\lambda}) \textbf{n}   & 0 < \zeta_n \leq \lambda \\
    0 & \text{otherwise}
  \end{cases} 
\end{equation}
The stiffness of our cohesive force model thus becomes $\tilde{k}_\text{coh} = 8 \text{Co} \tilde{R}_\text{eff} / \tilde{\lambda}^2$, so that the cohesive forces for a given physical system scale linearly with the cohesive number and the effective radius of the two colliding particles, which is consistent with the considerations of \cite{visser1989}, \cite{lick2004} and \cite{righetti2007}. \rone{The characteristics are meant to represent rough macroscopic particles, i.e. $D_p > 2\mu$m in saline water. A comprehensive translation of our modeling approach to various physical systems is given in Appendix \ref{app:comparison}.}
 \begin{table}
   \begin{center}
 \def~{\hphantom{0}}
   \begin{tabular}{c| c | c | c| c| c| c}
       $L_x/D_{50} \times L_y/D_{50} \times L_z/D_{50}$ & $\rho_p / \rho_f$ & $s / D_{50}$ & $\max\{D\}/{D_{50}}$ & $\min\{D\}/{D_{50}}$ &$\Rey = \frac{\sqrt{g'}D_{50}^{3/2}}{\nu_f}$  & $N_p$    \\
       \hline                                                                  
       $13.1 \times 40.0 \times 13.1$                  & 2.6               & 0.365        & 2.4           & 0.6           &1.35                                          & 1261     \\
   \end{tabular}
   \caption{Parameters for simulations of a large ensemble, where $L_x$, $L_y$, $L_z$ indicate the domain size, and $s$ represents the standard deviation of the grain size.}
   \label{tab:examples}
   \end{center}
 \end{table}

To rewrite the unresolved lubrication forces in dimensionless form, we substitute dimensional quantities in \eqref{eq:lubrication} by 
\begin{subequations}\label{eq:nondim_lub}
\begin{equation}\label{eq:nondim_Flub}
  \textbf{F}_{\text{lub}} =  m_{50} g' \,  \tilde{\textbf{F}}_{\text{lub}} = \frac{\rho_f \pi D_{50}^3}{6} \,  g'\, \tilde{\textbf{F}}_{\text{lub}}   
\end{equation}
and
\begin{equation}\label{eq:nondim_gn}
  \textbf{g}_n = u_s \tilde{\textbf{g}}_n = \sqrt{g' D_{50}}  \tilde{\textbf{g}}_n \qquad .
\end{equation}
\end{subequations}
Combining \eqref{eq:lubrication} with \eqref{eq:nondim_lub} then yields  
\begin{subequations}\label{eq:nondim_lubrication}
 \begin{alignat}{3}
  \tilde{\textbf{F}}_{\text{lub}} & = -36  \frac{\nu_f}{\sqrt{g' D_{50}} D_{50}}\frac{\tilde{R}_\text{eff}^2 \tilde{\textbf{g}}_n}{\max(\tilde{\zeta}_n, \tilde{\zeta}_\text{min})}                                               
   \end{alignat}
\end{subequations}
so that we obtain the dimensionless form 
\begin{equation}
   \tilde{\textbf{F}}_{\text{lub}} =   
     \begin{cases}
 -\frac{36}{\Rey} \frac{\tilde{R}_\text{eff}^2 \tilde{\textbf{g}}_n}{\max(\tilde{\zeta}_n, \tilde{\zeta}_\text{min})}   & 0 < \zeta_n \leq 2 h \\
    0 & \text{otherwise} \qquad .
  \end{cases}   
\end{equation}

Hence lubrication forces scale with the inverse of the particle Reynolds number $\Rey=D_{50} u_s / \nu_f$, while cohesive forces scale with the cohesive number $\text{Co} = \text{max}(|| \textbf{F}_{\text{coh},50} ||)/ (m_{50} g'). $ By quantifying these two dimensionless similarity parameters, the physical system is thus fully specified.

\subsection{Cohesive force model validation}\label{sec:fcrit}

The key idea behind the cohesive number introduced in \eqref{eq:cohesive_number} in \S \ref{sec:rescale} is to define a critical threshold of $\text{Co}=1$, for which the cohesive forces balance the specific weight of a particle of size $D_{50}$. To test and verify this behavior, we consider a simple test case of two particles in quiescent fluid. Particle $p$ is held fixed, and particle $q$ is placed right below it, cf. figure \ref{fig:binary_contact_scenario}. Particle $q$ wants to settle as a result of gravity, whereas the cohesive force acts to keep it attached to particle $p$. The governing Reynolds number is based on the quantities introduced in \S \ref{sec:rescale}, which yields $\Rey = u_s D / \nu_f$. We chose the Reynolds number for the test case from the experimental observations of \cite{lick2004}, who reported that cohesive forces start to alter the erosion behavior of silicate particles with a density of $\rho_p = 2,650 \, \text{kg/m}^3$ and a diameter in the range of $140 \, \mu \text{m} \leq D_p  \leq  390 \, \mu \text{m}$ which is submerged in water with a density and kinematic viscosity of $\rho_f = 1,000 \, \text{kg/m}^3$ and $\nu_f = 10^{-6} \text{m}^2 / \text{s}$, respectively. Choosing a representative value of $D_p  =242 \, \mu \text{m}$ yields $\Rey = u_s D / \nu_f = 15.1$. Since only two particles are involved, we define $D_{50}=\frac{1}{2}(D_p + D_q)$. Initially, particle $q$ is at rest and the size of the vertical gap with respect to the fixed particle $p$ is set to $\zeta_n=\frac{1}{2}\lambda$, where we chose $D_p / \lambda = 20$. Particle $q$ is then free to move. Note that for this case of particle $q$ interacting with the fixed particle $p$, the effective radius becomes $R_\text{eff} = R_q$.  The median particle size $D_{50}$ is discretized by 20 grid cells per diameter. In order to investigate whether or not particle $q$ will detach from particle $p$, a relatively small computational domain of $L_x \times L_y \times L_z = 2.5 D_{50} \times 5 D_{50} \times 2.5 D_{50}$ suffices, with gravity acting in the  negative$y$-direction. No-slip walls are imposed in the $y$-direction, whereas the $x$- and $z$-directions are being treated as periodic.

For $\text{Co} < 1$, the weight of particle $q$ is larger than the maximum of the cohesive force at $\zeta_n = \frac{1}{2}\lambda$, so that we expect particle $q$ to detach from particle $p$. This is confirmed by figure \ref{fig:steady_contact}a, which shows the gap size versus nondimensional time $t/\tau_s$. In addition, the detachment happens more slowly as the cohesive number approaches its critical value. For the critical cohesive number $\text{Co} = 1$, the particles remain stationary at a constant gap size $\zeta_n = \frac{1}{2}\lambda$. Increasing the cohesive number even further causes particle $q$ to move closer towards particle $p$, as the maximum of the attractive forces increases. Since the cohesive force approaches zero as the gap decreases, particle $q$ will find an equilibrium position within the interval $0 \leq \zeta_n \leq \frac{1}{2}\lambda$ at which the cohesive force is balanced by the weight of the particle. Also note that even if we do not explicitly resolve the nanoscale over which cohesive forces typically act, the smearing of the cohesive potential over $\lambda$ bonds particles at a constant gap size of $\zeta_n \leq D_p / 40$, which is sufficiently small to reproduce physically realistic behavior. Particles can remain a finite distance apart during flocculation under the influence of divergent forces \citep{israelachvili1992, thornton2017}, and they can experience friction while in direct contact during collisions.

\setlength{\unitlength}{1cm}
\begin{figure}
\begin{picture}(7,3.9)
  \put(0.0 ,   0  ){\includegraphics[width=0.47\textwidth]{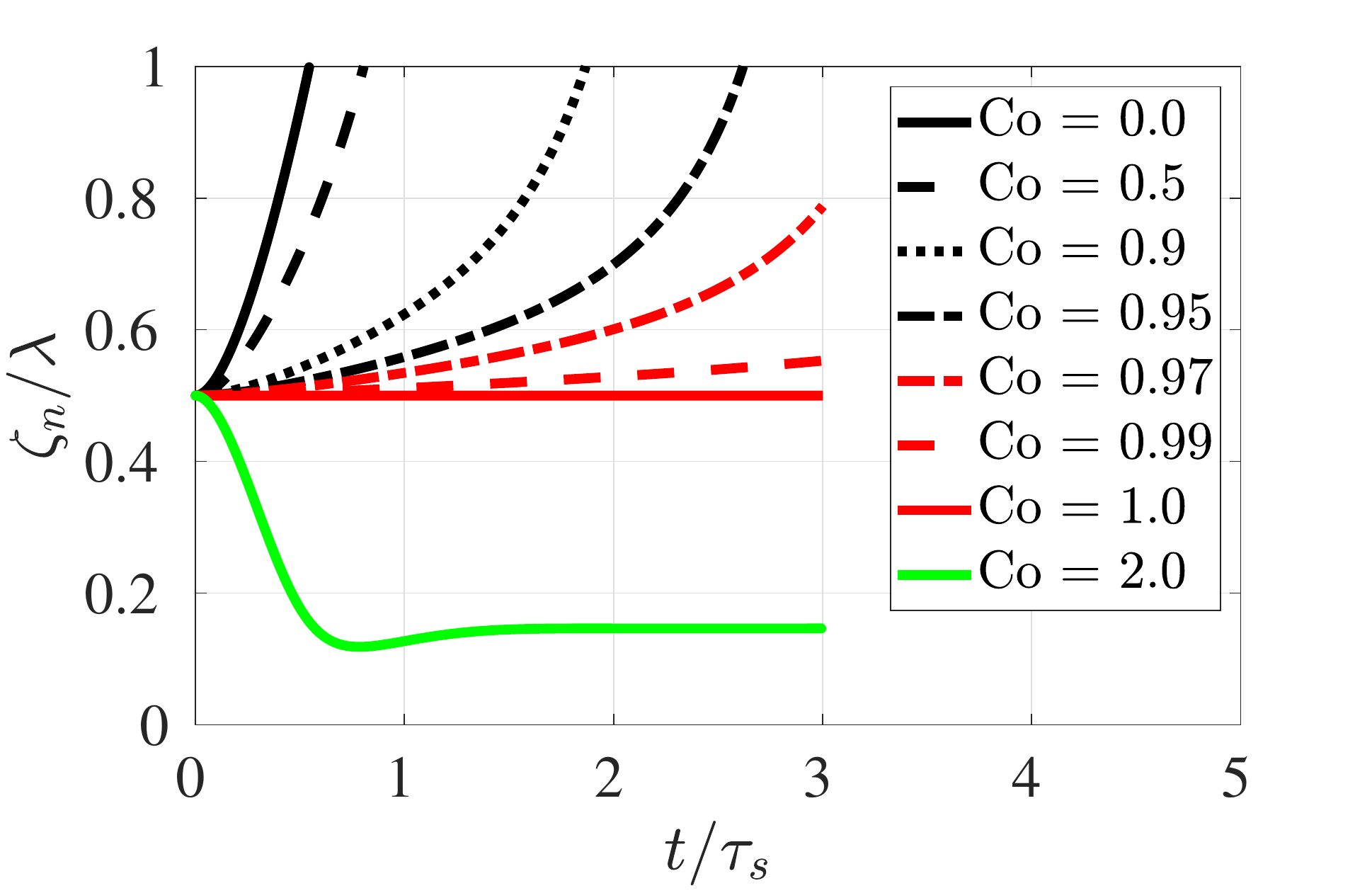}}
  \put(6.5 ,   0  ){\includegraphics[width=0.53\textwidth]{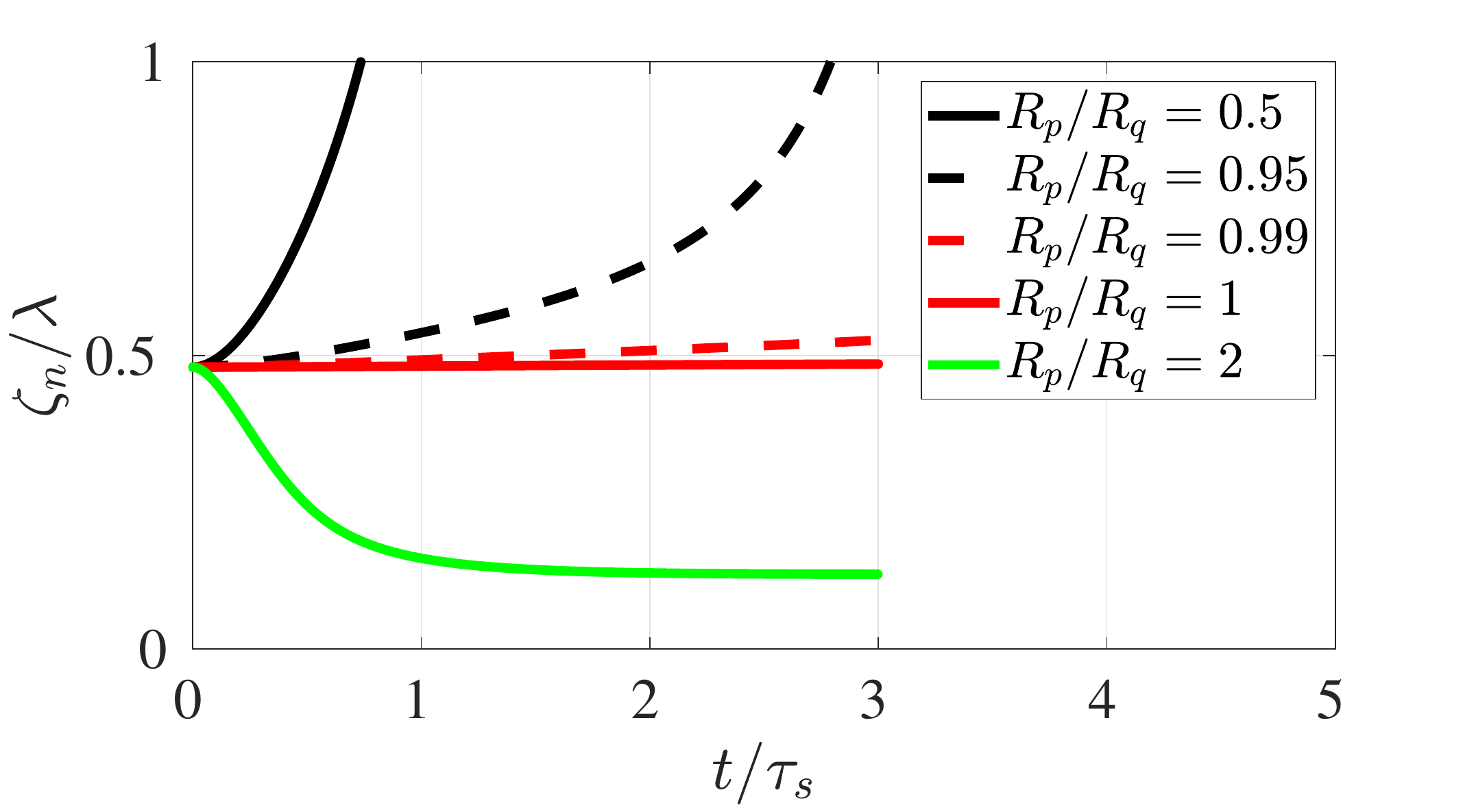}}

    \put( 0,  3.5)     {a) }
    \put( 6.5,3.5)     {b) }    
\end{picture}
        \caption{Gap size versus time: (a) different runs with varying cohesive number. (b) different runs with varying ratio of the two particle radii with $\text{Co}=1$. 
        }
    \label{fig:steady_contact}
\end{figure}

Corresponding tests can be carried out for polydisperse particles. Here we set the cohesive number to $\text{Co}=1$ while varying the ratio of the two particle radii $R_p / R_q$. The median grain size $D_{50} = \frac{1}{2}(D_p + D_q)$ serves a basis for calculating the similarity parameters $\Rey$ and $\text{Co}$. The results are shown in figure \ref{fig:steady_contact}b. If the radius ratio is smaller than unity, particle $q$ is larger than particle $p$ and its weight causes particle $q$ to detach. On the other hand, if particle $q$ is smaller than particle $p$ they stay in contact at a constant gap size $\zeta_n<\frac{1}{2}\lambda$. This test case validates the arguments underlying \eqref{eq:cohesive_forces_dimensionless} for polydisperse particles. 

The above analysis demonstrates that our computational model allows for the precise control of cohesive forces. Since \cite{biegert2017} showed that the present computational approach also yields excellent agreement with experimental data for cohesionless particles, we expect it to reproduce the settling dynamics of cohesive particles with high fidelity.

%% file: 04_drafting_kissing_tumbling.tex
%
\subsection{Drafting, Kissing, Tumbling}\label{sec:DKT}
\setlength{\unitlength}{1cm}
\begin{figure}
\begin{picture}(7,6.8)
  \put(-1.0 ,   0  ){\includegraphics[width=0.5\textwidth]{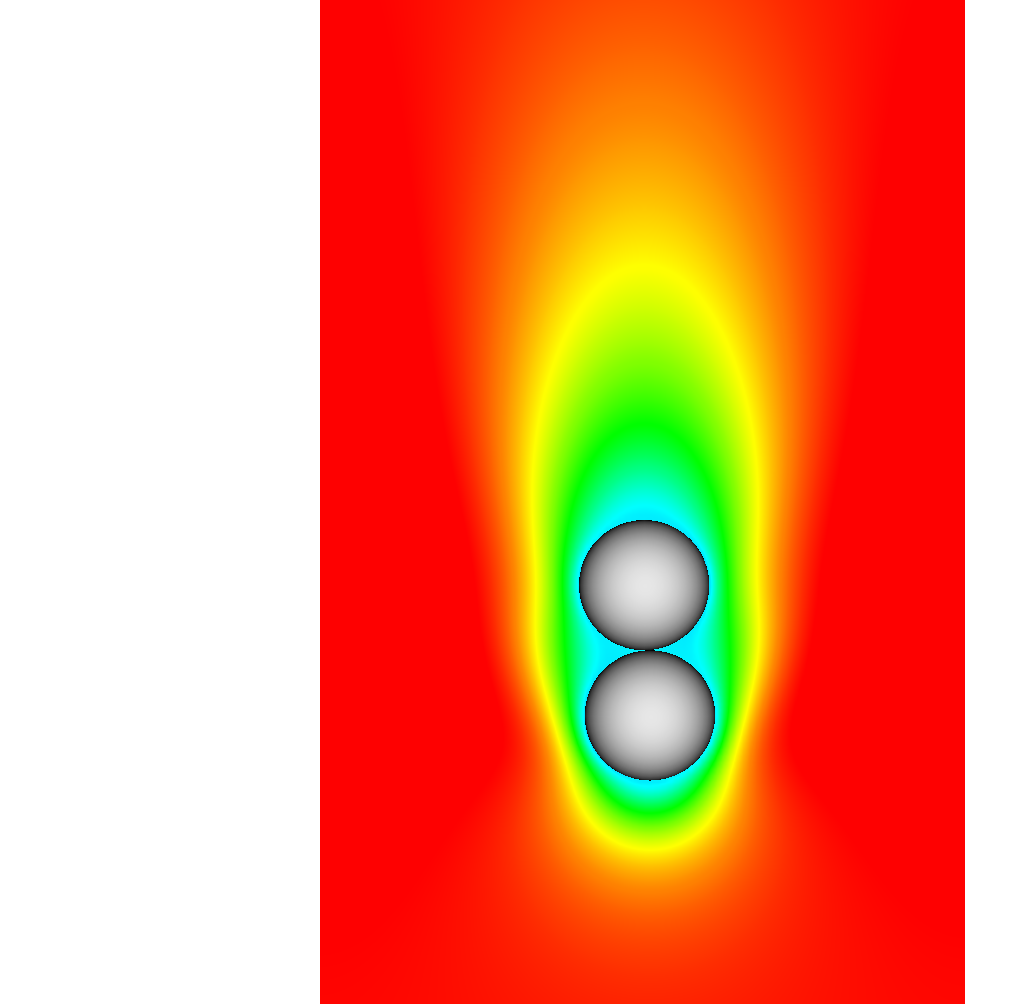}}
  \put( 6.5 ,   0  ){\includegraphics[width=0.5\textwidth]{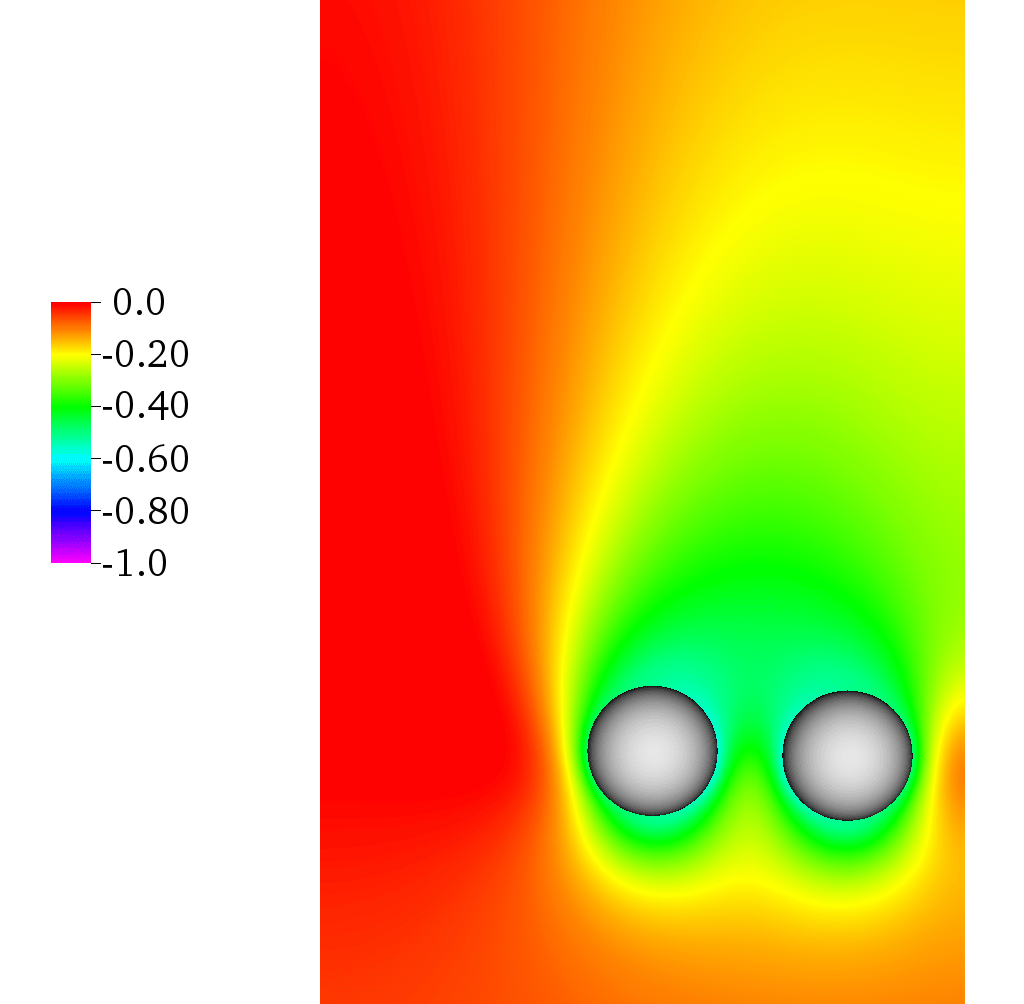}}

    \put( 0.5,6.5)     {a) }
    \put( 7.8,6.5)     {b) }    
    \put( 6.8,5.0)     {$u/u_s$ }  
    \put( 7.1,2.0)   {$\textbf{g}$ }      
    \thicklines
    \put( 6.8,2.5)     { \vector(0,-2){1}}
\end{picture}
        \caption{Cohesionless particles undergoing DKT. a) Kissing at $t / \tau_s = 10$, and b) tumbling at $t / \tau_s = 50$. Contours show the downward fluid velocity component.
        }
    \label{fig:DKT}
\end{figure}
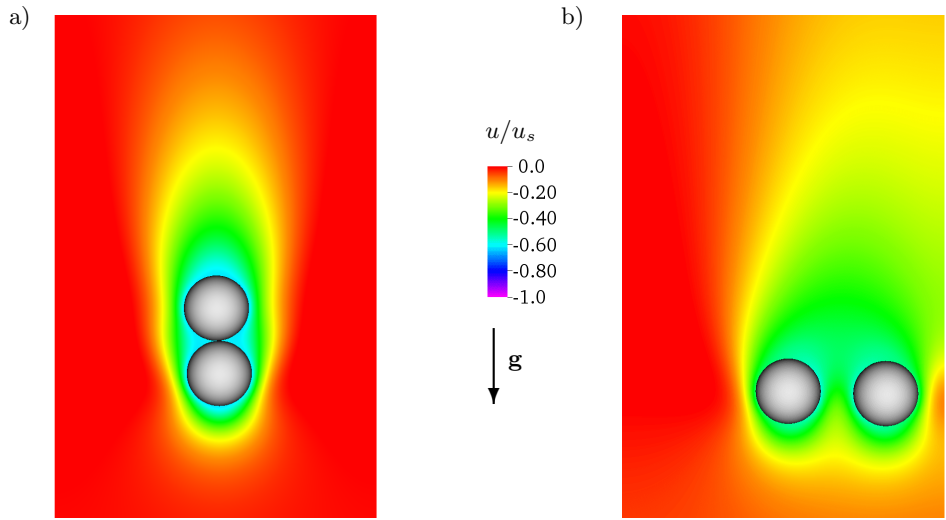
In order to assess the influence of cohesive forces under simplified conditions, we focus on the classical Drafting-Kissing-Tumbling (DKT) experiment of two particles settling under gravity, which has been explored in depth for cohesionless grains \citep[e.g.][]{fortes1987,glowinski2001}. The initial configuration is shown in figure \ref{fig:binary_contact_scenario}. \rone{Two particles with a density greater than the ambient fluid are placed} above each other in a tank of quiescent fluid, with an initial gap size that is substantially larger than the distance of the short-range lubrication and cohesive forces. As the particles are released and settle under the influence of gravity, the trailing particle is drafted by the wake of the leading particle, so that it experiences reduced drag. The two particles touch (or kiss, figure \ref{fig:DKT}a) and subsequently rearrange themselves (tumble) into a side-by-side configuration. In the absence of cohesive effects, hydrodynamic forces eventually push them apart, and they start to separate laterally (figure \ref{fig:DKT}b). 

\setlength{\unitlength}{1cm}
\begin{figure}
\begin{picture}(7,10)
  \put(0.0 ,   5.0  ){\includegraphics[width=0.5\textwidth]{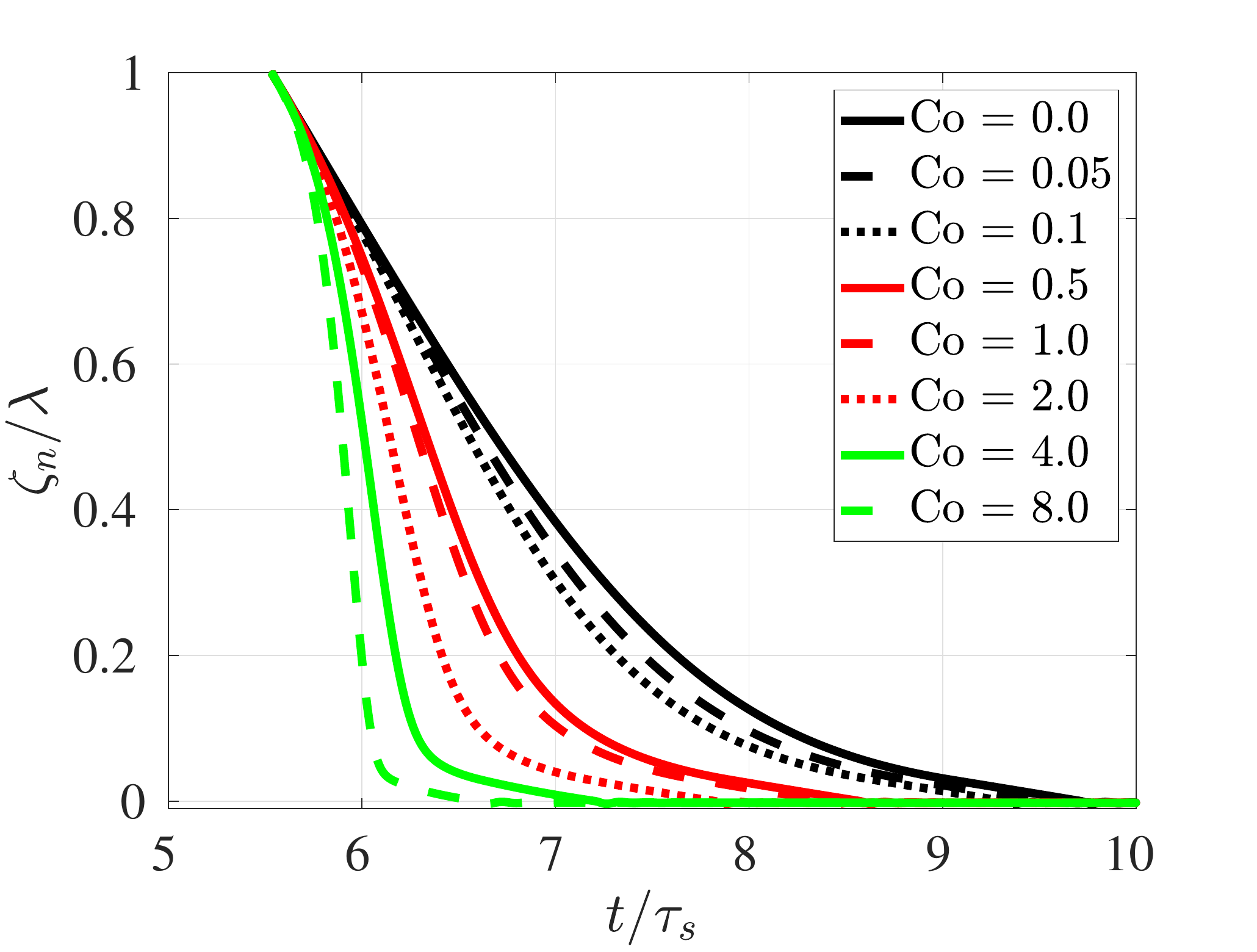}}
  \put(7.0 ,   5.0  ){\includegraphics[width=0.5\textwidth]{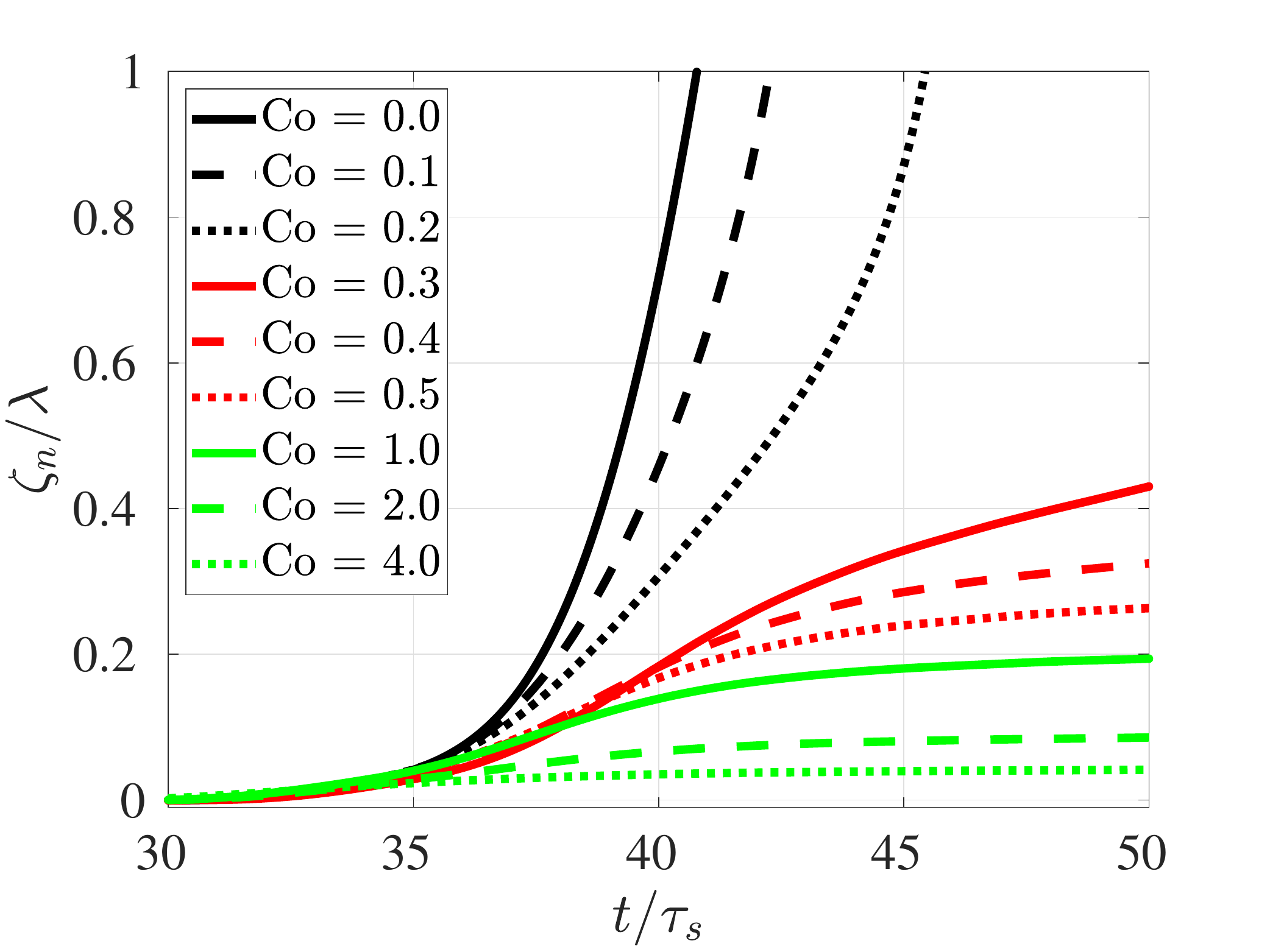}}
  \put(3.5 ,   0    ){\includegraphics[width=0.5\textwidth]{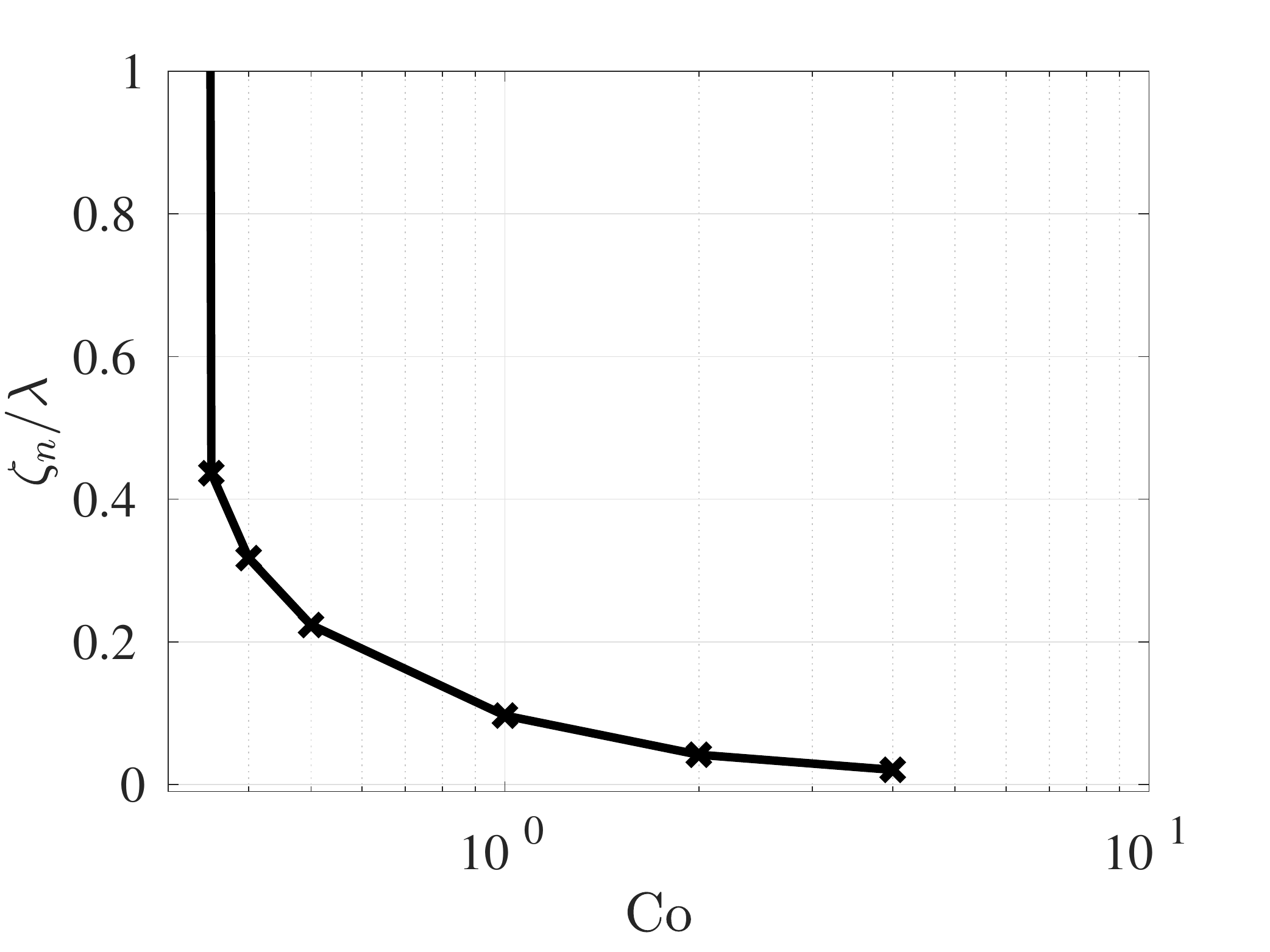}}  

    \put( 0,  9.5)     {a) }
    \put( 7.0,9.5)     {b) }    
    \put( 3.5,4.5)     {c) }        
\end{picture}
        \caption{DKT results for equal-sized spheres with different cohesive numbers showing  separation width $\zeta_n$ for (a) drafting phase, (b) tumbling phase, (c) steady-state. 
        }
    \label{fig:DKT_mono}
\end{figure}
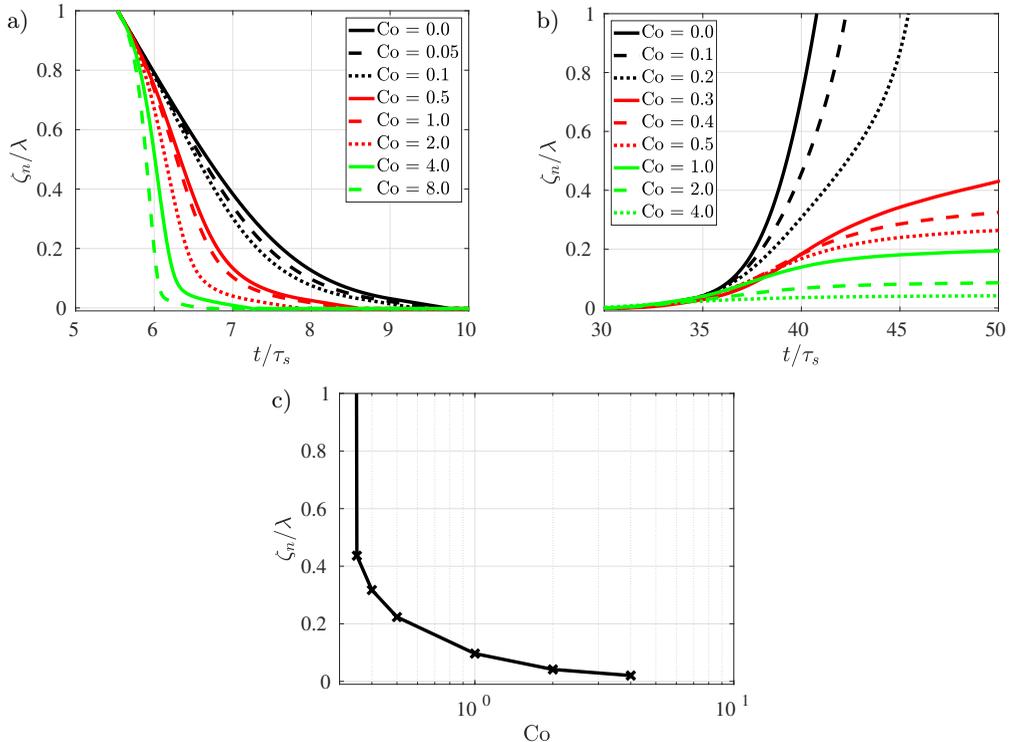

To investigate the DKT scenario for cohesive particles, we employ the same physical setup and the same numerical parameters as described in \S \ref{sec:fcrit}, i.e. $\Rey = 15.1$, $D_{p} / h = 20$, and $D_{50} / \lambda = 20$. \rone{Since we are dealing with grains that should resemble the characteristic of natural silt, the Reynolds number of our system is substantially lower compared to the studies of \cite{fortes1987} and \cite{glowinski2001}. Decreasing the Reynolds number results in slower dynamics of the interacting particles and increases the time scales to observe the DKT-motion. Hence, we choose a rather long computational domain} of $L_x \times L_y \times L_z = 120 D_{50} \times 5 D_{50} \times 5 D_{50}$, with gravity acting in the negative $x$-direction. The boundary conditions assume periodicity in $x$, and free-slip in $y$ and $z$, and everything is at rest initially. Due to the low Reynolds number, the domain length in the $x$-direction is sufficiently large for particles to establish a steady state configuration. We begin by considering two equal-sized spheres that are initially placed at $\textbf{x}_p = (x_q + D_{50} + 2h, y_q + 0.5h, z_q)^T$, which is sufficiently close to trigger the DKT behavior, but far enough apart for lubrication and cohesive forces to be unimportant initially. We remark that the simulation results do not depend on the exact initial conditions, as long as $ \zeta_n \leq D_p$, so that DKT is initiated. The initial horizontal offset of $0.5h$ in the $y$-coordinate triggers the physical instability leading to the particle rearrangement during the kissing phase.

The results for the monodisperse case are presented in figures \ref{fig:DKT_mono}a and b, which show the gap width $\zeta_n$ of the two particles as a function of time. Consistent with the classical observations by \cite{fortes1987}, for $\text{Co}=0$ the cohesionless particles approach each other (figure \ref{fig:DKT_mono}a), touch for about 20 time units and then separate in a tumbling behavior (figure \ref{fig:DKT_mono}b). Increasing the cohesive forces speeds up the drafting phase, and slows down or prevents the subsequent separation of the particles. The fact that the particles remain in steady state contact for $\text{Co} < 1$ indicates that hydrodynamic forces are not as effective in pulling them apart as gravity was in \S \ref{sec:fcrit}. Figure \ref{fig:DKT_mono}c demonstrates that already a cohesive number value of $\text{Co}=0.35$ suffices to maintain the steady-state bond between the particles.

\setlength{\unitlength}{1cm}
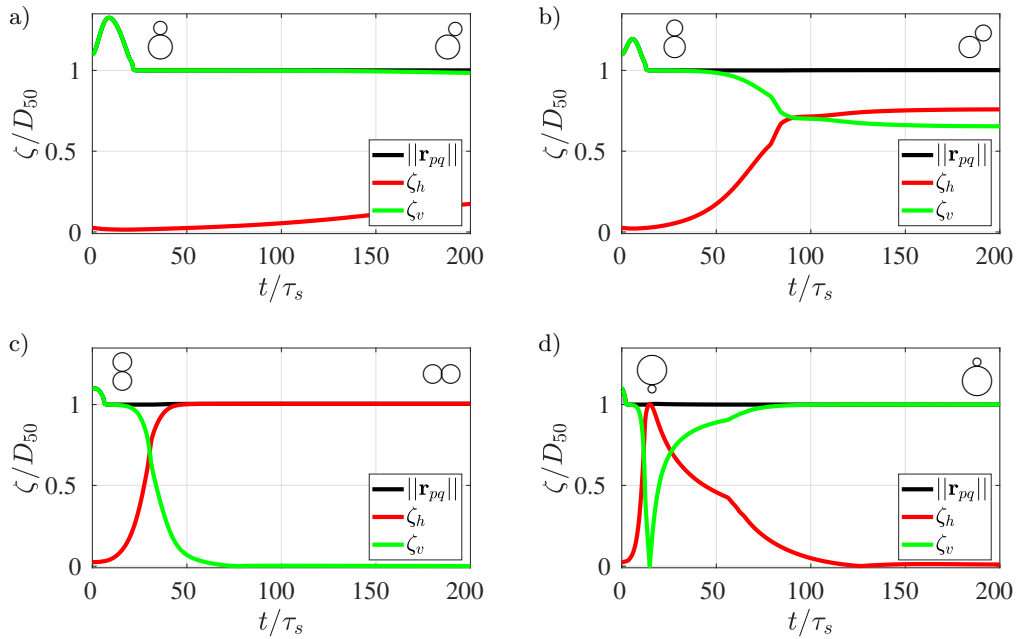
\begin{figure}
\begin{picture}(7,8.4)
  \put(0.0 ,   4.4  ){\includegraphics[width=0.5\textwidth]{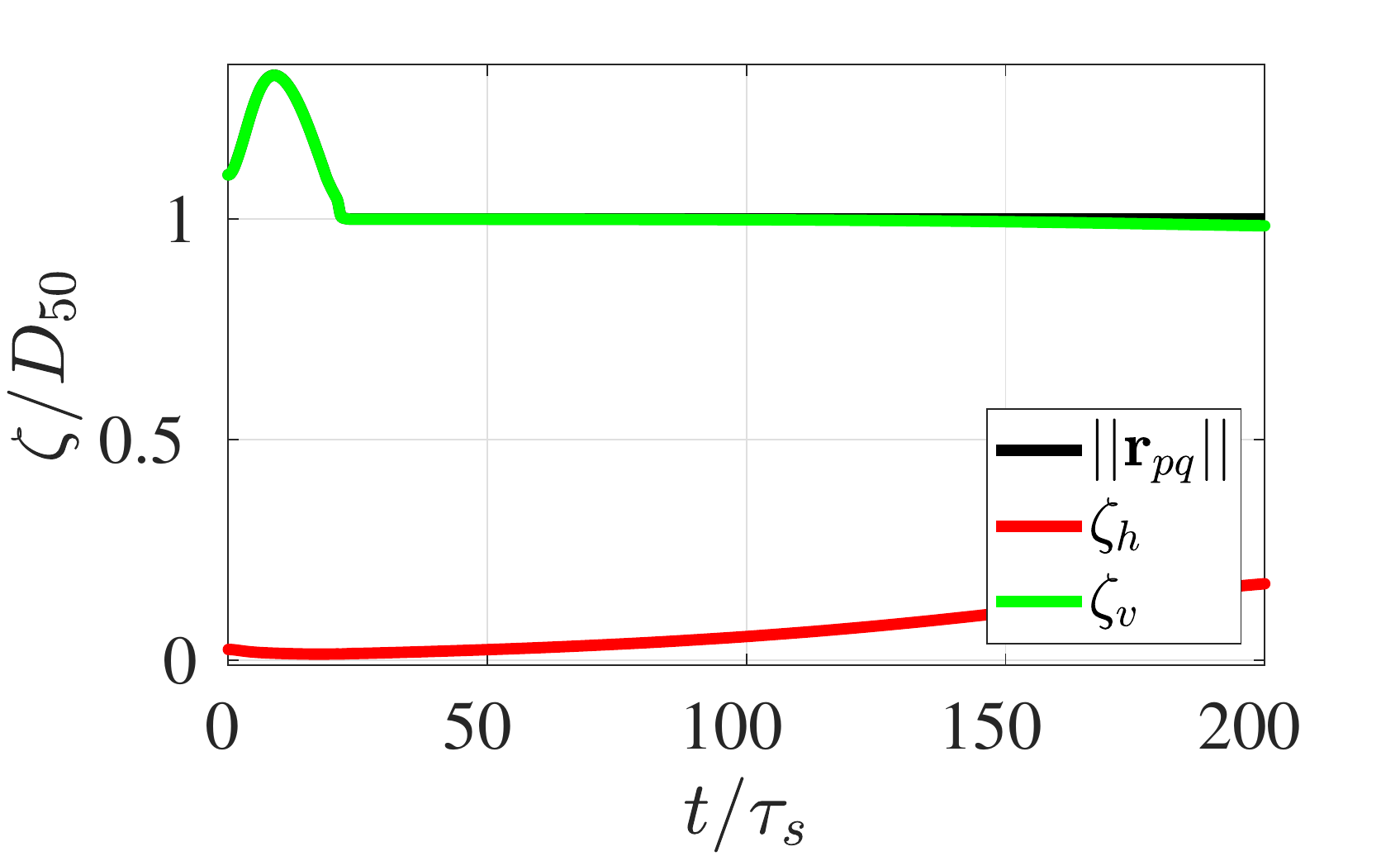}}
  \put(7.0 ,   4.4  ){\includegraphics[width=0.5\textwidth]{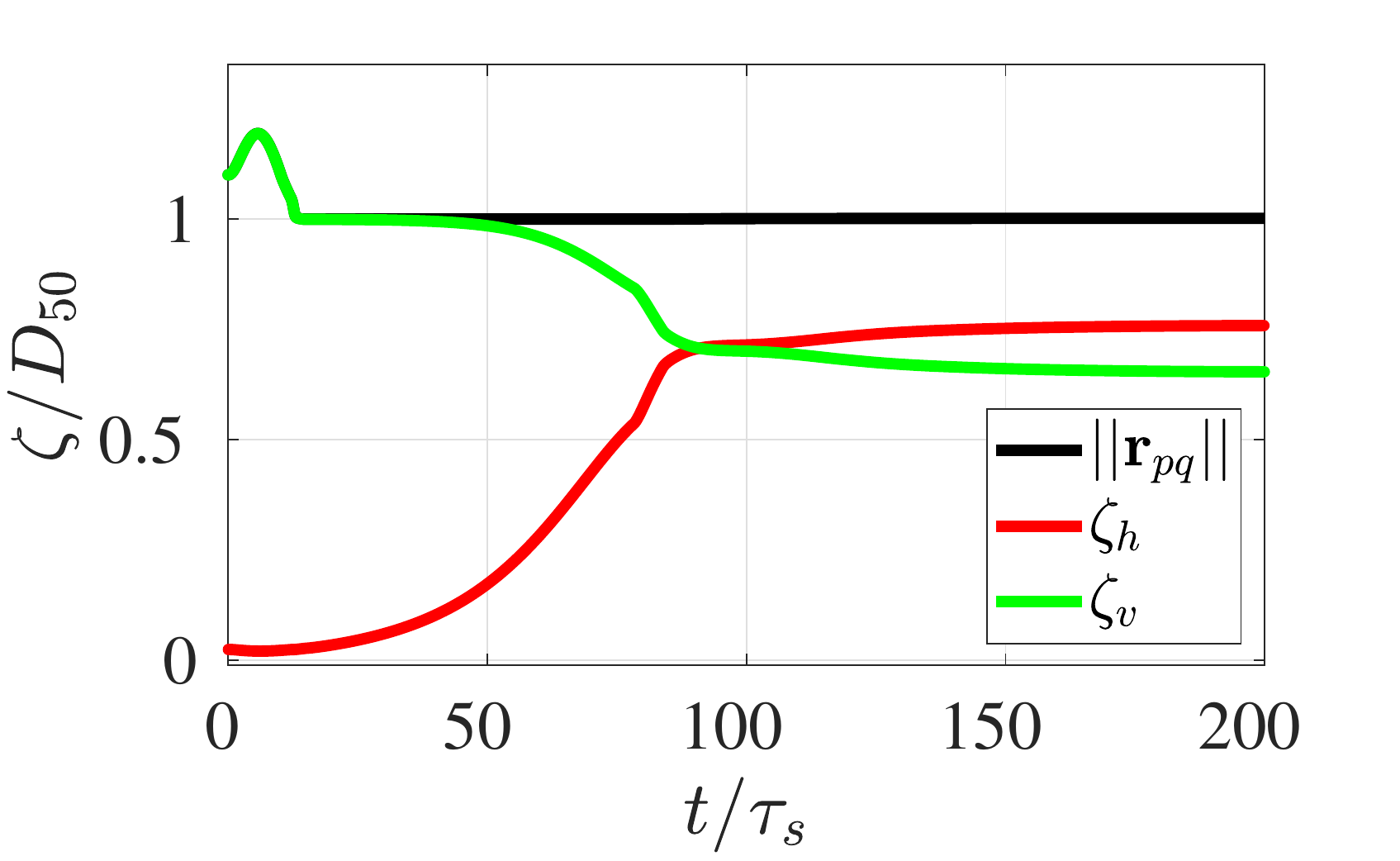}}
  \put(0.0 ,   0    ){\includegraphics[width=0.5\textwidth]{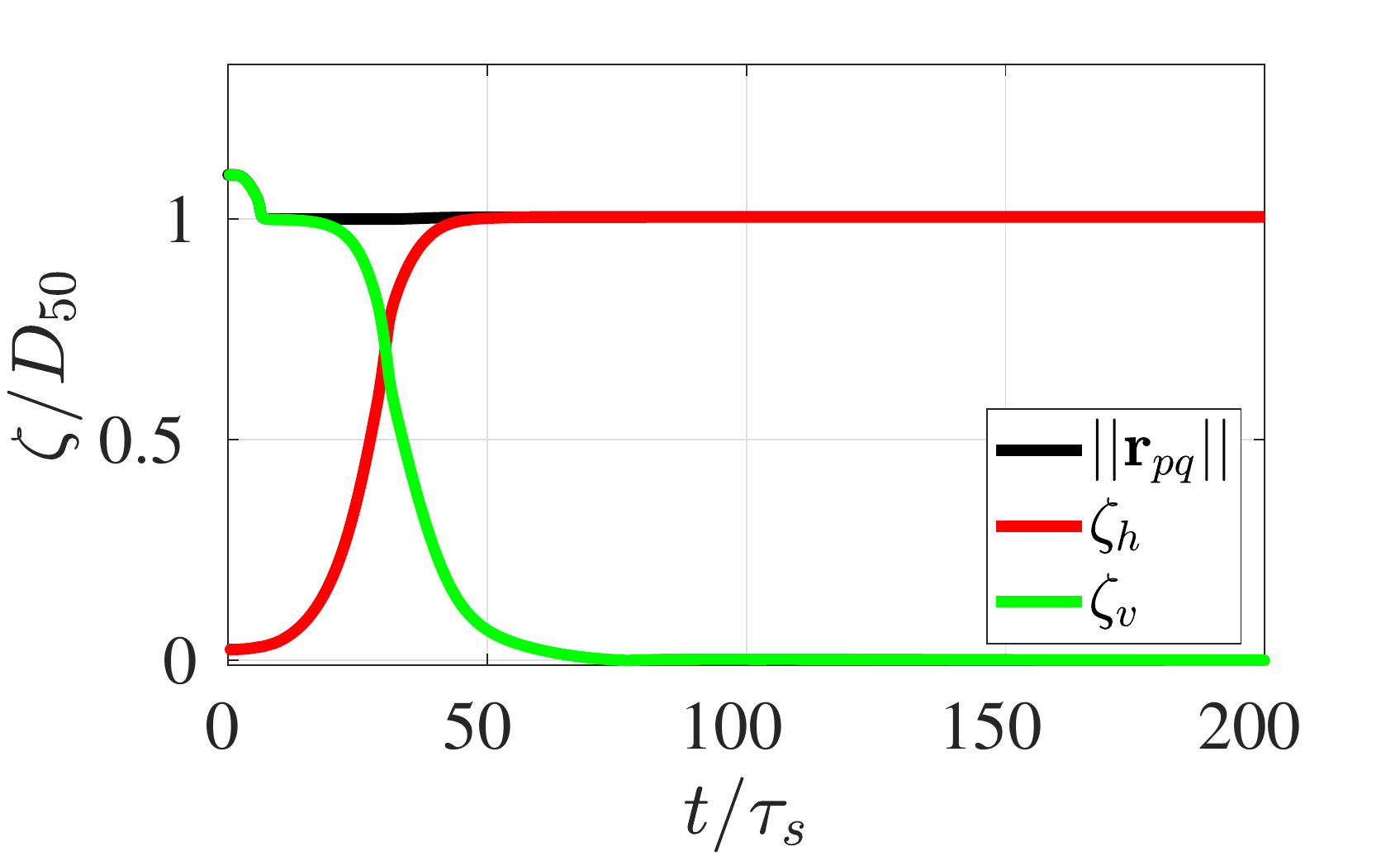}}  
  \put(7.0 ,   0    ){\includegraphics[width=0.5\textwidth]{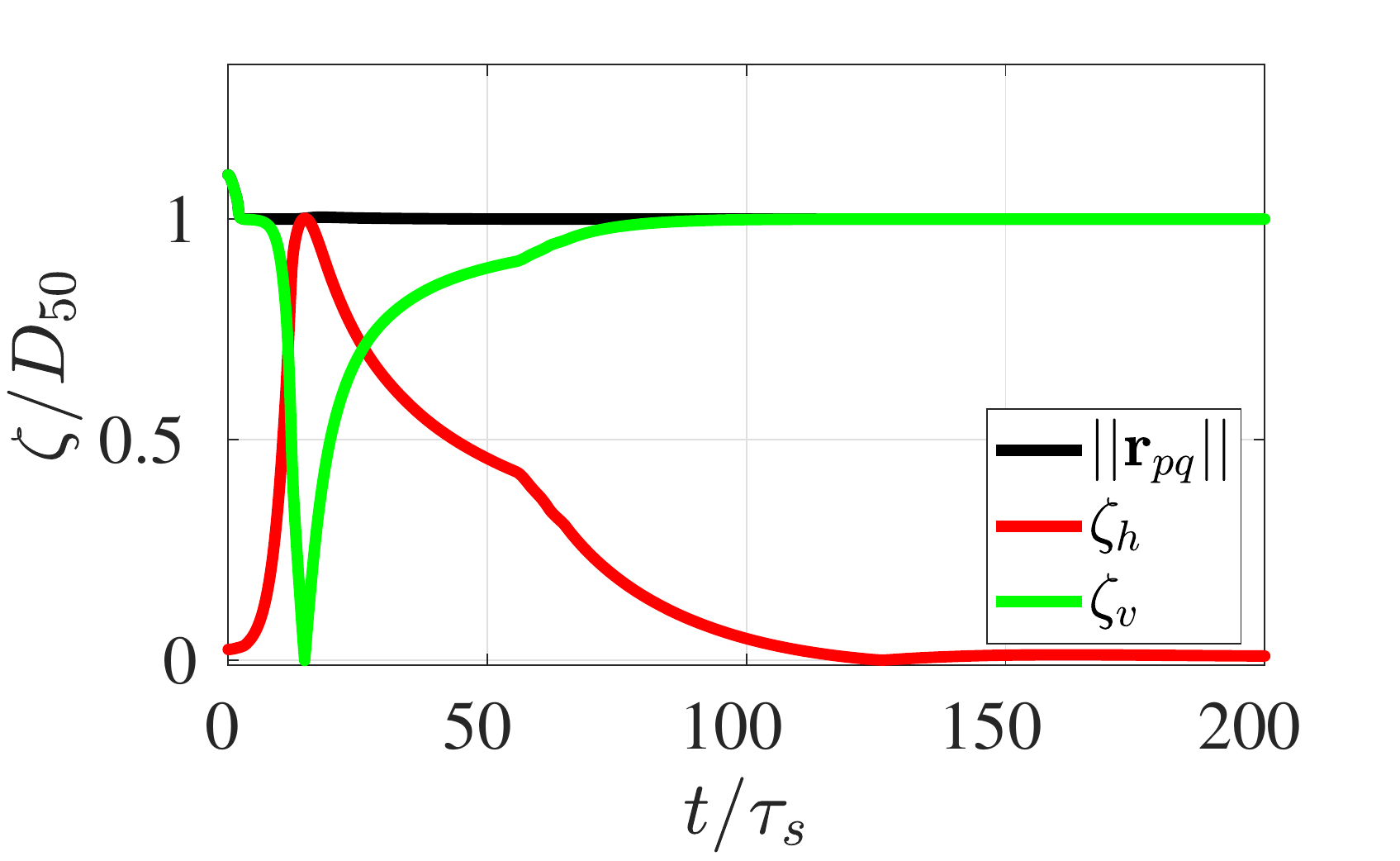}}    

    \put( 0,  8.1)     {a) }
    \put( 7.0,8.1)     {b) }    
    \put( 0  ,3.8)     {c) }        
    \put( 7  ,3.8)     {d) }
    \put( 2.0,7.8)     {\circle{0.3}}
    \put( 2.0,8.05)    {\circle{0.18}}    
    \put( 5.8,7.8)     {\circle{0.3}}    
    \put( 5.9,8.04)    {\circle{0.18}}   %
    \put( 8.8,7.8)     {\circle{0.28235}}
    \put( 8.8,8.05)    {\circle{0.19765}}    
    \put( 12.7 ,7.8)    {\circle{0.28235}}    
    \put( 12.88,7.98)  {\circle{0.19765}}   %
    \put( 1.5,3.4)     {\circle{0.24}}
    \put( 1.5,3.65)    {\circle{0.24}}    
    \put( 5.6,3.5)     {\circle{0.24}}    
    \put( 5.84,3.5)    {\circle{0.24}}   %
    \put( 8.5,3.3 )     {\circle{0.096}}
    \put( 8.5,3.55)    {\circle{0.384}}    
    \put( 12.8,3.4 )   {\circle{0.384}}    
    \put( 12.8,3.65)   {\circle{0.096}}   %
    
\end{picture}
        \caption{DKT results for spheres of different size with $\text{Co} =1$ and different ratios of $R_p/R_q$. (a) $R_p/R_q=0.6$, (b) $R_p/R_q=0.7$, (c) $R_p/R_q=1$, and (d) $R_p/R_q=4$. Circles indicate the initial and final particle configuration, respectively.            
        }
    \label{fig:DKT_poly}
\end{figure}

\setlength{\unitlength}{1cm}
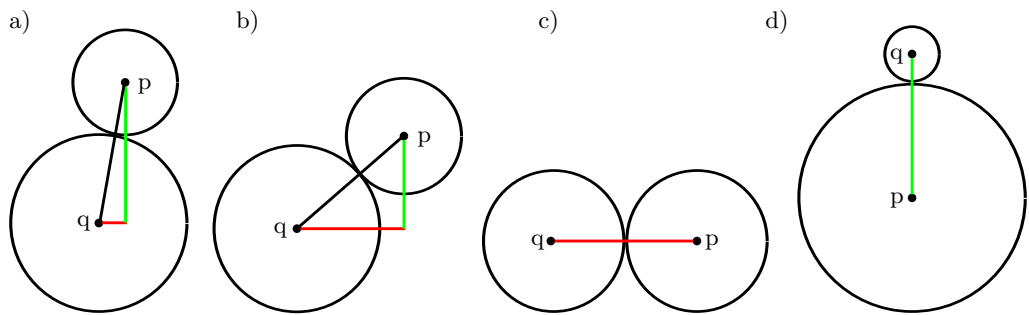
\begin{figure}
\begin{picture}(7,4)
  \put(0.0 ,   0  ){\includegraphics[width=1.0\textwidth]{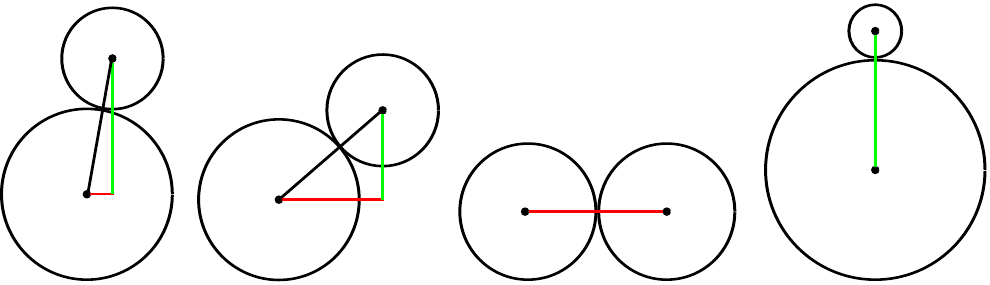}}
    \put( 0   ,3.8)     {a) }        
    \put( 3   ,3.8)     {b) }            
    \put( 7   ,3.8)     {c) }                
    \put(10   ,3.8)     {d) }                    
    \put( 0.9 ,1.2)     {q  }        
    \put( 1.7 ,3.0)     {p  }     
    \put( 3.5 ,1.1)     {q  }        
    \put( 5.4 ,2.3)     {p  }   
    \put( 6.9 ,0.95)    {q  }        
    \put( 9.2 ,0.95)    {p  }   
    \put(11.65,3.4)     {q  }
    \put(11.65,1.5)     {p  }    
\end{picture}
        \caption{Sketch of the asymptotically steady settling configuration of polydisperse, cohesive particles, as illustrated in figure \ref{fig:DKT_poly}. (a) $R_p/R_q=0.6$, (b) $R_p/R_q=0.7$, (c) $R_p/R_q=1$, and (d) $R_p/R_q=4$. The color coding of the connecting lines corresponds to figure \ref{fig:DKT_poly}.
        }
    \label{fig:DKT_config}
\end{figure}

%
\subsection{Settling of cohesive particles of different size}\label{sec:DKT_polydisperse}
To investigate the impact of polydispersity on the settling behavior of cohesive particles, we repeat the above simulations for a fixed cohesive number value of $\text{Co} = 1$, while varying the ratio of the particle radii $R_p / R_q$ in the range $0.25 \leq R_p/R_q \leq 4$.  For $R_p / R_q < 1$, the lower particle is larger and tends to settle faster than the upper, trailing particle. However, already for a ratio of $R_p/R_q = 0.6$ the wake of the leading particle is sufficiently strong to draft the trailing particle into the DKT motion. Corresponding behavior was also reported for 2D-simulations of cohesionless settling circular disks by \cite{wang2014}, who found that there exists a critical value for $R_p/R_q$ to initiate DKT for a smaller particle trailing a larger one. To analyze the particle positions relative to each other, we define the vector connecting the particle centers as $\textbf{r}_{pq} = \textbf{x}_p - \textbf{x}_q$. We then plot the distance between these two centers, as well as the horizontal and vertical separation components 
\begin{subequations}\label{eq:particle_distance}
  \begin{equation}
  ||\textbf{r}_{pq}|| = \zeta_n + D_{50} = \sqrt{(x_p - x_q)^2+(y_p - y_q)^2+(z_p - z_q)^2} \qquad ,
  \end{equation}
  \begin{equation}
  \zeta_h = \sqrt{(y_p - y_q)^2+(z_p - z_q)^2} \qquad , 
  \end{equation}
  and
  \begin{equation}
  \zeta_v = \sqrt{(x_p - x_q)^2}  \qquad ,
  \end{equation}  
\end{subequations}
respectively. Note that particles touch when $||\textbf{r}_{pq}|| /D_{50}  = 1$. The temporal evolution of these quantities is shown in figure \ref{fig:DKT_poly}. For the ratio $R_p / R_q = 0.6$, which was found to be the approximate threshold for drafting, the particles initially separate but then quickly approach each other and touch. The horizontal distance between the particle centers is seen to increase slowly throughout the simulation (figure \ref{fig:DKT_poly}a), which suggests that the pair rotates into an oblique configuration, although the simulation time is too short for a quasi-steady state to be reached. For $R_p / R_q = 0.7$ this rotation occurs more rapidly, and a quasi-steady oblique configuration emerges (figure \ref{fig:DKT_poly}b). When the particle radii ratio is close to, but not equal to unity, we are in the regime for which cohesionless particles repeatedly undergo DKT interactions as reported by \cite{Shao2005}. 

For equal size particles the vertical distance between the centers decays to zero, while their horizontal distance approaches the particle diameter (figure \ref{fig:DKT_poly}c), indicating that the particles align horizontally while touching. This observation is consistent with the classical kissing behavior of the monodisperse case described in the literature \citep{fortes1987,glowinski2001}. Increasing the ratio even further, so that the trailing particle $p$ becomes larger than the leading particle $q$, causes the two particles to swap positions by rotating around each other, since particle $p$ has a bigger settling velocity than particle $q$ (figure \ref{fig:DKT_poly}d). They subsequently align approximately vertically with $\zeta_v \approx ||\textbf{r}_{pq}||$, and only a small horizontal separation width. Corresponding observations of larger trailing particles swapping positions with smaller leading ones were reported for cohesionless grains in both 2D and 3D simulations \citep{wang2014,liao2015}. It is interesting to note that even though particles do not bond for $R_p/R_q \leq 0.5$, they do form a lasting bond for the more disparate size ratio of $R_p/R_q = 4$ via this swapping mechanism. This demonstrates that the ability of the particles to form flocs strongly depends on their initial configuration. The quasi-steady settling configuration of cohesive particle pairs undergoing DKT is sketched in figure \ref{fig:DKT_config}. Particle pairs of very different sizes tend to align vertically (figure \ref{fig:DKT_config}d), while the alignment becomes increasingly horizontal as the particle sizes approach each other. 

\setlength{\unitlength}{1cm}
\begin{figure}
\begin{picture}(7,4.2)
  \put(0.0 ,   0  ){\includegraphics[width=\textwidth]{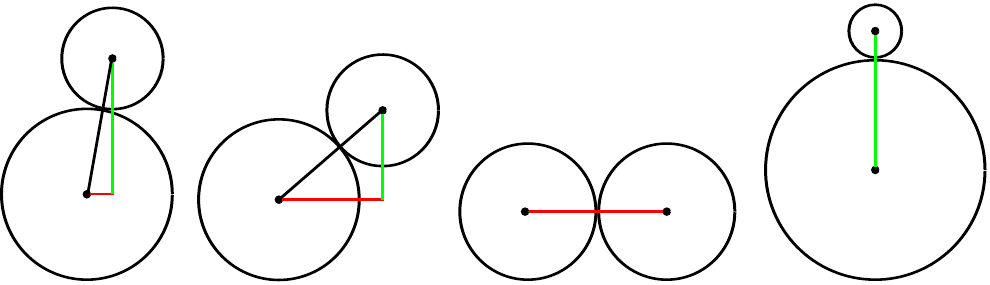}}

\end{picture}
        \caption{Settling velocity in the DKT scenario as a function of time for monodisperse cohesionless and polydisperse cohesive particles as shown in figures \ref{fig:DKT_poly} and \ref{fig:DKT_config}.
        }
    \label{fig:DKT_velocity}
\end{figure}

The mechanism governing the particle orientation has important implications for the settling speeds of the interacting particles. Figure \ref{fig:DKT_velocity} compares the situations addressed in figures \ref{fig:DKT_poly} and \ref{fig:DKT_config} by showing the settling velocity  $u_{pq} = 0.5(u_p + u_q)$  of the two interacting particles $p$ and $q$ as a function of time. Note that the settling velocity is normalized by $u_s$, which is identical for all of the cases shown here since we only change the ratio of $R_p/R_q$ but not the the sum $R_p + R_q$. This allows for a direct comparison of the settling velocities for the different radii ratios. The figure shows only the first passage of the two particles through the periodic domain, in order to exclude any effects from perturbations that might be caused by the particle wake flows. Due to the rather long domain employed for this study, this is equivalent to more than $140 \tau_s$ even for the fastest settling particles. As soon as the monodisperse, cohesionless grains tumble apart (at $t / \tau_s \approx 40$), their settling speed decreases. As expected, introducing cohesive forces increases the settling velocity, as the particles remain in contact which reduces their total drag. This is true for both monodisperse and polydisperse particle configurations. For polydisperse particles, the settling speed is ultimately governed by the larger, leading particle. The kink for $R_p / R_q = 4$ at $t/\tau_s \approx 11$ reflects the situation of the two particles swapping their leading/trailing configuration. Subsequently, these two particles acquire the largest settling velocity among all of the cases presented here. As $R_p / R_q$ approaches unity, the settling velocity decreases. These results clearly illustrate the effect of polydispersity on particle settling speeds.

\setlength{\unitlength}{1cm}
\begin{figure}
\begin{picture}(7,4.2)
  \put(0.0 ,   0    ){\includegraphics[width=0.5\textwidth]{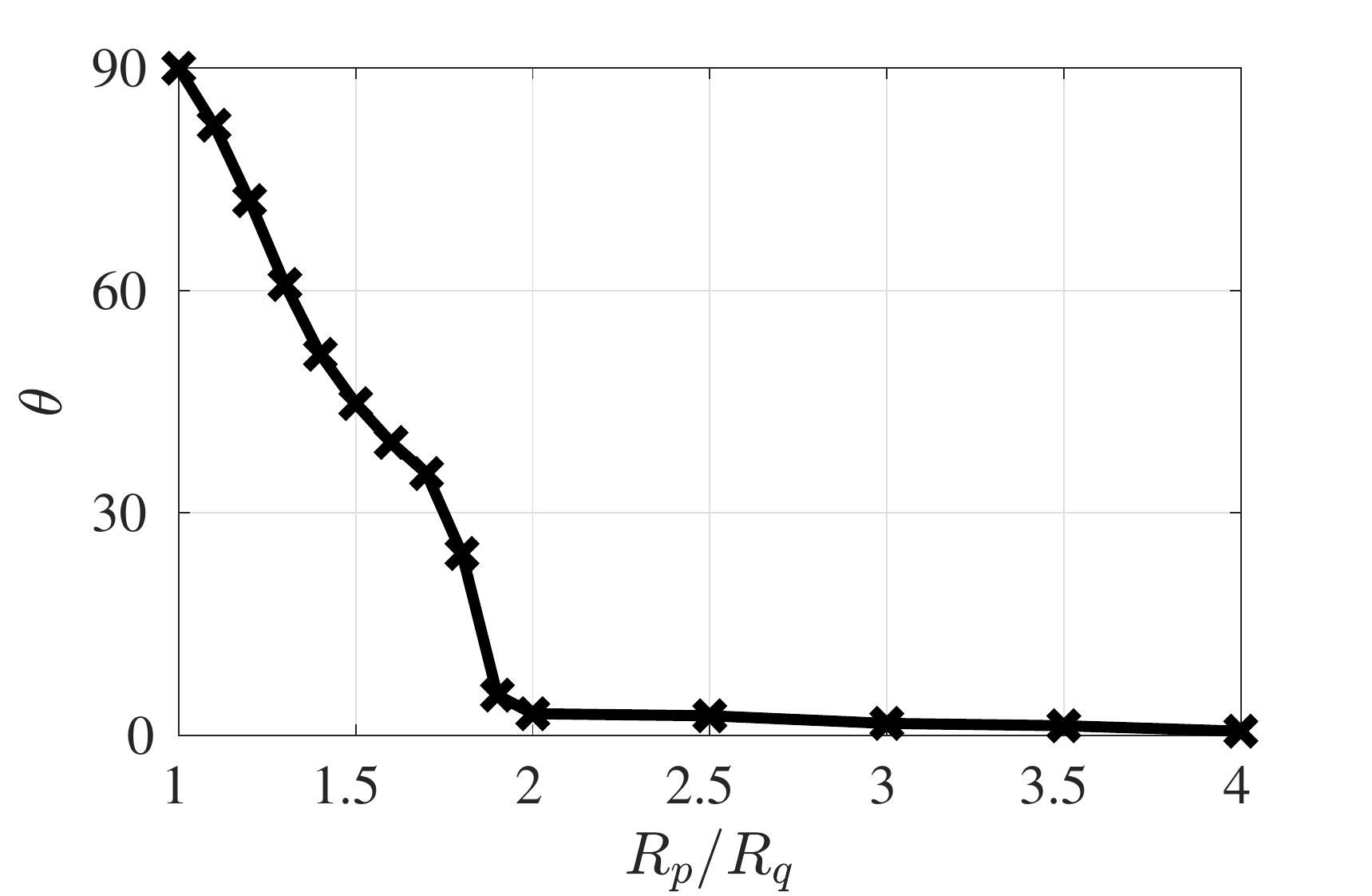}}  
  \put(7.0 ,   0    ){\includegraphics[width=0.5\textwidth]{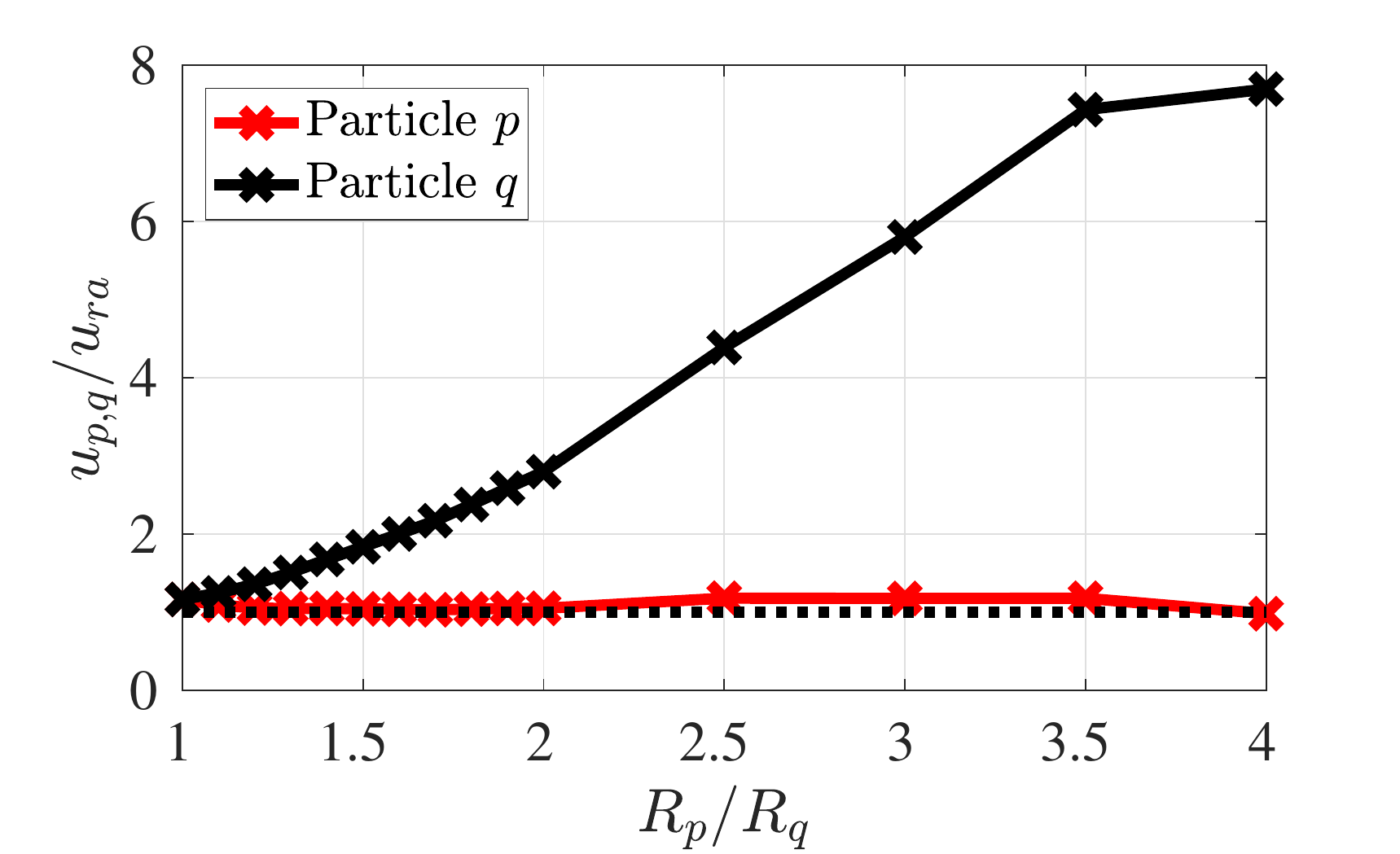}}    
    \put( 0  ,3.8)     {a) }        
    \put( 7  ,3.8)     {b) }  
\end{picture}
        \caption{Quasi-steady configuration of cohesive particle pairs undergoing DKT for different particle radii ratios $R_p/R_q$. (a) orientation angle $\theta$ ($\theta = 90^\circ$: horizontal alignment, $\theta = 0^\circ$: vertical alignment), (b) settling speed. The dotted horizontal line indicates the undisturbed settling velocity.
        }
    \label{fig:DKT_angle}
\end{figure}

The relationship between the quasi-steady particle alignment and the settling speed is displayed in figure~\ref{fig:DKT_angle} for the parameter range $1 \leq R_p /  R_q \leq 4$. When this ratio exceeds two, the particles are aligned approximately vertically, while oblique configurations are observed for smaller ratios, as indicated by the orientation angle ${\cos \theta = \zeta_v / ||\textbf{r}_{pq}||}$. The orientation angle is seen to decrease approximately linearly between $1 \leq R_p/R_q \leq 2$.

To investigate whether the particle settling is accelerated, we estimate the undisturbed settling velocity $u_{ra}$ by using Rayleigh's drag equation (Appendix \ref{app:char_vel}). Figure \ref{fig:DKT_angle}b displays the settling velocities $u_p$ and $u_q$ normalized by their respective undisturbed settling velocity $u_{ra}$. Equal-sized particles, i.e. $R_p / R_q = 1$, are seen to settle with a velocity that is slightly higher than their undisturbed settling velocity. For increasing ratios $R_p/R_q$, we find that the settling of the smaller particle $q$ is substantially accelerated by the stable bond, whereas the larger particle $p$ still settles approximately with its undisturbed settling velocity $u_{ra}$. Hence the center of mass of a cohesive particle pair with a stable bond settles more rapidly than that of two cohesionless particles. 

\subsection{Sensitivity of cohesive force range}\label{sec:sensitivity_lambda}
\setlength{\unitlength}{1cm}
\begin{figure}
\begin{picture}(7,3.7)
  \put(0.0 ,   0  ){\includegraphics[width=0.455\textwidth]{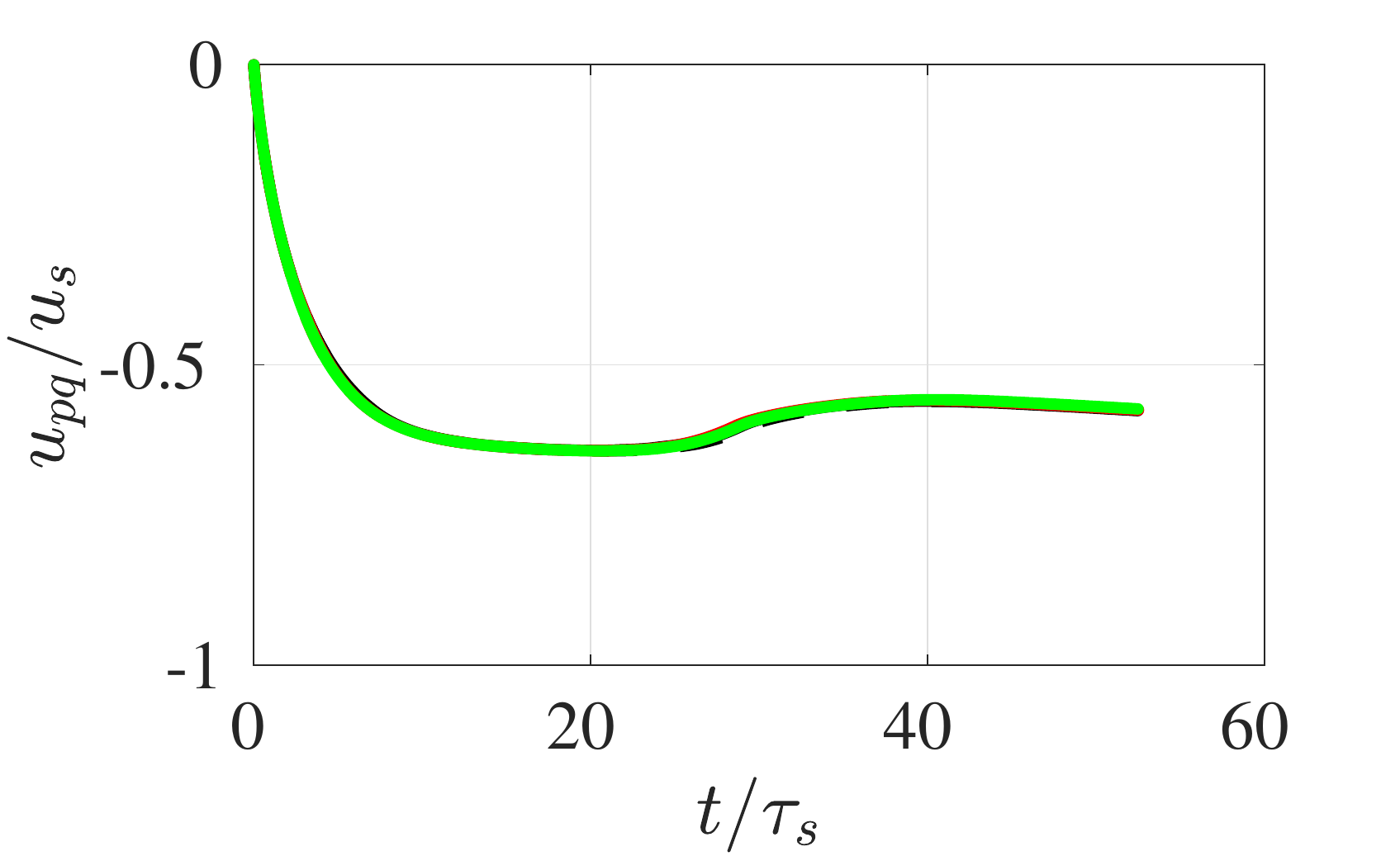}}
  \put(6.5 ,   0  ){\includegraphics[width=0.545\textwidth]{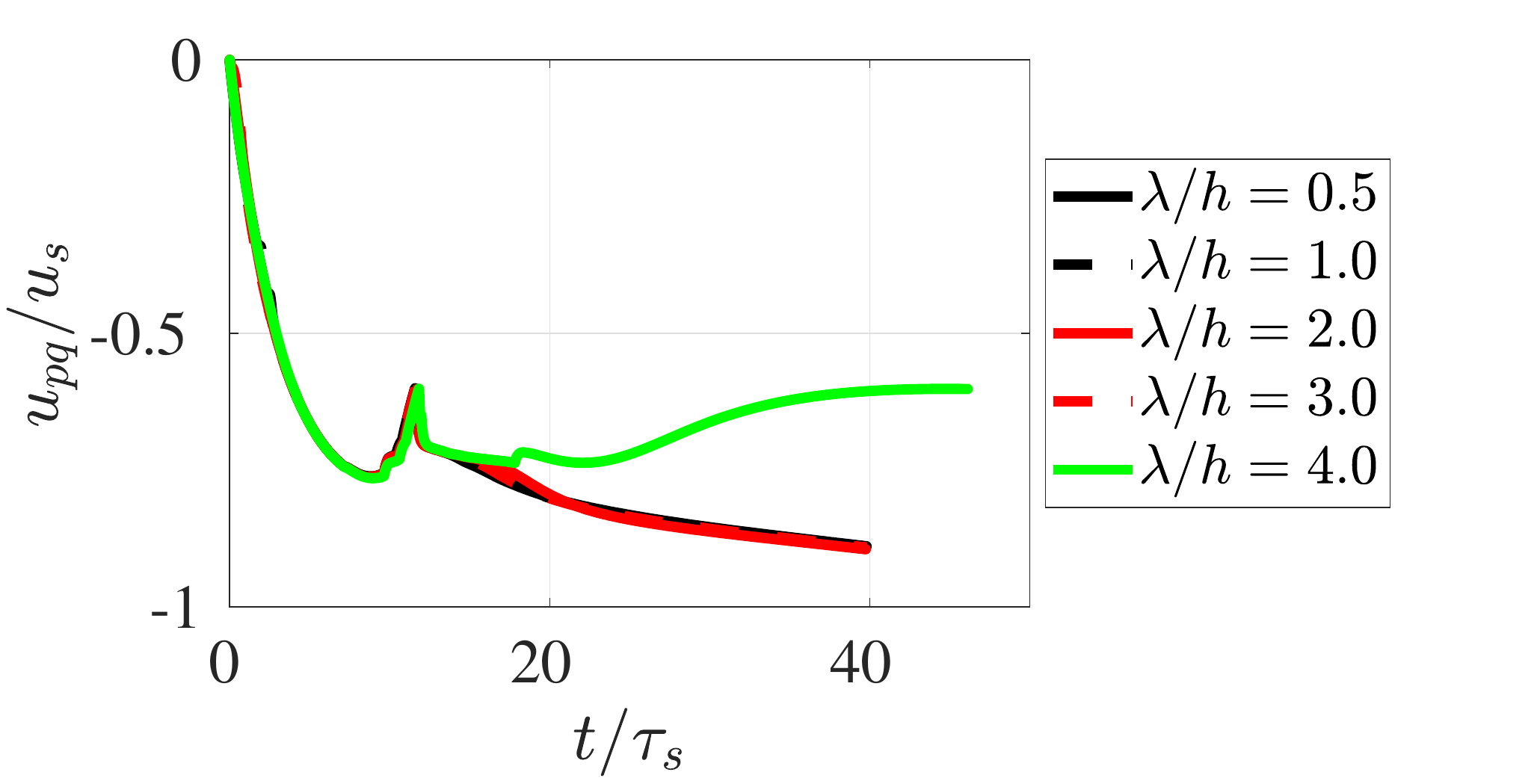}}

    \put( 0,  3.5)     {a) }
    \put( 6.5,3.5)     {b) }    
\end{picture}
        \caption{Influence of the cohesive force range $\lambda$ on the settling velocity of two interacting particles. (a) $R_p / R_q = 1$ and (b) $R_p / R_q = 4$.
        }
    \label{fig:sensitivity_lambda}
\end{figure}

While the model proposed in \S \ref{sec:cohesion} replaces the empirical constants $A_H$ and $\zeta_0$ by the physically meaningful dimensionless cohesive number $\text{Co}$, it still contains the dependency on the cohesive force range $\lambda$. To test the sensitivity of the simulation results on this parameter, we conducted simulations with $R_p / R_q = 1$ and $R_p / R_q = 4$ and different values of $\lambda$. The results are shown in figure \ref{fig:sensitivity_lambda}. For the equal-sized particles, the settling velocities for all values of $\lambda$ collapse on a single curve. The same is true for $R_p / R_q = 4$ and $\lambda \leq 3 h$. However, for $\lambda = 4h$ the particles start to separate at $t / \tau_s \approx 20$. The reason for this detachment is that $\lambda$ is now equal to the radius $R_q$ of the smaller particle, so that the range of the cohesive force is no longer much smaller than the particle radius, which violates the assumptions underlying the model of \S \ref{sec:cohesion}. In summary, we find that as long as $\lambda$ is significantly smaller than the smallest particle radius, the simulation results are independent of $\lambda$.\rone{The analysis presented in this section also provides evidence that decreasing the steady state separation distance by increasing the Cohesive number does not affect the hydrodynamics of the two interacting particles. }

%% file: 05_settling.tex
 \subsection{Computational setup}\label{sec:setup_settling}
In order to explore the influence of cohesive forces on the sedimentation process of a large, polydisperse ensemble of particles, we reproduce the experiments by \cite{teslaa2015}, albeit on a smaller spatial scale. These authors investigated the hindered settling of silt particles, with diameters in the range $2 \mu \text{m} \leq D_p \leq 63 \mu \text{m}$. To this end, we place a polydisperse mixture \rtwo{with a homogenous particle volume fraction of $15 \%$ in a tank of quiescent fluid} (figure \ref{fig:initial_distribution}a). As before, it is convenient to define the reference velocity based on the buoyancy velocity $u_{s} = \sqrt{g'D_{50}}$, where $D_{50}$ denotes the median grain size of the entire particle size distribution. The characteristic time scale based on the buoyancy velocity and the median diameter then becomes $\tau_s = D_{50} / u_{s}$.

Consistent with the experiments, we choose a Reynolds number of $\Rey = 1.35$. As in the experiments of \cite{teslaa2015}, the computational grain sizes obey a cumulative log-normal distribution $\frac{1}{2}+\frac{1}{2}\erf \left[\frac{\ln D_p - \mu}{\sqrt{2} \sigma} \right]$ around the median diameter $D_{50}$, with the arithmetic moments ${\mu = -1.33}$ and ${\sigma= 0.34}$ (figure \ref{fig:initial_distribution}c). \rtwo{This yields a total of 1,261 particles with maximum size ratio of $\max\{D\} / \min\{D\}=4$, which is smaller than it was in the experiments of \cite{teslaa2015}. Our computational approach of particle-resolved simulations, however, requires us to resolve even the smallest grain size by at least eight grid cells per diameter \citep{uhlmann2005}. Hence, it was concluded from \S \ref{sec:binary_interaction} that a size ratio of four is plenty to account for polydispersity but remain computationally feasible.}  The small deviations from the analytical log-normal distribution stem from the fact that we slightly rearranged \rtwo{the particle distribution} due the initial random particle placement. This became necessary as the rather small computational domain yielded large variations in volume fraction over the vertical extent of the domain. To smooth out the horizontally averaged volume fraction profile, we applied a two-step procedure: first, we removed larger particles from $y$-locations with higher concentrations, and subsequently we replaced them with the exact same volume of a few smaller particles in $y$-locations with lower concentrations at random $x$- and $z$-positions. This procedure yields an almost uniform particle volume fraction $\phi_v = V_p / V_0 \approx 0.155$ (figure \ref{fig:initial_distribution}b), where $V_p$ and $V_0$ denote the volume occupied by the particles and the computational domain size, respectively. Note that this rearrangement of particles would not have been required for a much larger tank and many more particles, but this would have been prohibitively expensive computationally. The computational domain is of size $L_x \times L_y \times L_z = 13.1 D_{50} \times 40.0 D_{50}\times 13.1 D_{50}$, with gravity pointing in the  negative $y$-direction. We assume periodic boundary conditions in the $x$- and $z$-directions, respectively, along with a no-slip condition at the bottom wall and a free-slip condition at the top wall. The median particle size is discretized by $D_{50}/h = 18.25$ grid cells. 
 \setlength{\unitlength}{1cm}
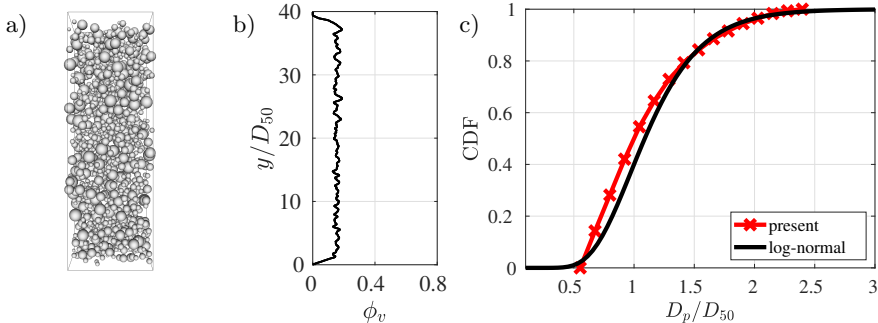
\begin{figure}
\begin{picture}(7,4.2)
  \put(1.00,   0.1  ){\includegraphics[width=0.205\textwidth]{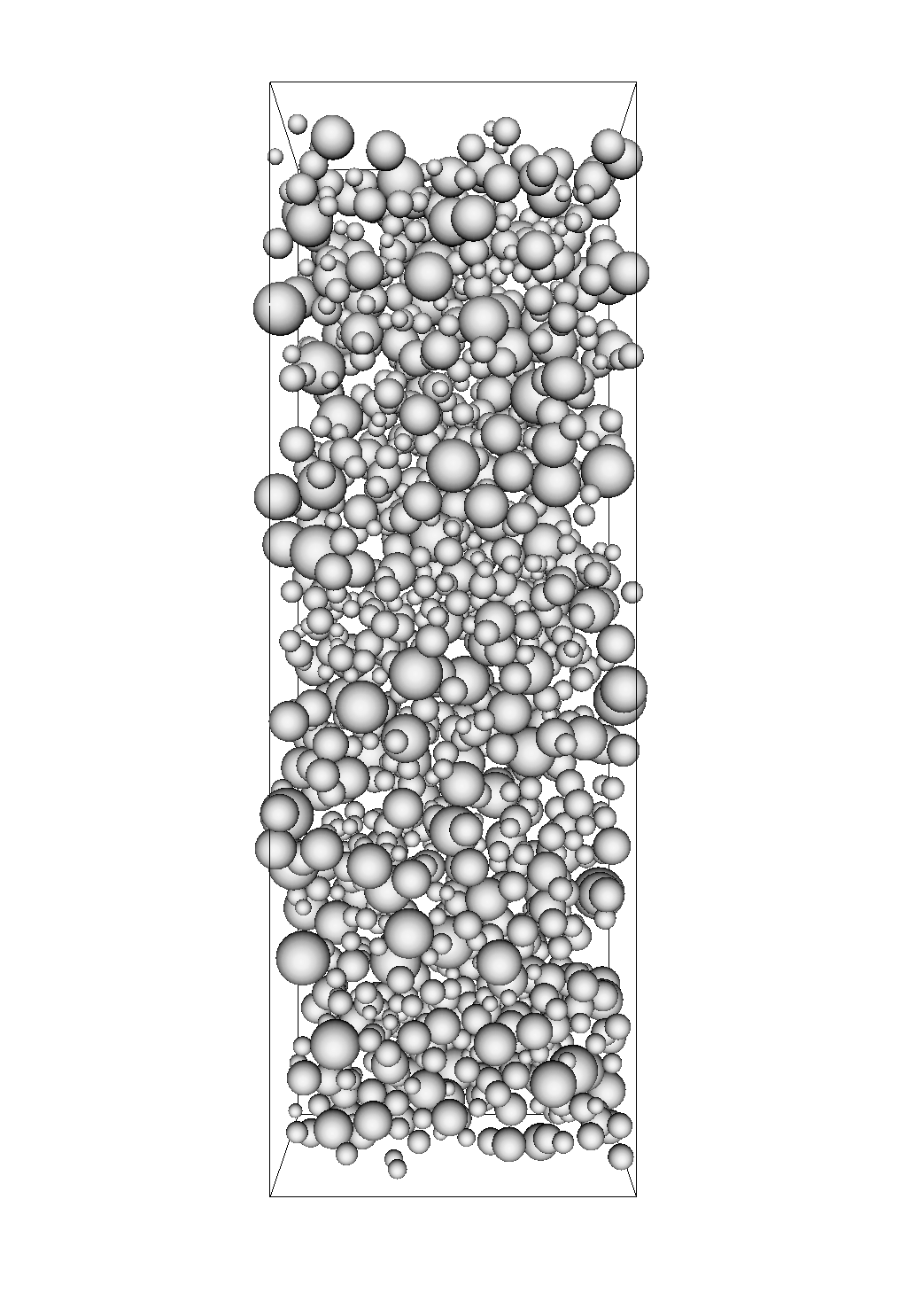}}
  \put(4.2 ,  -0.28 ){\includegraphics[width=0.205\textwidth]{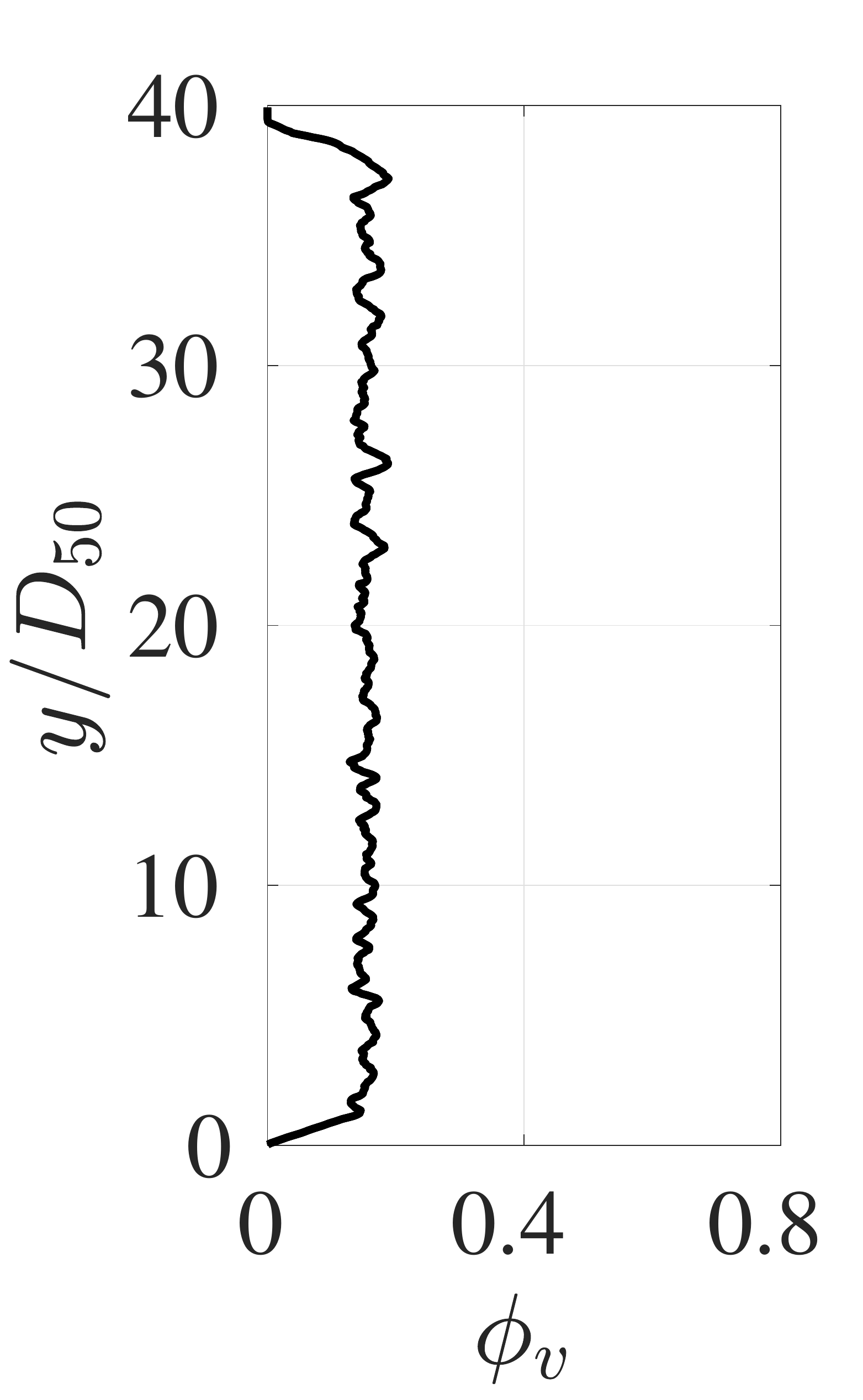}}
  \put(7.00,  -0.25 ){\includegraphics[width=0.45 \textwidth]{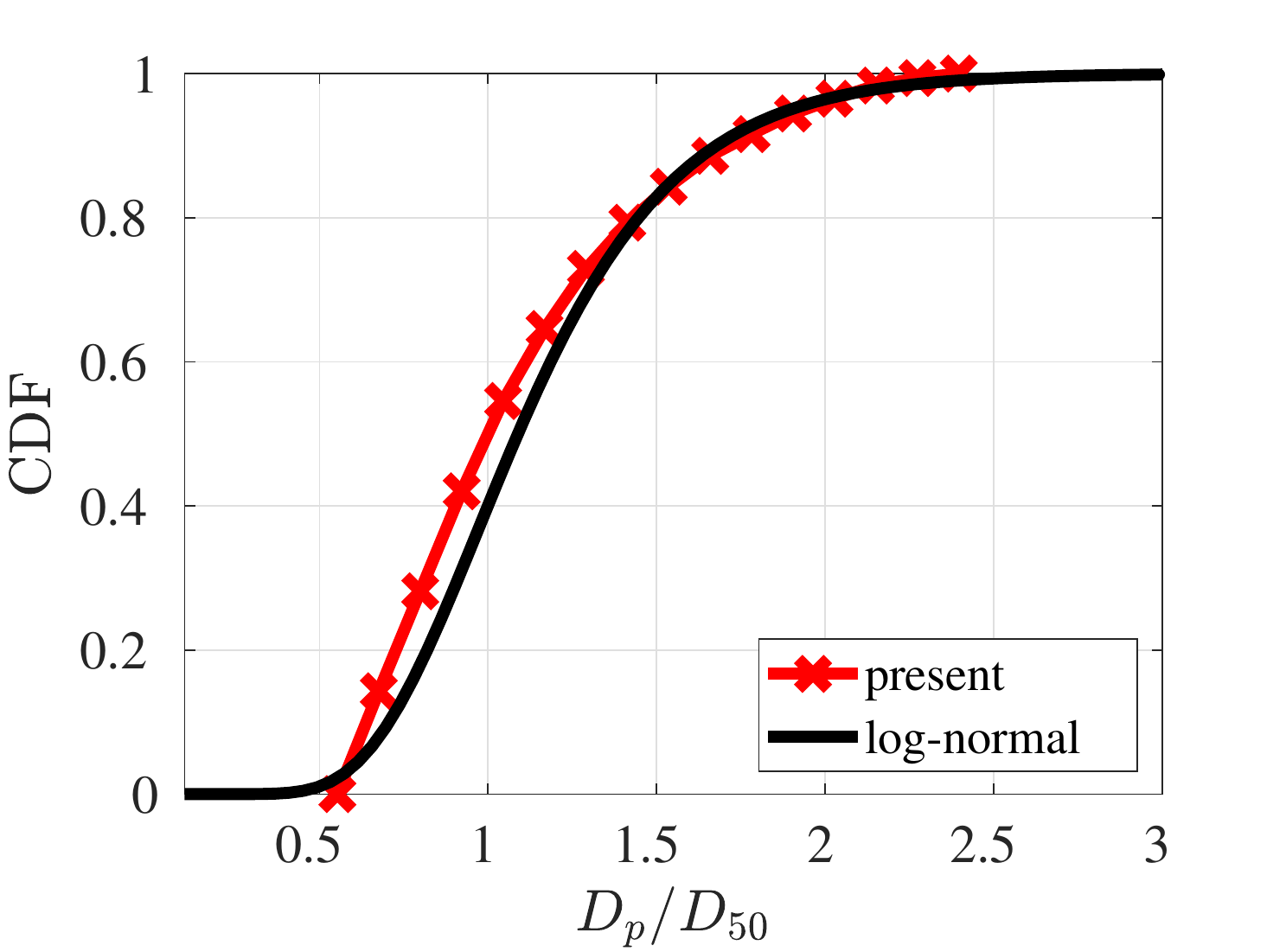}}
    \put( 1.0,  3.6)     {a) }
    \put( 4.0,  3.6)     {b) }    
    \put( 7.0,  3.6)     {c) }       
\end{picture}
        \caption{(a) Initial particle distribution, (b) initial particle volume fraction distribution, and (c) cumulative distribution function of the particle diameter, along with the log-normal distribution.
        }
    \label{fig:initial_distribution}
\end{figure}
 \begin{table}
   \begin{center}
 \def~{\hphantom{0}}
   \begin{tabular}{c| c | c | c| c| c| c}
       $L_x/D_{50} \times L_y/D_{50} \times L_z/D_{50}$ & $\rho_p / \rho_f$ & $s / D_{50}$ & $\max\{D\}/{D_{50}}$ & $\min\{D\}/{D_{50}}$ &$\Rey = \frac{\sqrt{g'}D_{50}^{3/2}}{\nu_f}$  & $N_p$    \\
       \hline                                                                  
       $13.1 \times 40.0 \times 13.1$                  & 2.6               & 0.365        & 2.4           & 0.6           &1.35                                          & 1261     \\
   \end{tabular}
   \caption{Parameters for simulations of a large ensemble, where $L_x$, $L_y$, $L_z$ indicate the domain size, and $s$ represents the standard deviation of the grain size.}
   \label{tab:settling_parameters}
   \end{center}
 \end{table}

Three simulations were performed for different values of the Cohesion number $\text{Co}$: (i) cohesionless grains with a cohesive number $\text{Co} = \max(|| \textbf{F}_{\text{coh},50} ||) / m_{50} g' = 0$, (ii) mildly cohesive sediment with $\text{Co} = 1$, and (iii) strongly cohesive sediment with $\text{Co} = 5$. For all simulations, the particles are released from rest in quiescent fluid, and subsequently settle under the influence of gravity. The key simulation parameters are listed in table \ref{tab:settling_parameters}. The particle collisions are inelastic with $e_\text{dry} = 0.97 < 1$, and they experience friction through \eqref{eq:lin_tan}. 

\setlength{\unitlength}{1cm}
\begin{figure}
\begin{picture}(7,18)
  \put( 0.0, 11.7 ){\includegraphics[                       height=0.3\textheight]{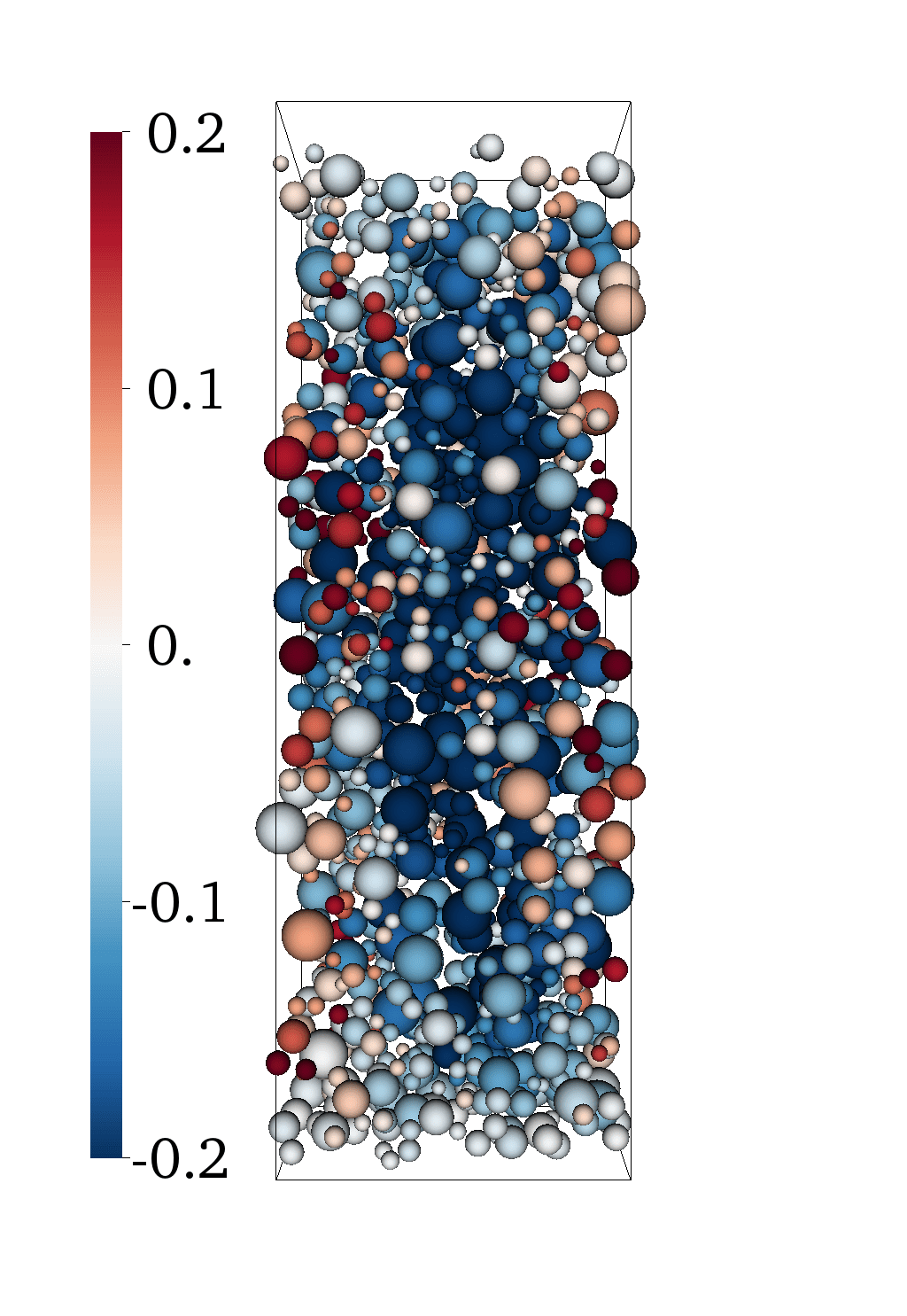}}
  \put( 3.1, 11.7 ){\includegraphics[trim={9cm 0 0 0},clip, height=0.3\textheight]{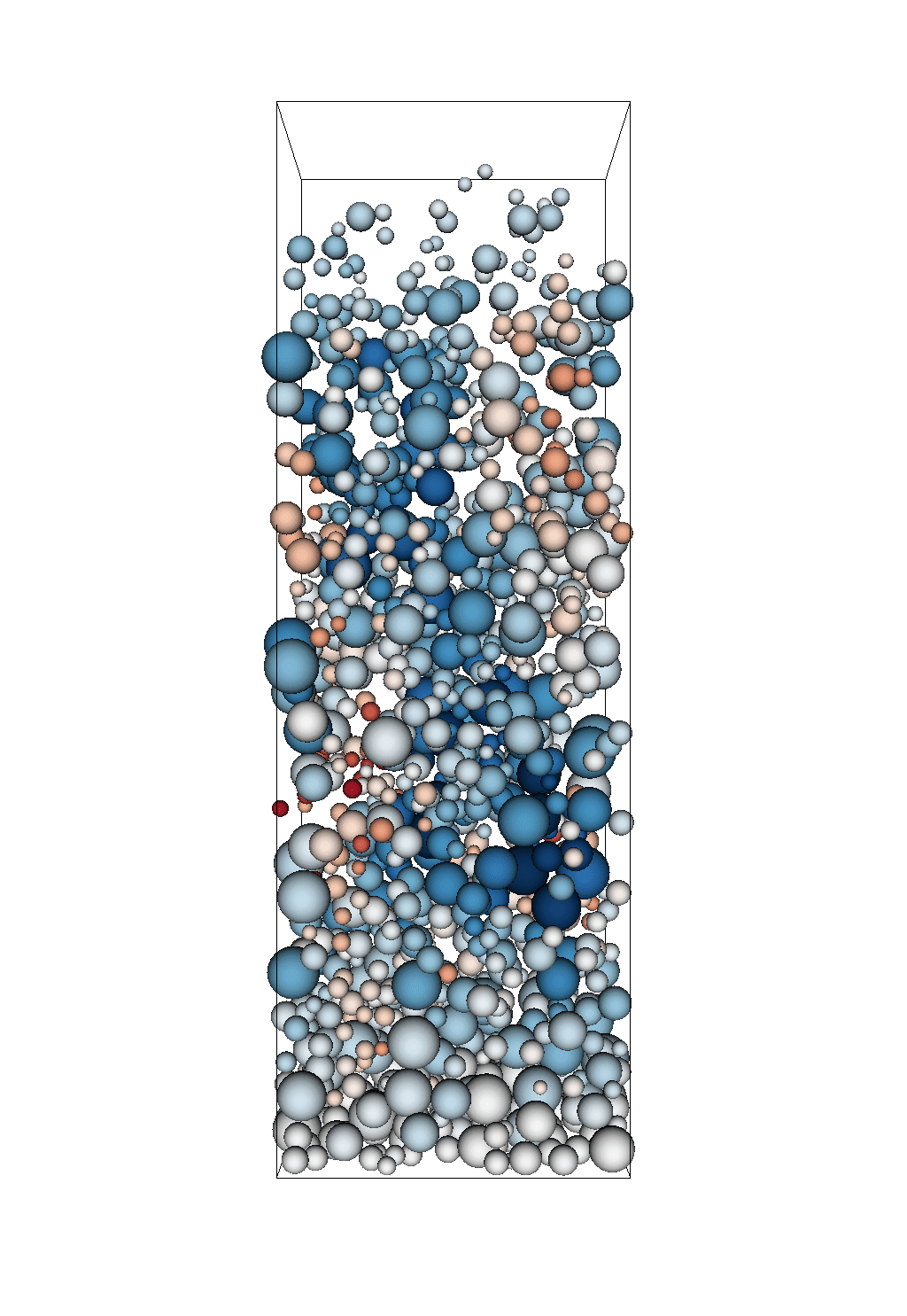}}  
  \put( 5.2, 11.7 ){\includegraphics[trim={9cm 0 0 0},clip, height=0.3\textheight]{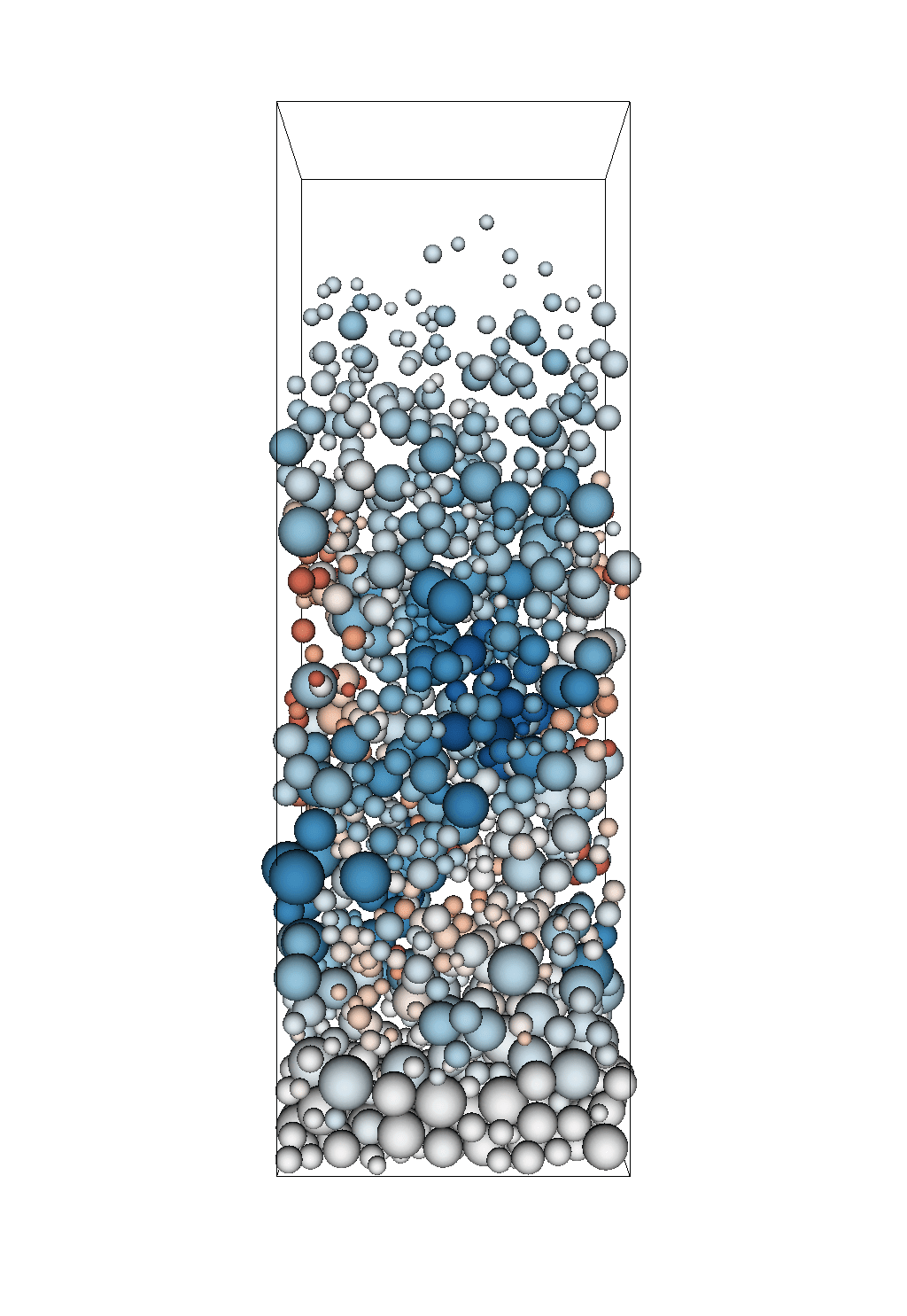}}   
  \put( 7.3, 11.7 ){\includegraphics[trim={9cm 0 0 0},clip, height=0.3\textheight]{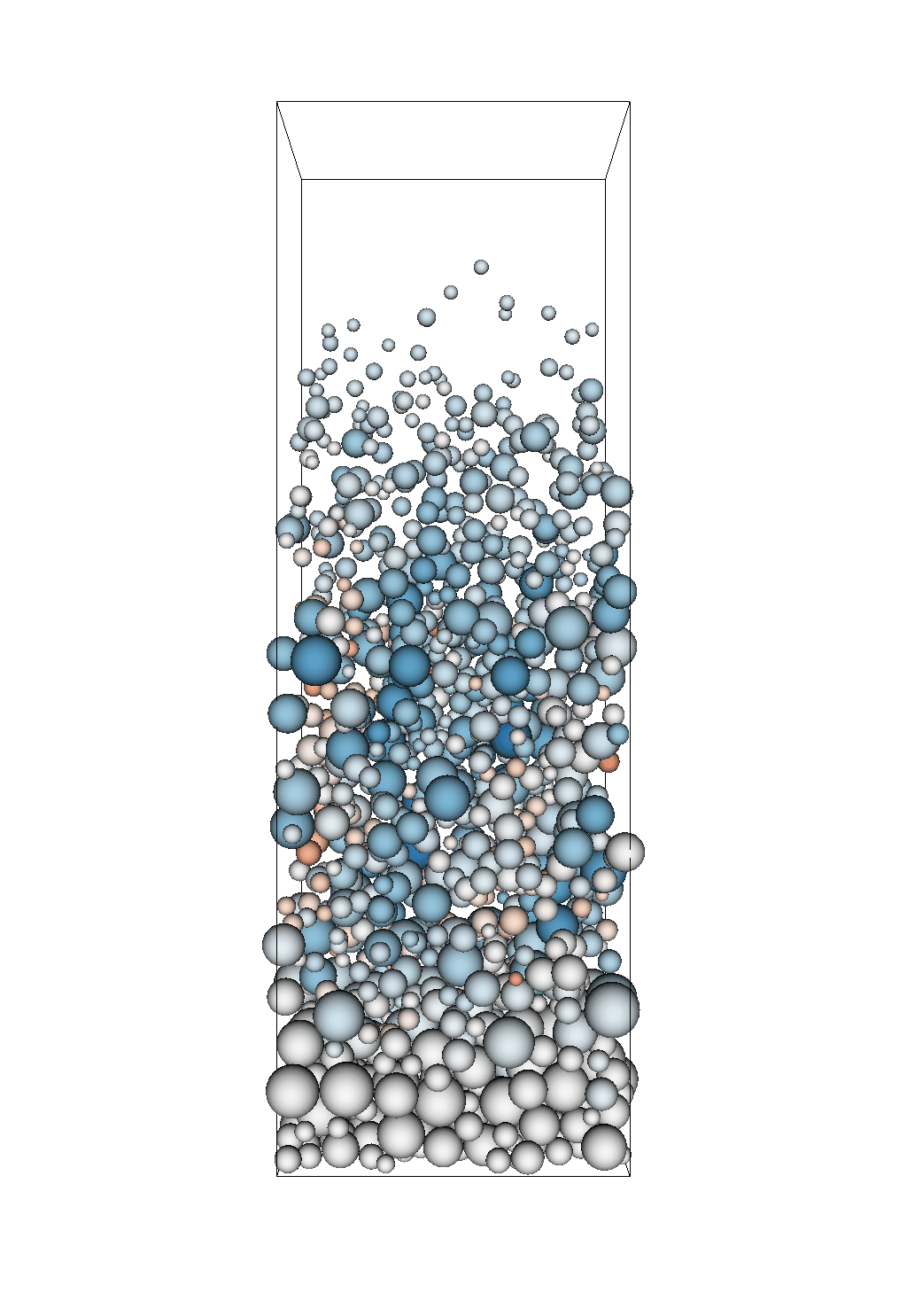}}   
  \put( 9.4, 11.7 ){\includegraphics[trim={9cm 0 0 0},clip, height=0.3\textheight]{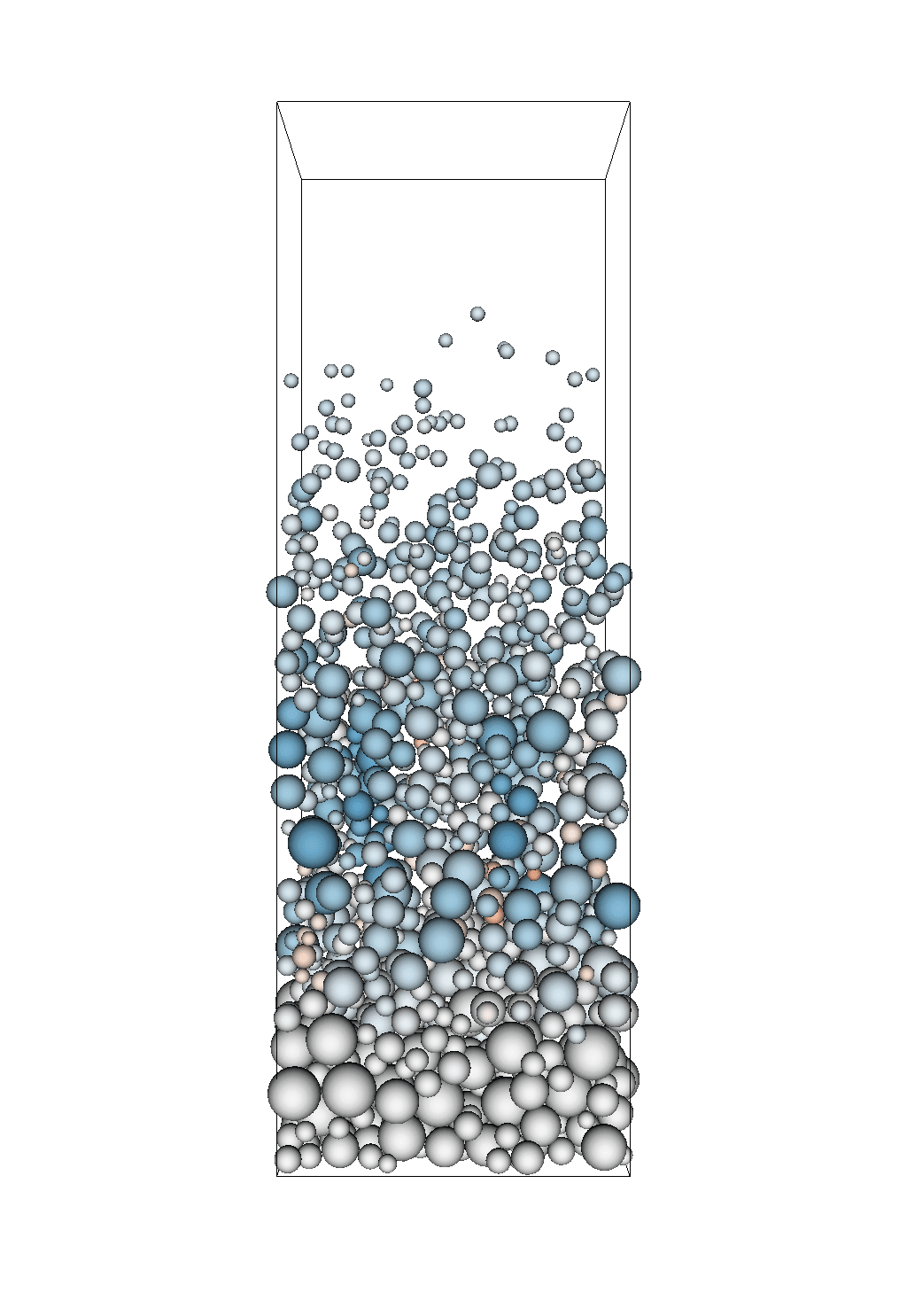}}   
  \put(11.5, 11.7 ){\includegraphics[trim={9cm 0 0 0},clip, height=0.3\textheight]{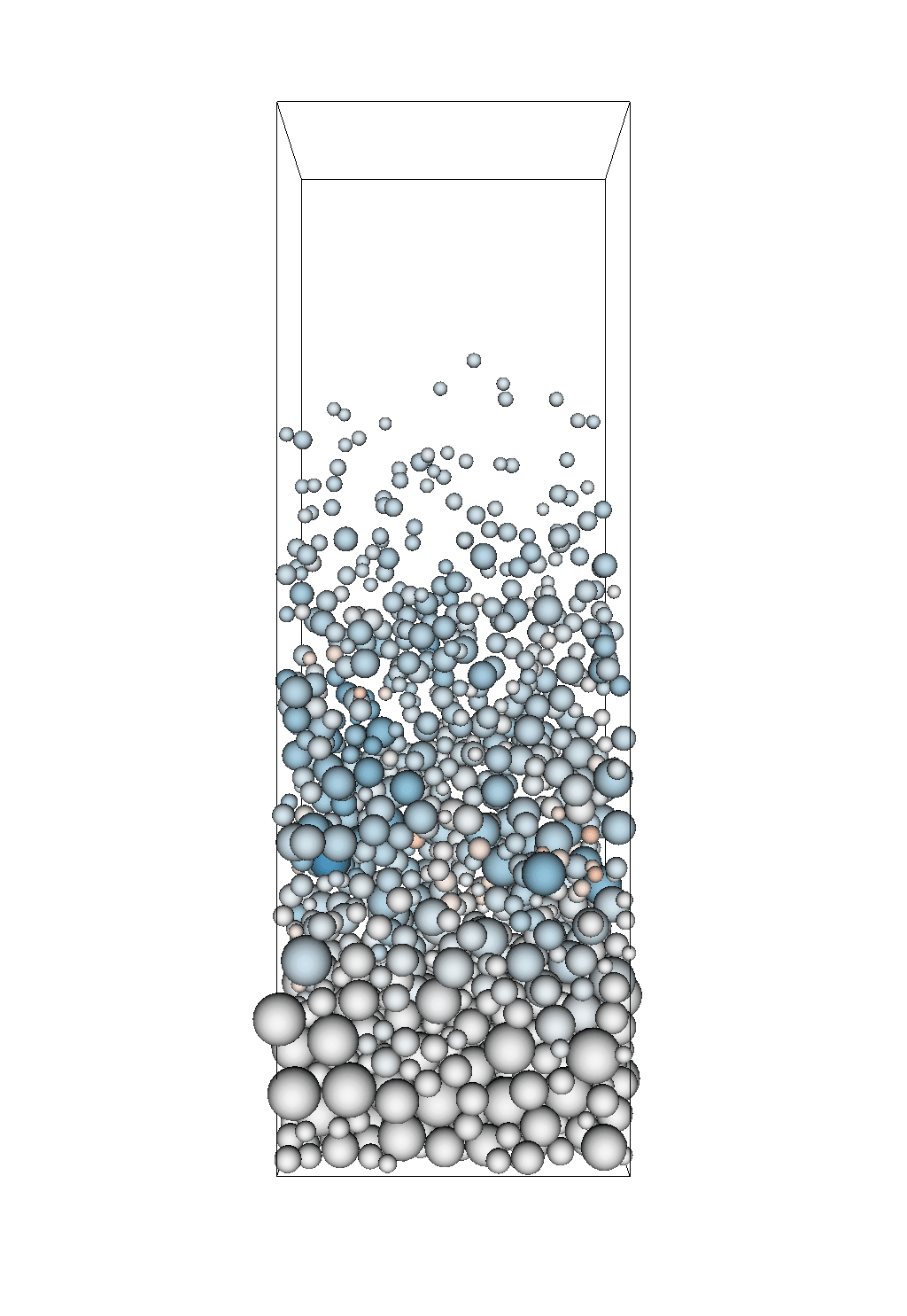}}   
  \put( 0.0,  5.9 ){\includegraphics[                       height=0.3\textheight]{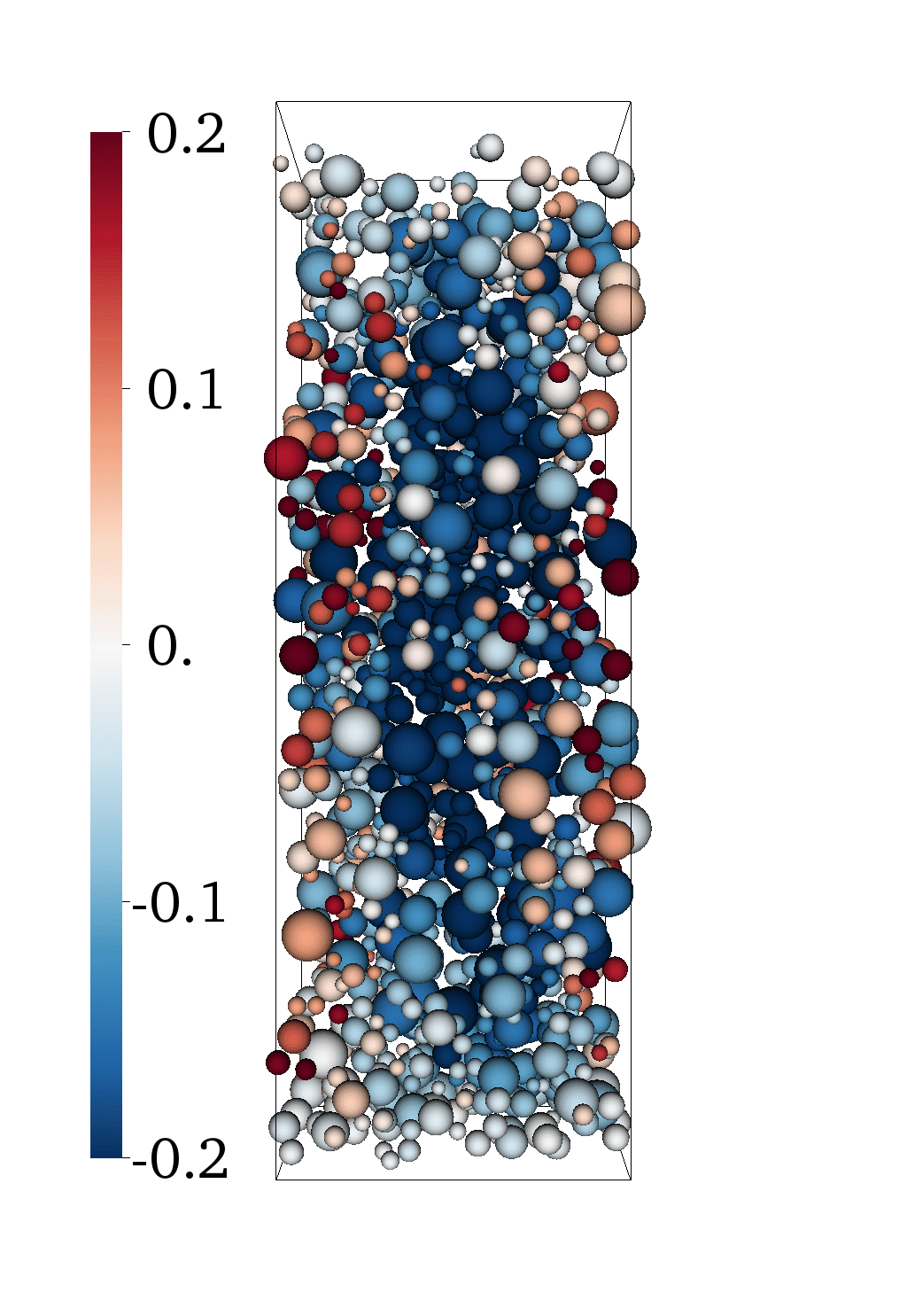}}
  \put( 3.1,  5.9 ){\includegraphics[trim={9cm 0 0 0},clip, height=0.3\textheight]{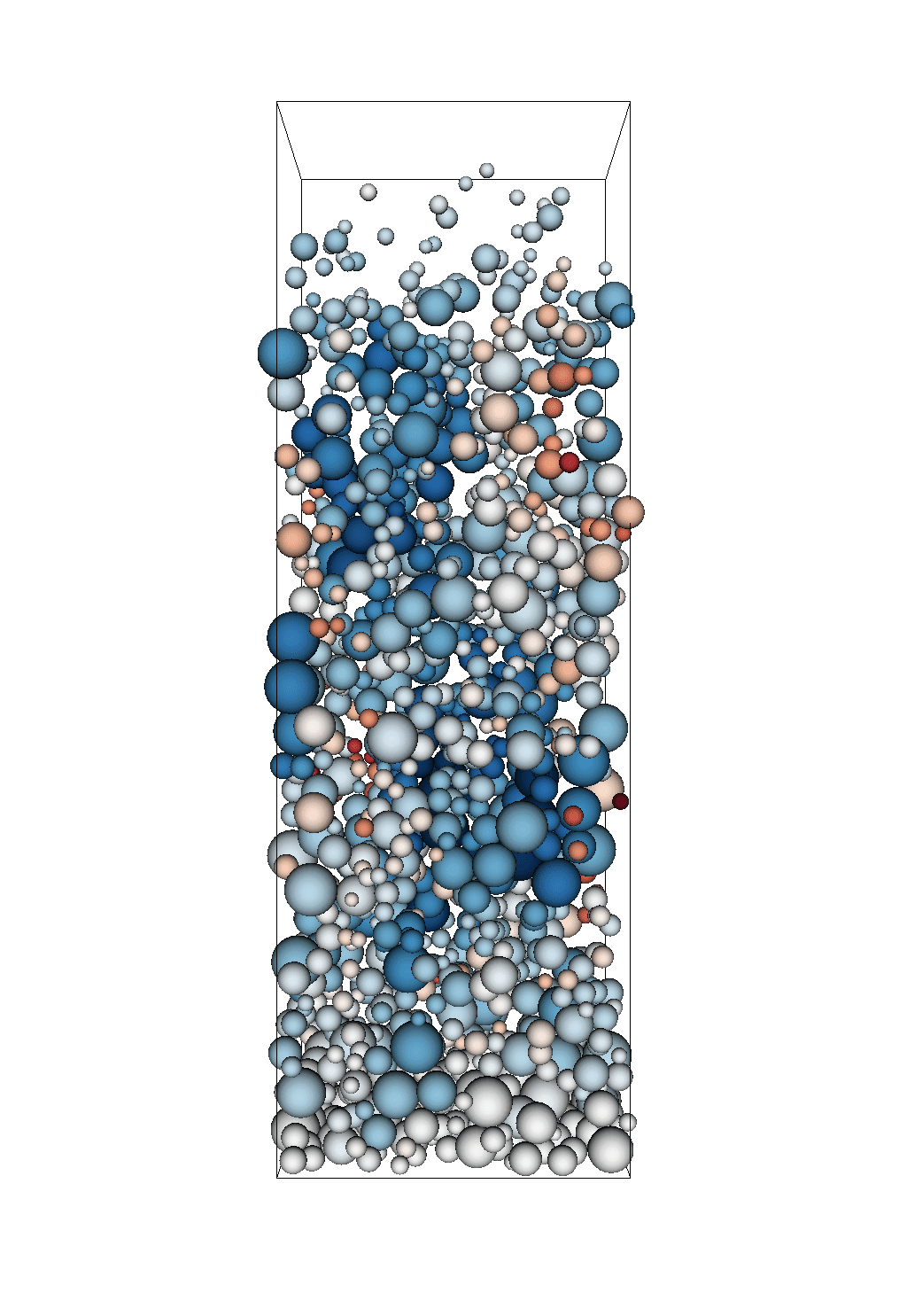}}  
  \put( 5.2,  5.9 ){\includegraphics[trim={9cm 0 0 0},clip, height=0.3\textheight]{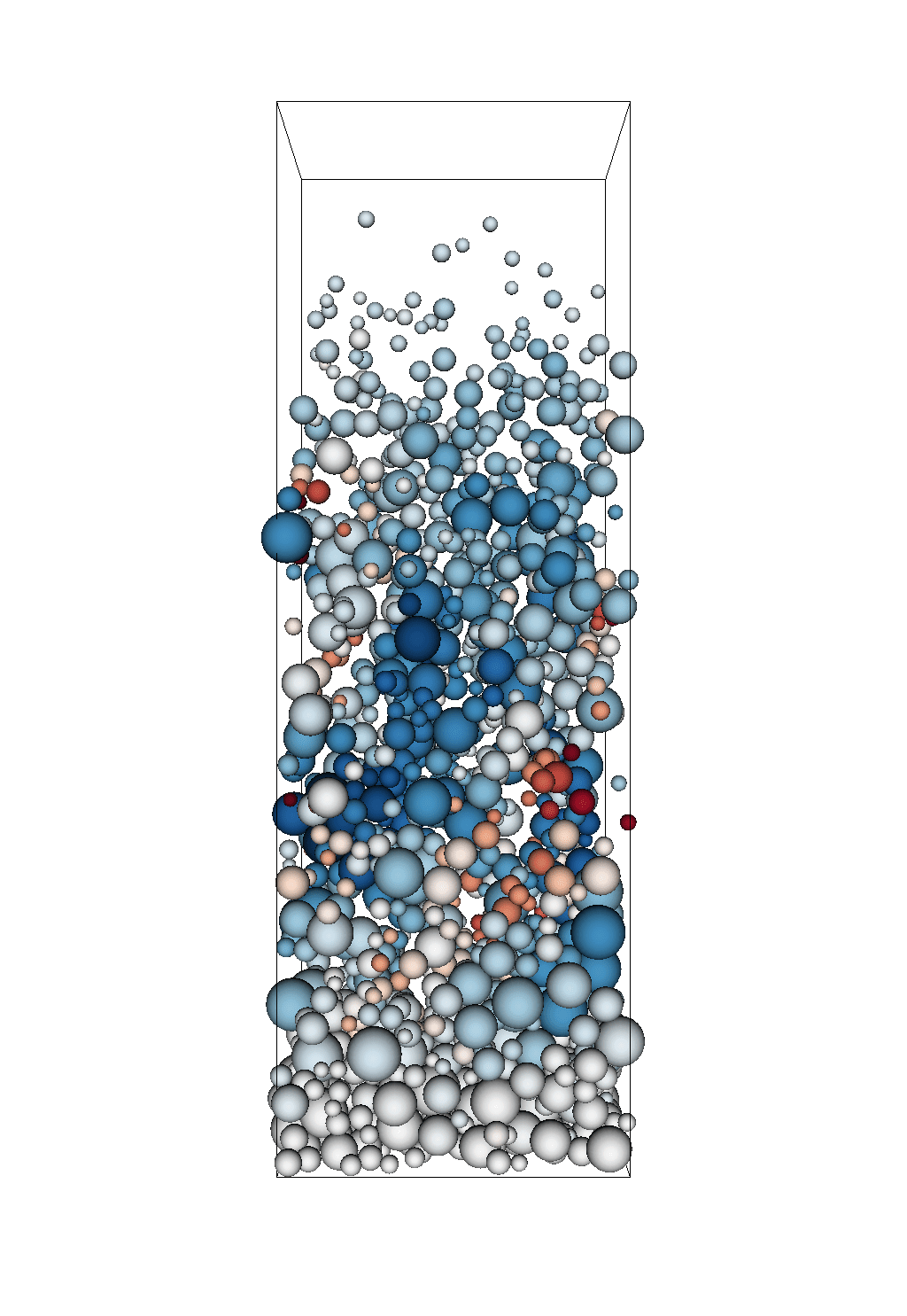}}   
  \put( 7.3,  5.9 ){\includegraphics[trim={9cm 0 0 0},clip, height=0.3\textheight]{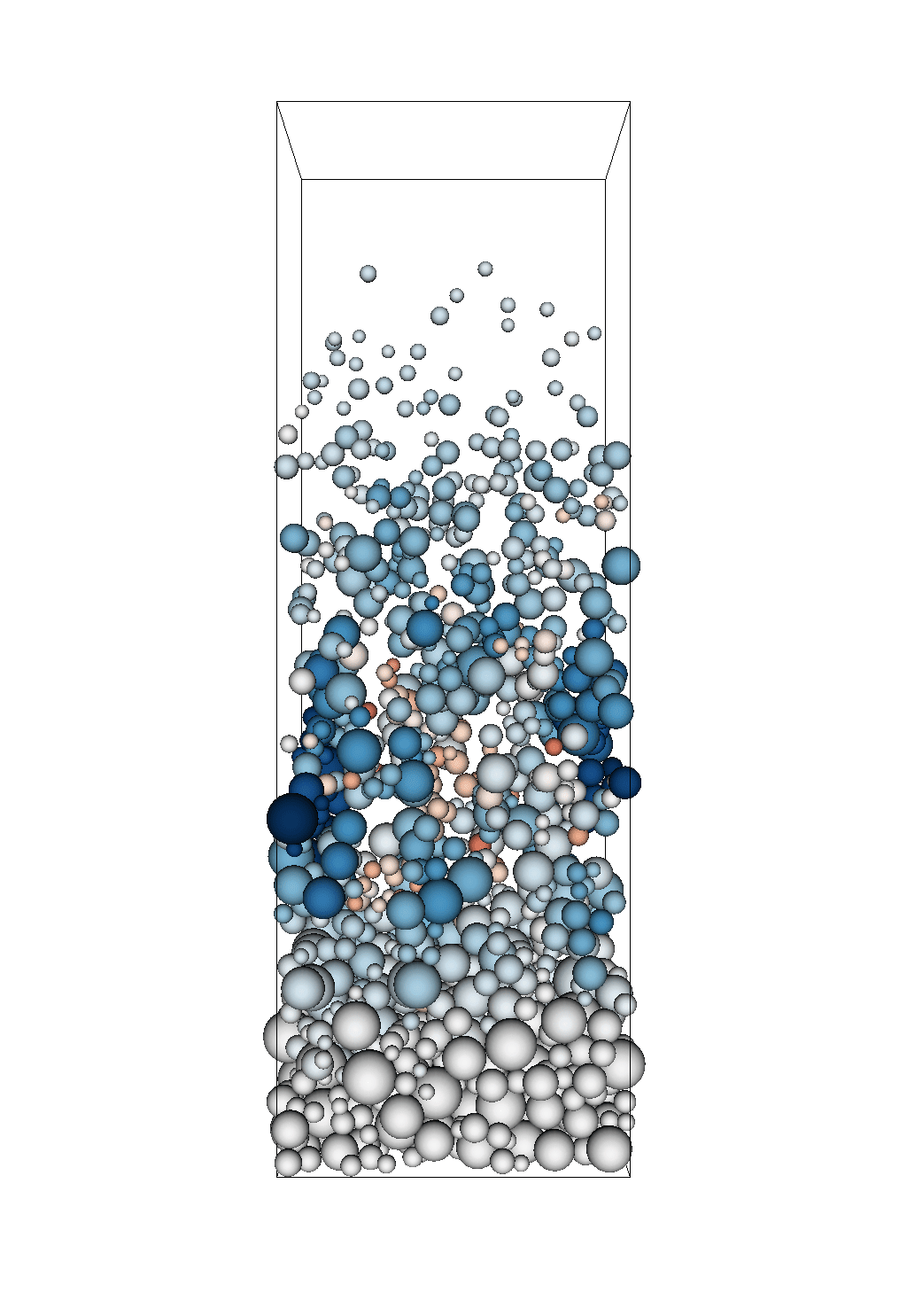}}     
  \put( 9.4,  5.9 ){\includegraphics[trim={9cm 0 0 0},clip, height=0.3\textheight]{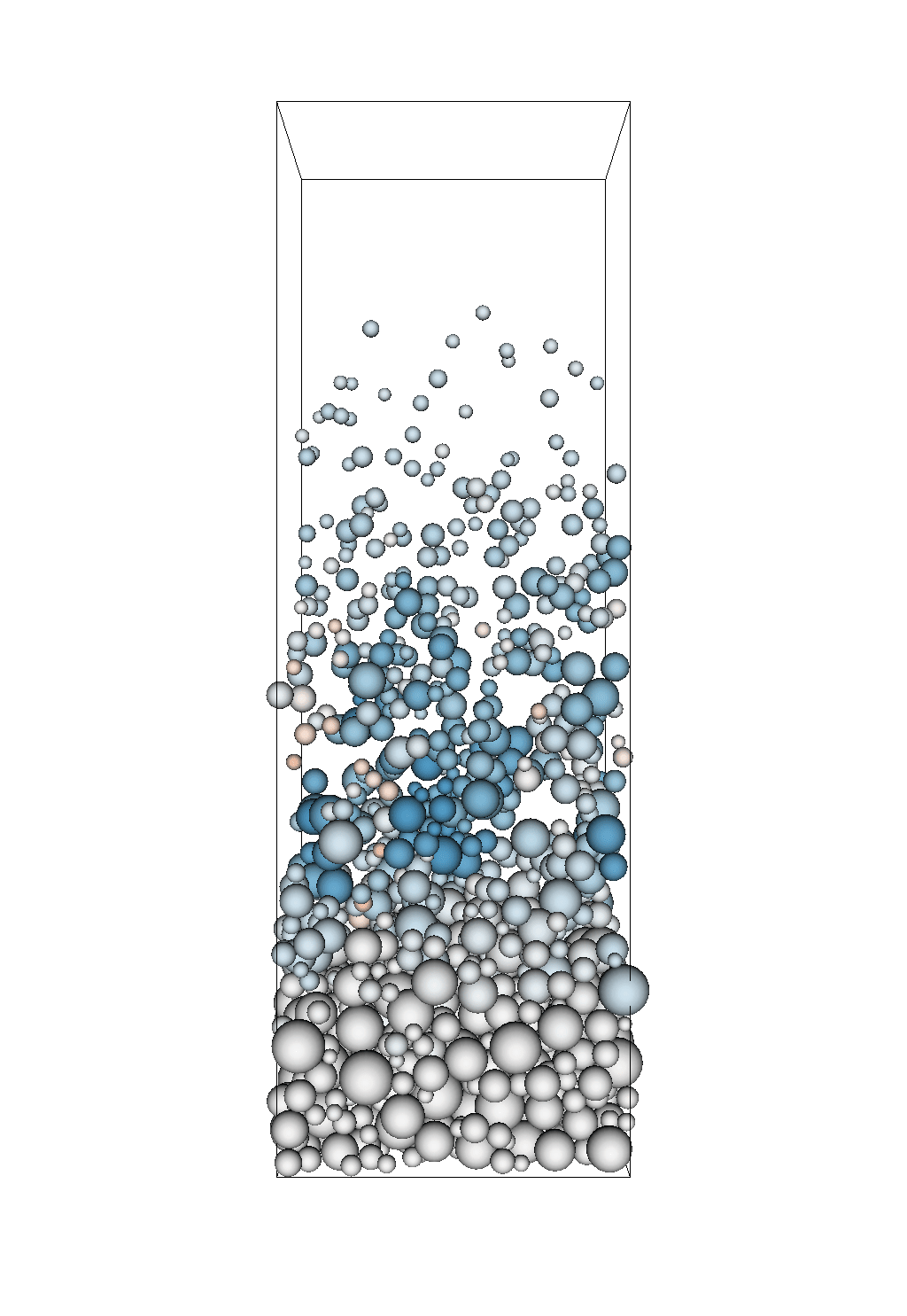}}       
  \put(11.5,  5.9 ){\includegraphics[trim={9cm 0 0 0},clip, height=0.3\textheight]{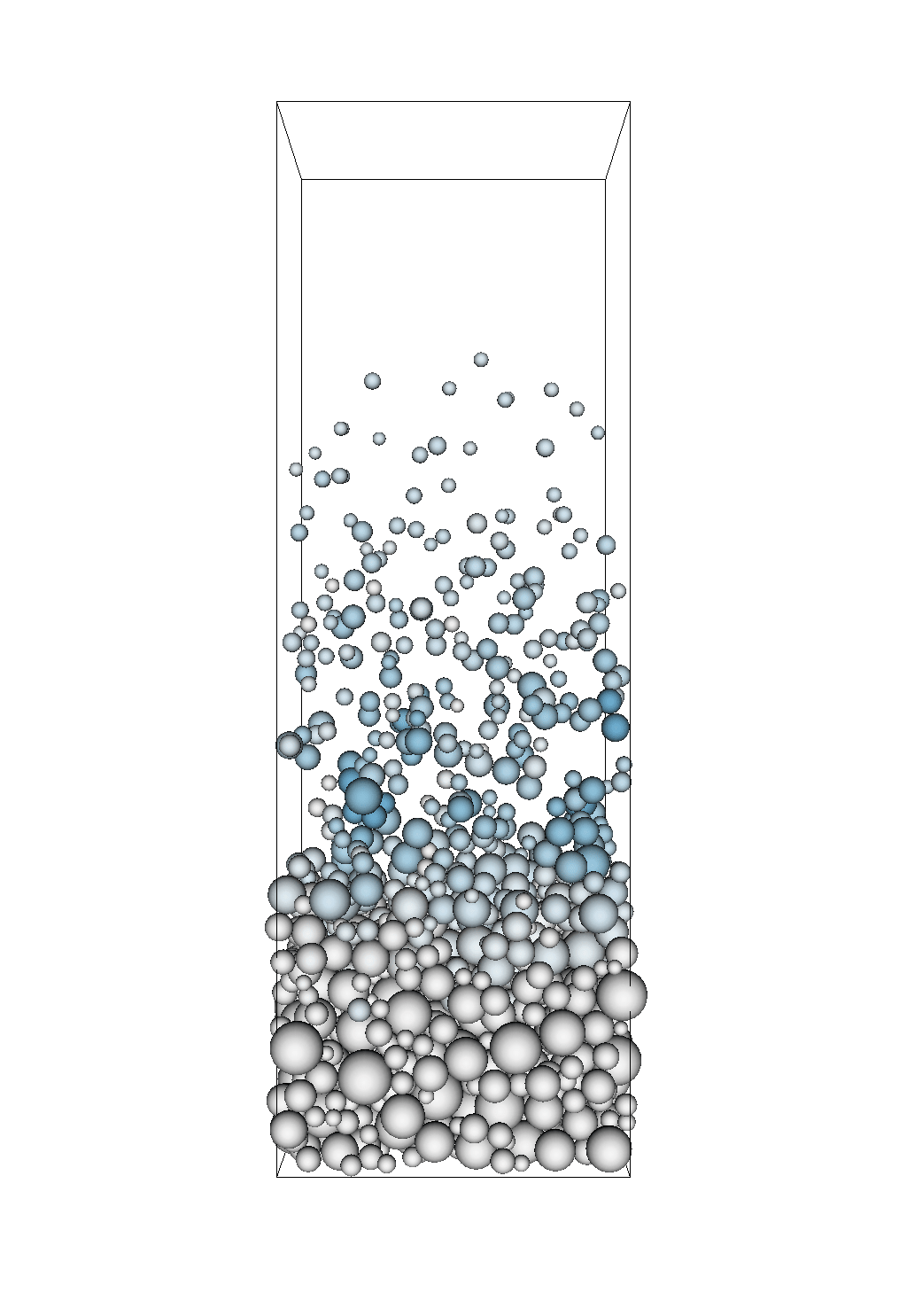}}      
  \put( 0.0, -0.0 ){\includegraphics[                       height=0.3\textheight]{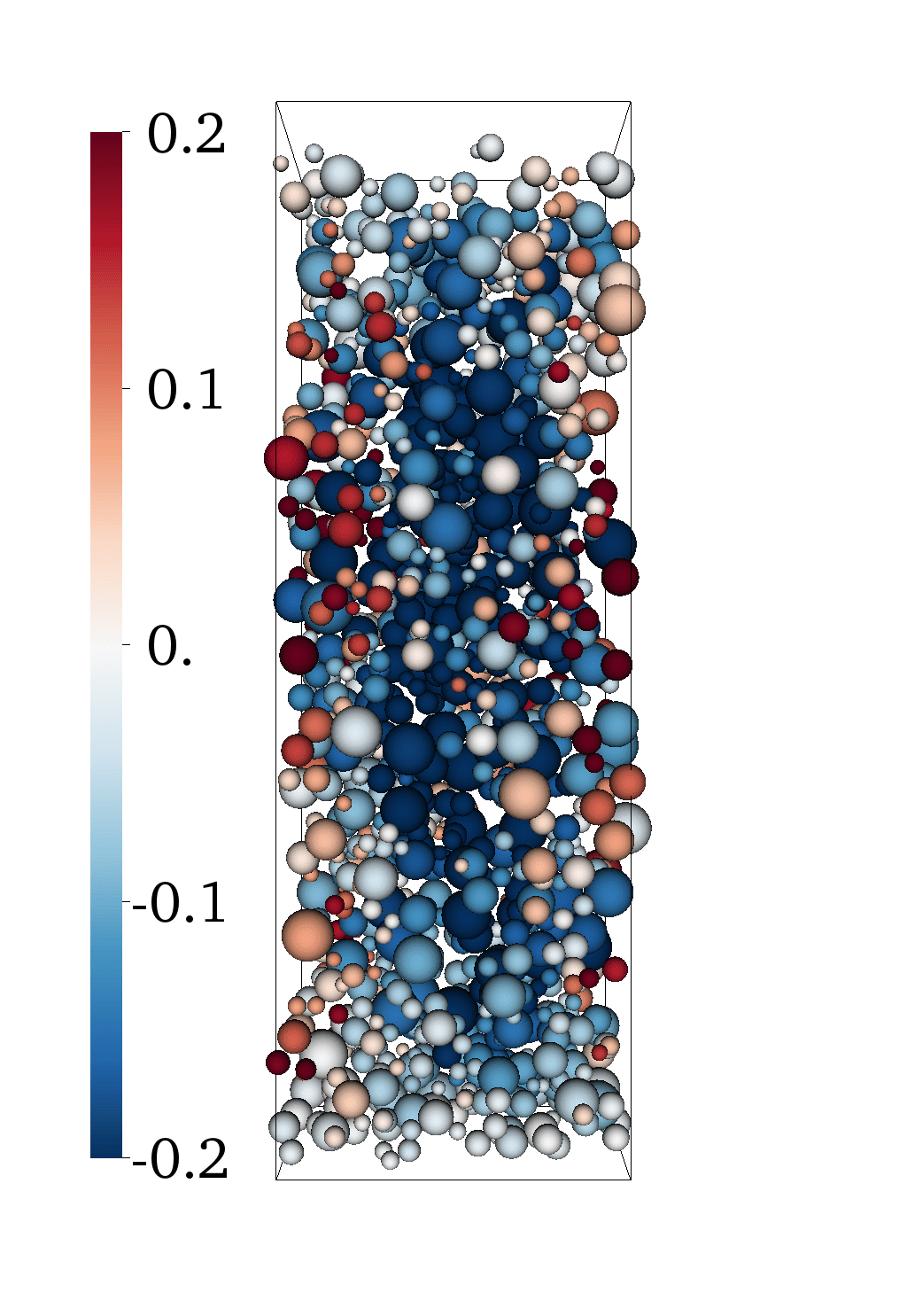}}
  \put( 3.1, -0.0 ){\includegraphics[trim={9cm 0 0 0},clip, height=0.3\textheight]{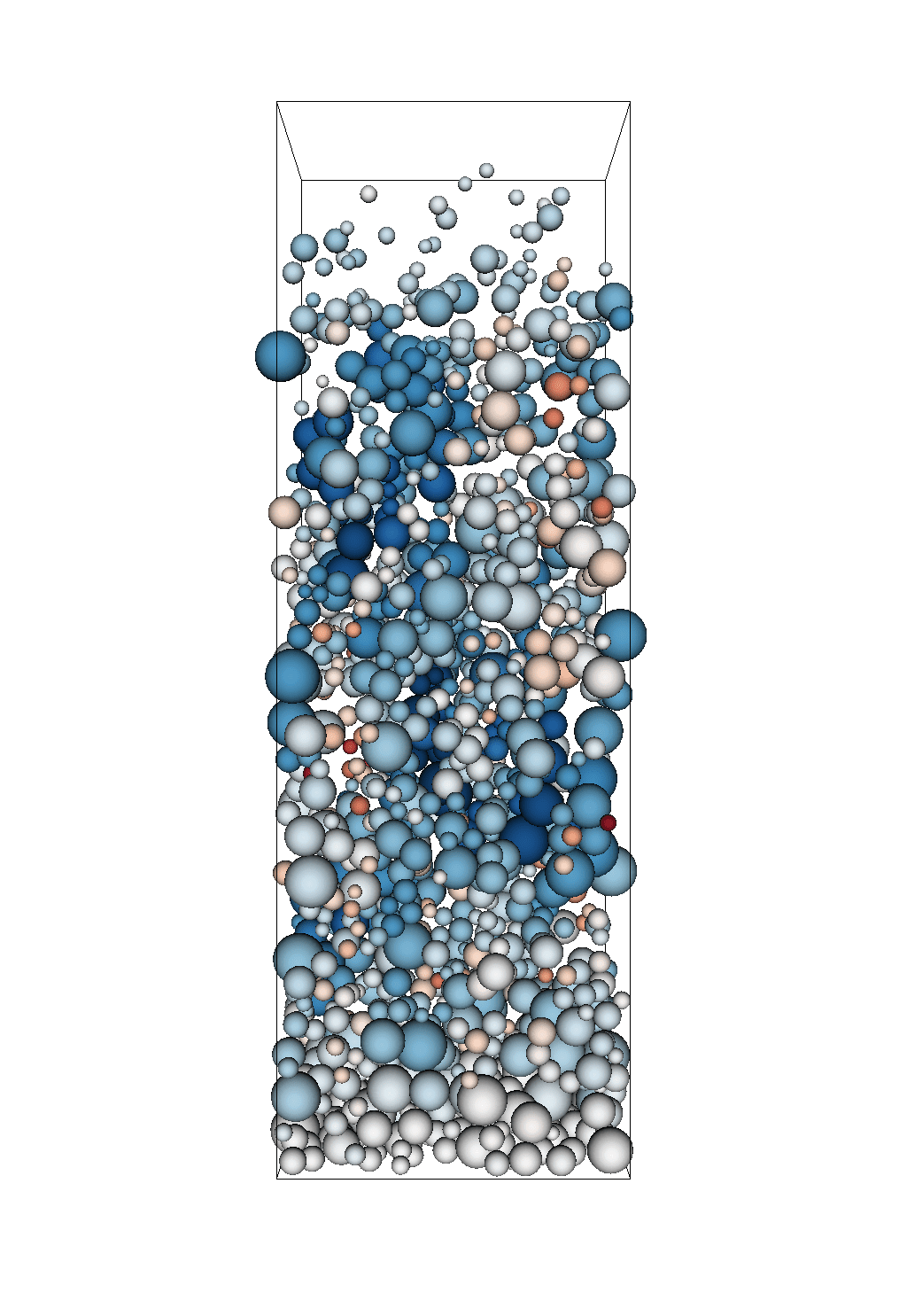}}  
  \put( 5.2, -0.0 ){\includegraphics[trim={9cm 0 0 0},clip, height=0.3\textheight]{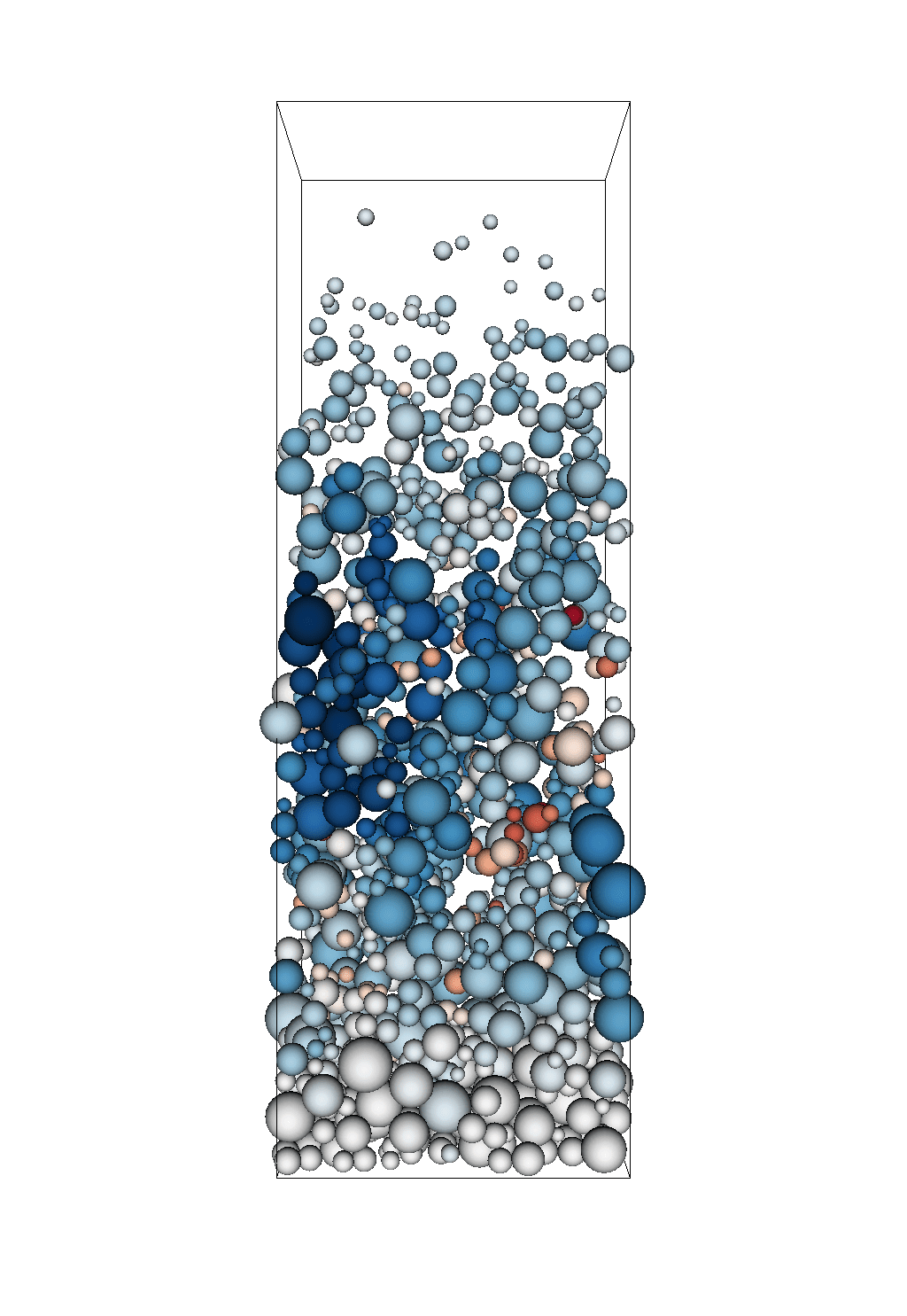}}  
  \put( 7.3, -0.0 ){\includegraphics[trim={9cm 0 0 0},clip, height=0.3\textheight]{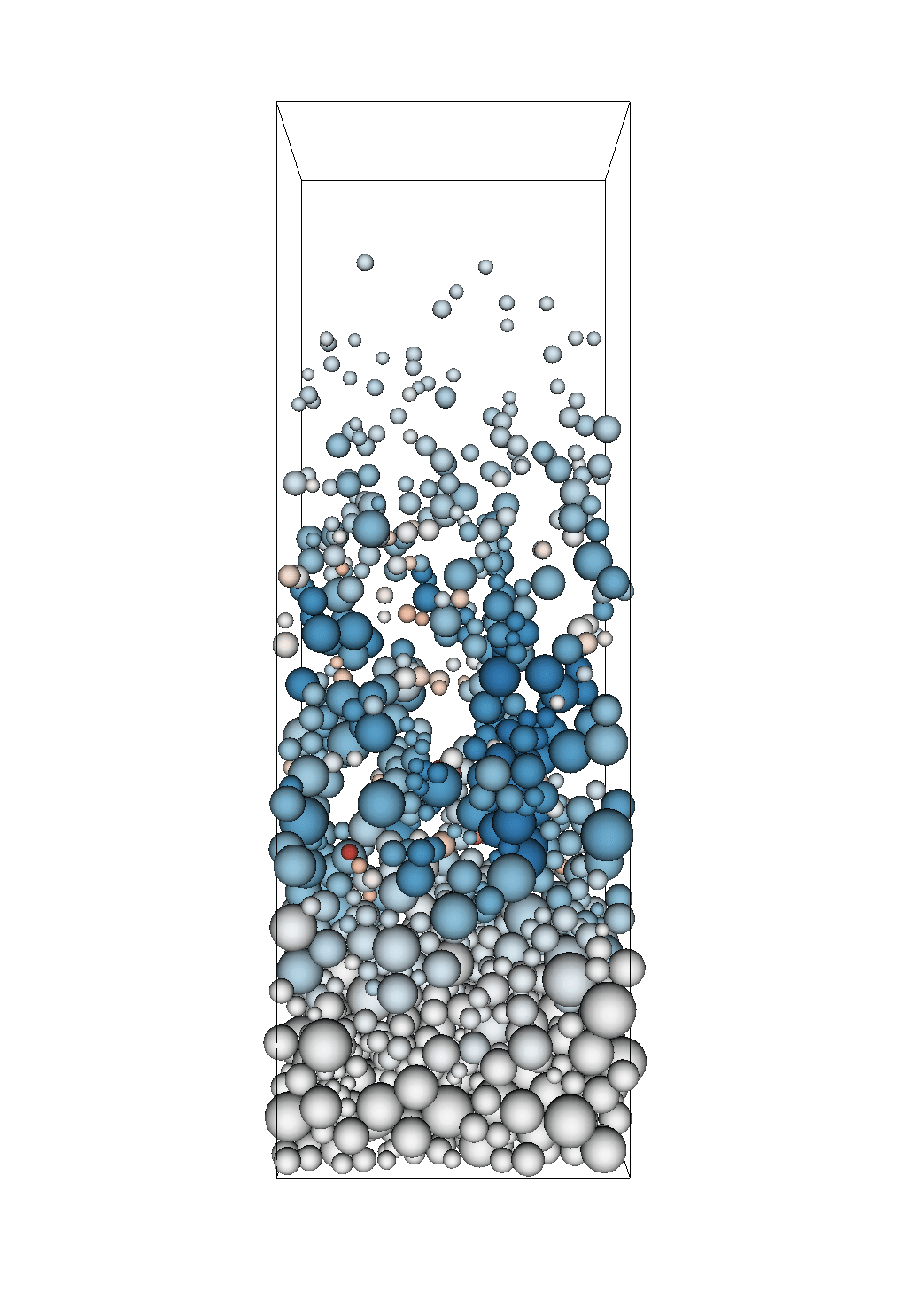}}  
  \put( 9.4, -0.0 ){\includegraphics[trim={9cm 0 0 0},clip, height=0.3\textheight]{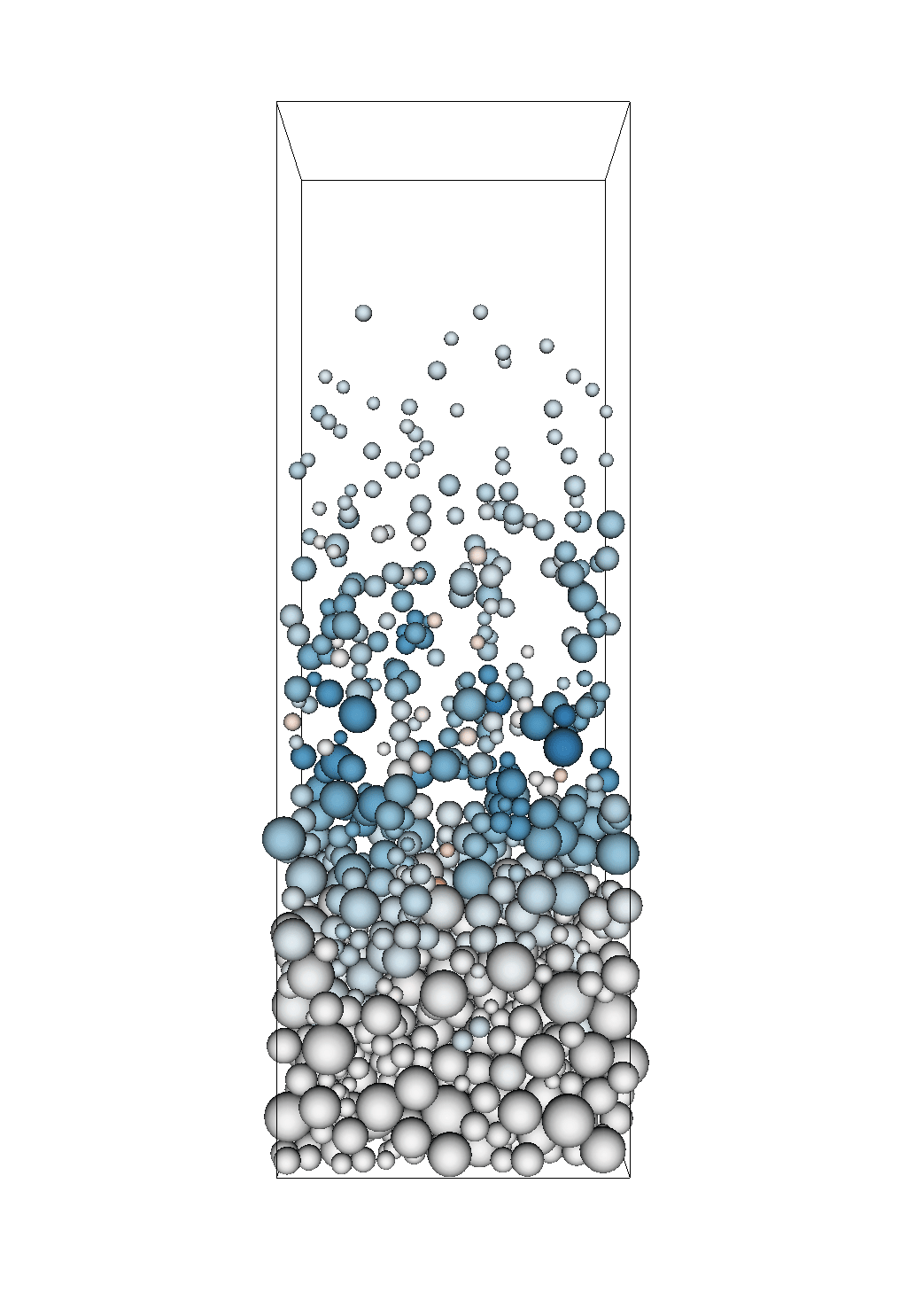}}  
  \put(11.5, -0.0 ){\includegraphics[trim={9cm 0 0 0},clip, height=0.3\textheight]{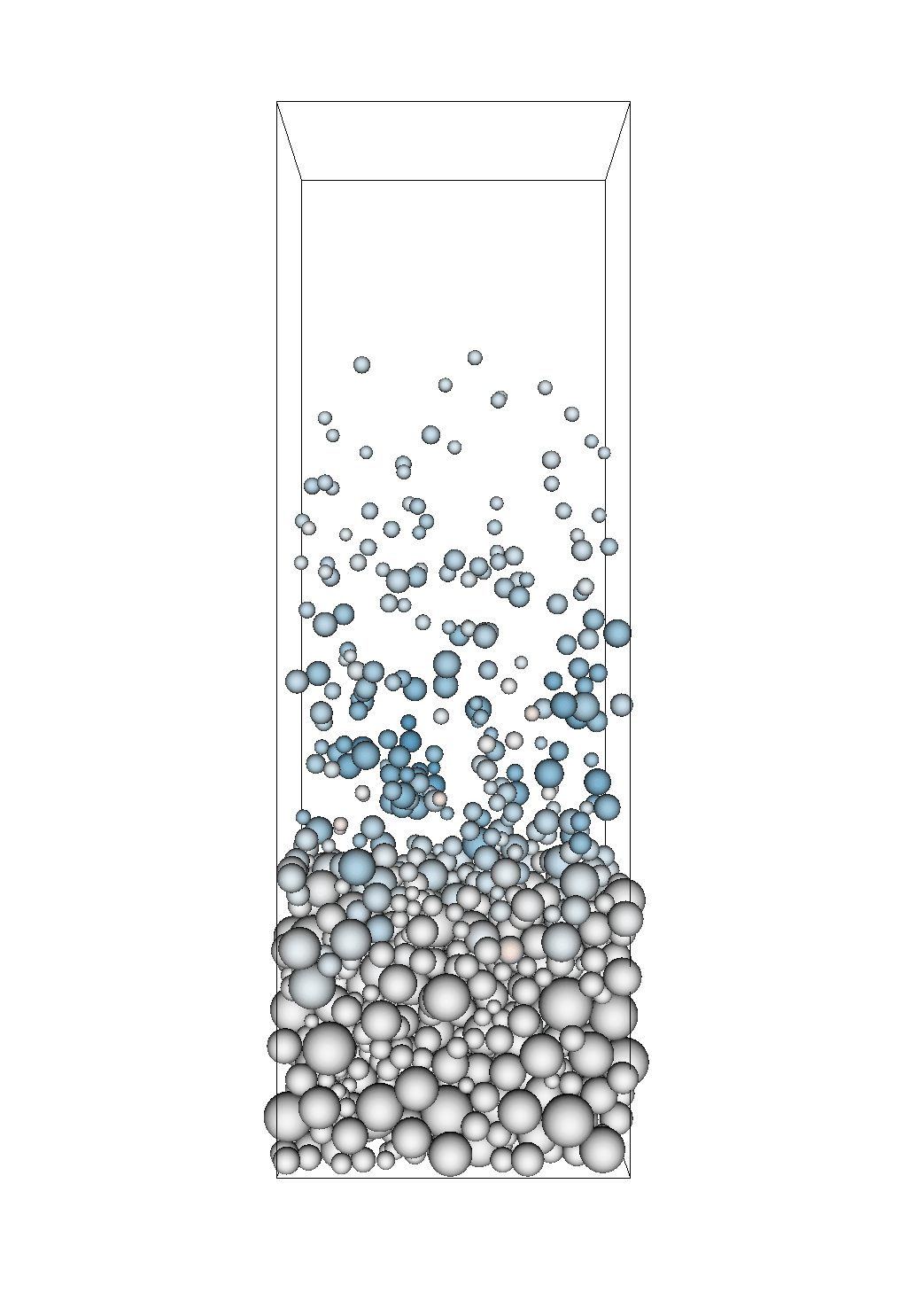}}  
    \put( 0.1, 17.6)     {$v_p / u_{s}$}                             
    \put( 2.1, 17.7)     {a) }
    \put( 4.1, 17.7)     {b) }    
    \put( 6.1, 17.7)     {c) }        
    \put( 8.2, 17.7)     {d) }            
    \put(10.3, 17.7)     {e) }                
    \put(12.4, 17.7)     {f) }      
    \put( 0.1, 11.8)     {$v_p / u_{s}$}                             
    \put( 2.1, 11.9)     {g) }
    \put( 4.1, 11.9)     {h) }    
    \put( 6.1, 11.9)     {i) }        
    \put( 8.2, 11.9)     {j) }            
    \put(10.3, 11.9)     {k) }                
    \put(12.4, 11.9)     {l) }      
    \put( 0.1, 6.0)     {$v_p / u_{s}$}                             
    \put( 2.1, 6.1)     {m) }
    \put( 4.1, 6.1)     {n) }    
    \put( 6.1, 6.1)     {o) }        
    \put( 8.2, 6.1)     {p) }            
    \put(10.3, 6.1)     {q) }                
    \put(12.4, 6.1)     {r) }       

\end{picture}
        \caption{Particle configurations during the settling process. Top row: $Co=0$, middle row: $Co=1$, bottom row: $Co=5$. Left column: $t=17.6 \tau_s$, which corresponds to the time at which the particle phase has its maximum kinetic energy. From left to right, the columns are separated by time intervals of $72.5 \tau_s$. The gray shading reflects the vertical particle velocity. The cohesive sediment is seen to settle more rapidly than its noncohesive counterpart (see also supplementary movie).}
    \label{fig:settling}
\end{figure}
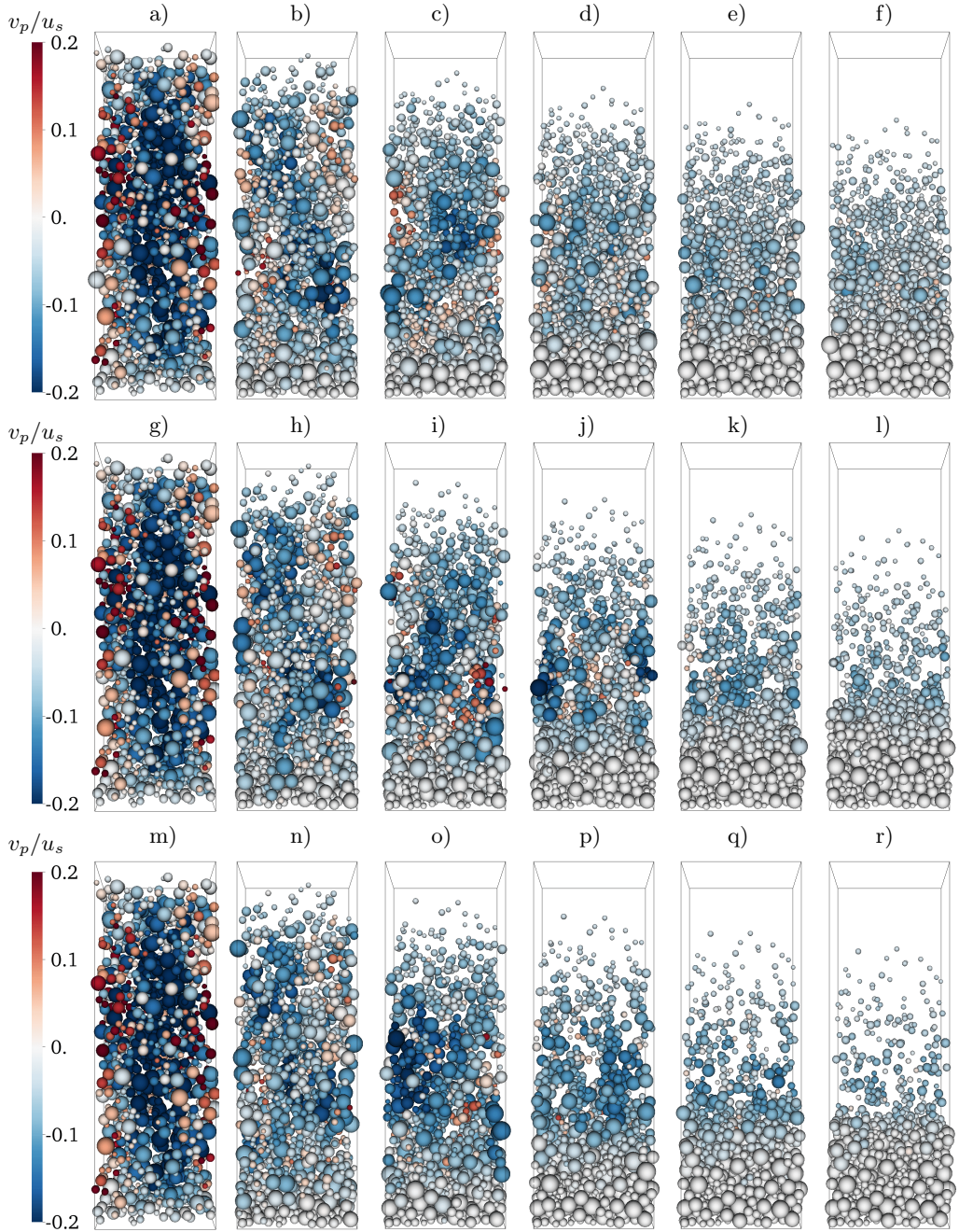
\setlength{\unitlength}{1cm}
\begin{figure}
\begin{picture}(7,7.2)  
  \put(0.00,  0.0 ){\includegraphics[width=0.33\textwidth]  {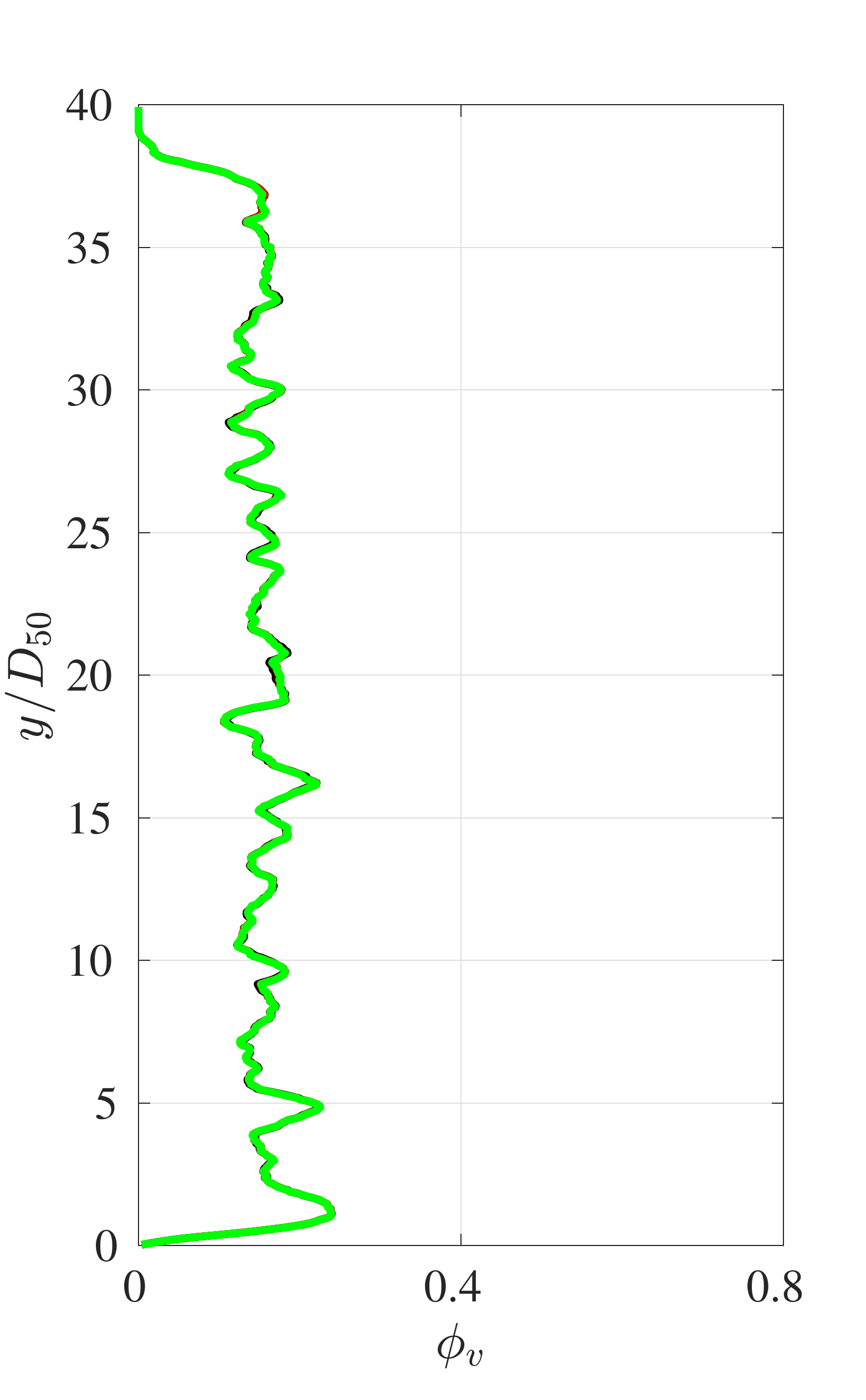}}
  \put(4.5 ,  0.0 ){\includegraphics[width=0.33\textwidth]  {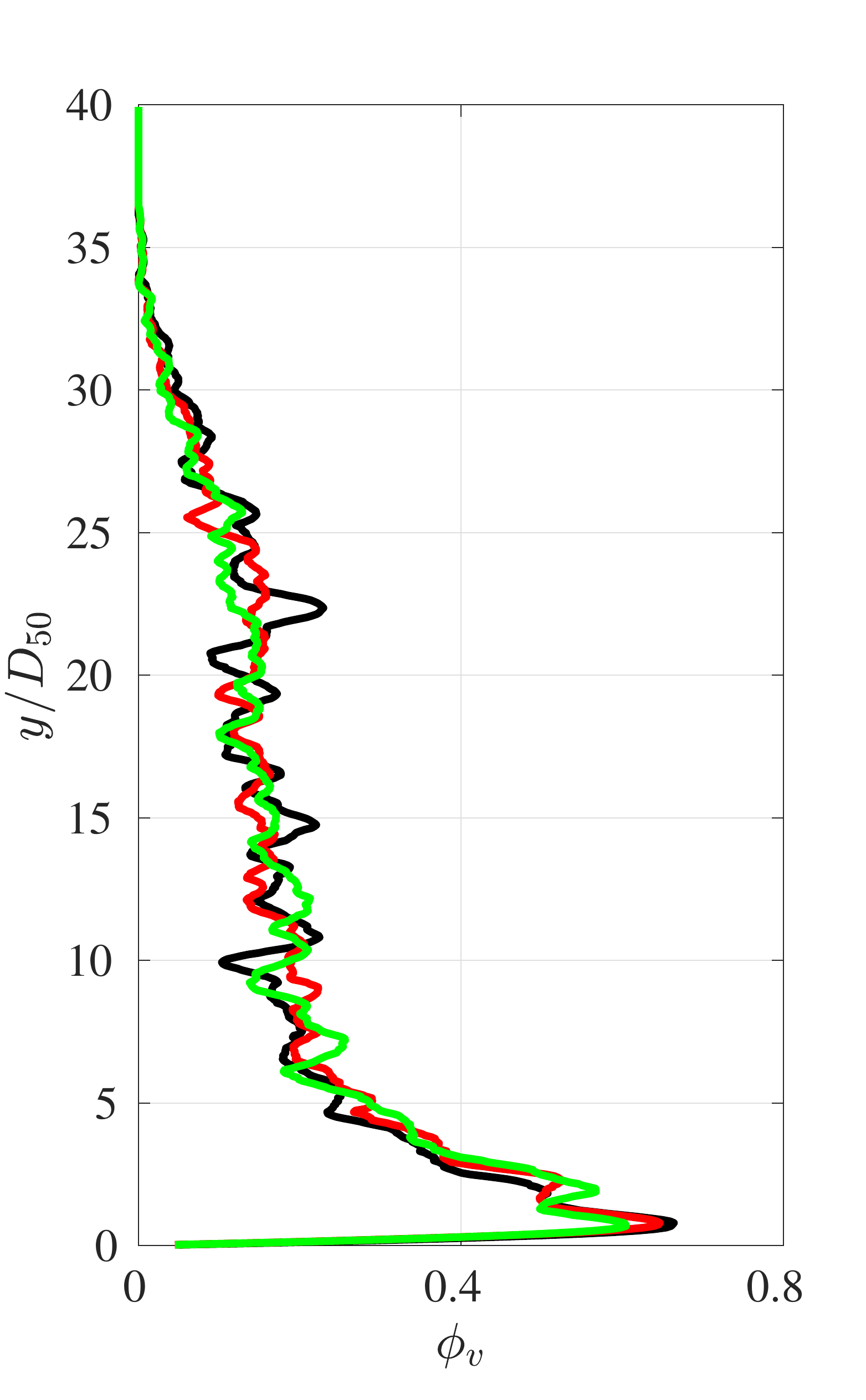}}  
  \put(9.00, -0.05 ){\includegraphics[width=0.325\textwidth]{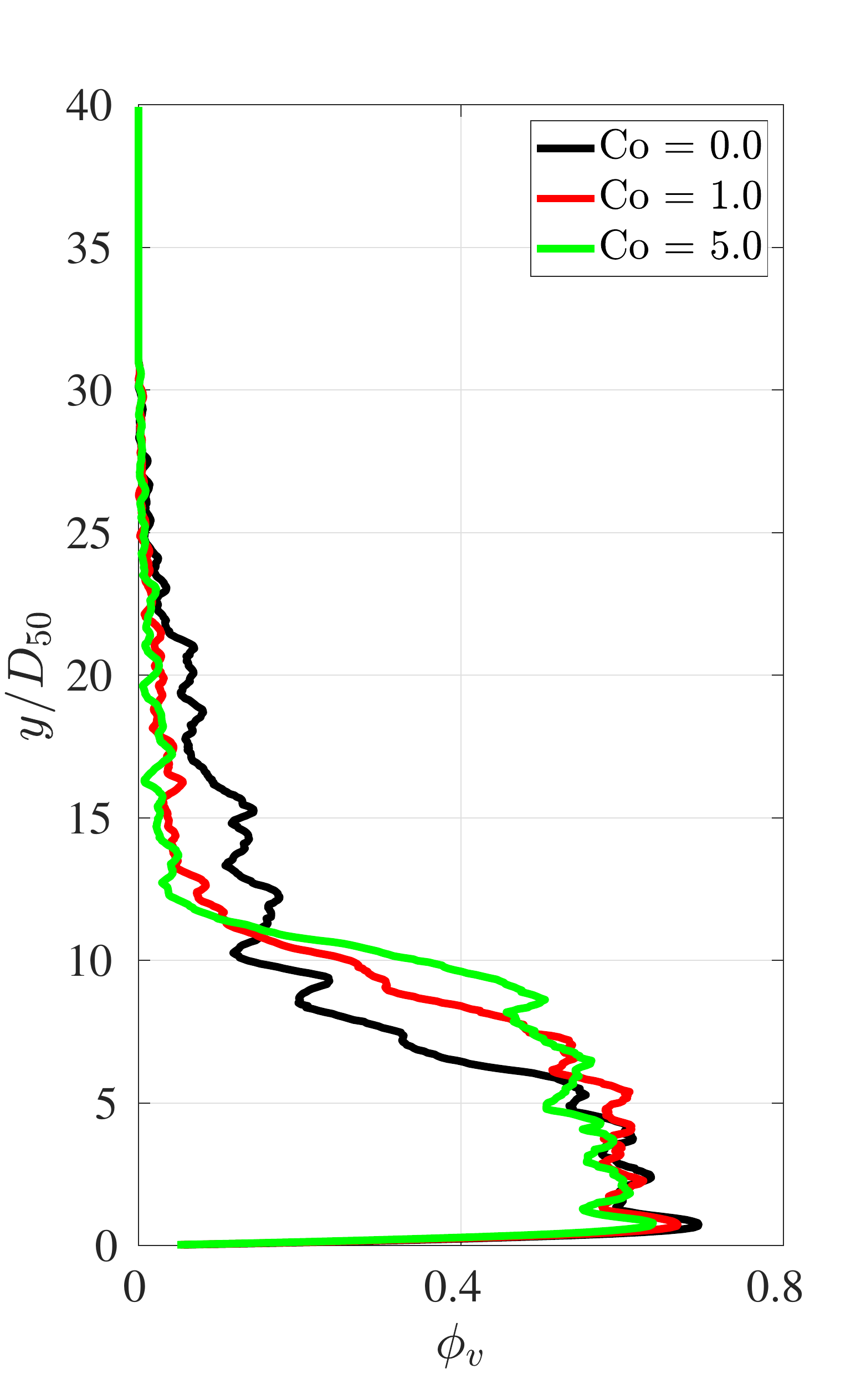}}  
    \put(0.0,  6.8)     {a) }
    \put(4.5,  6.8)     {b) }       
    \put(9.0,  6.8)     {c) }           
\end{picture}
        \caption{Horizontally averaged particle volume fractions during the settling process. (a) at $t=17.6 \tau_s$, (b) at $t=162.6 \tau_s$, and (c) at $t=380.1 \tau_s$. }
    \label{fig:settling_concentration}
\end{figure}
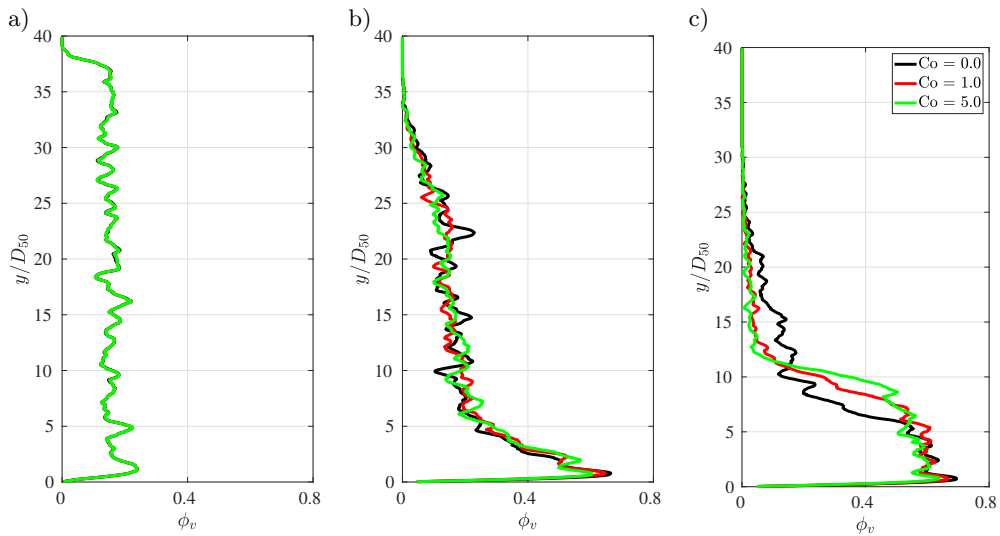
\setlength{\unitlength}{1cm}
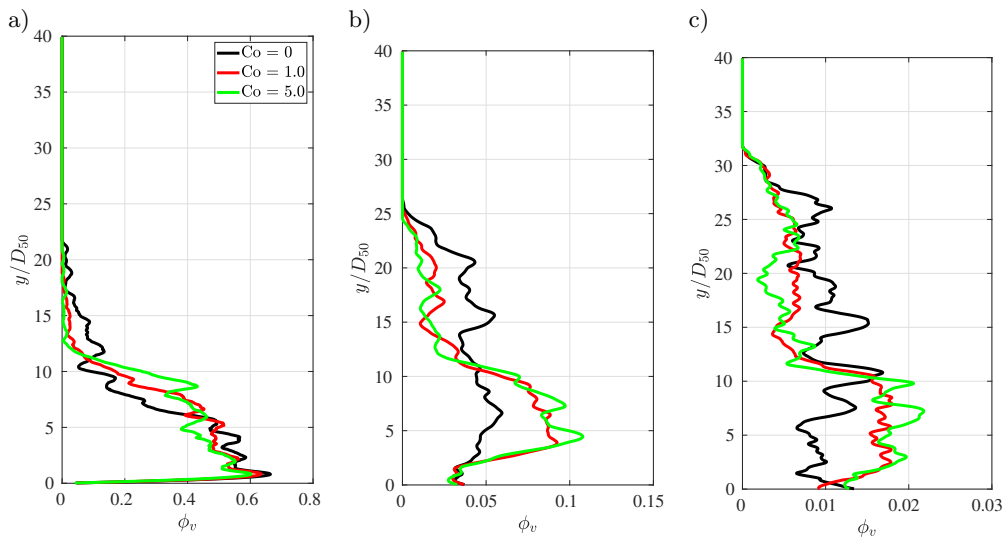
\begin{figure}
\begin{picture}(7,7.2)
  \put(0.00,   0.0 ){\includegraphics[width=0.33\textwidth]  {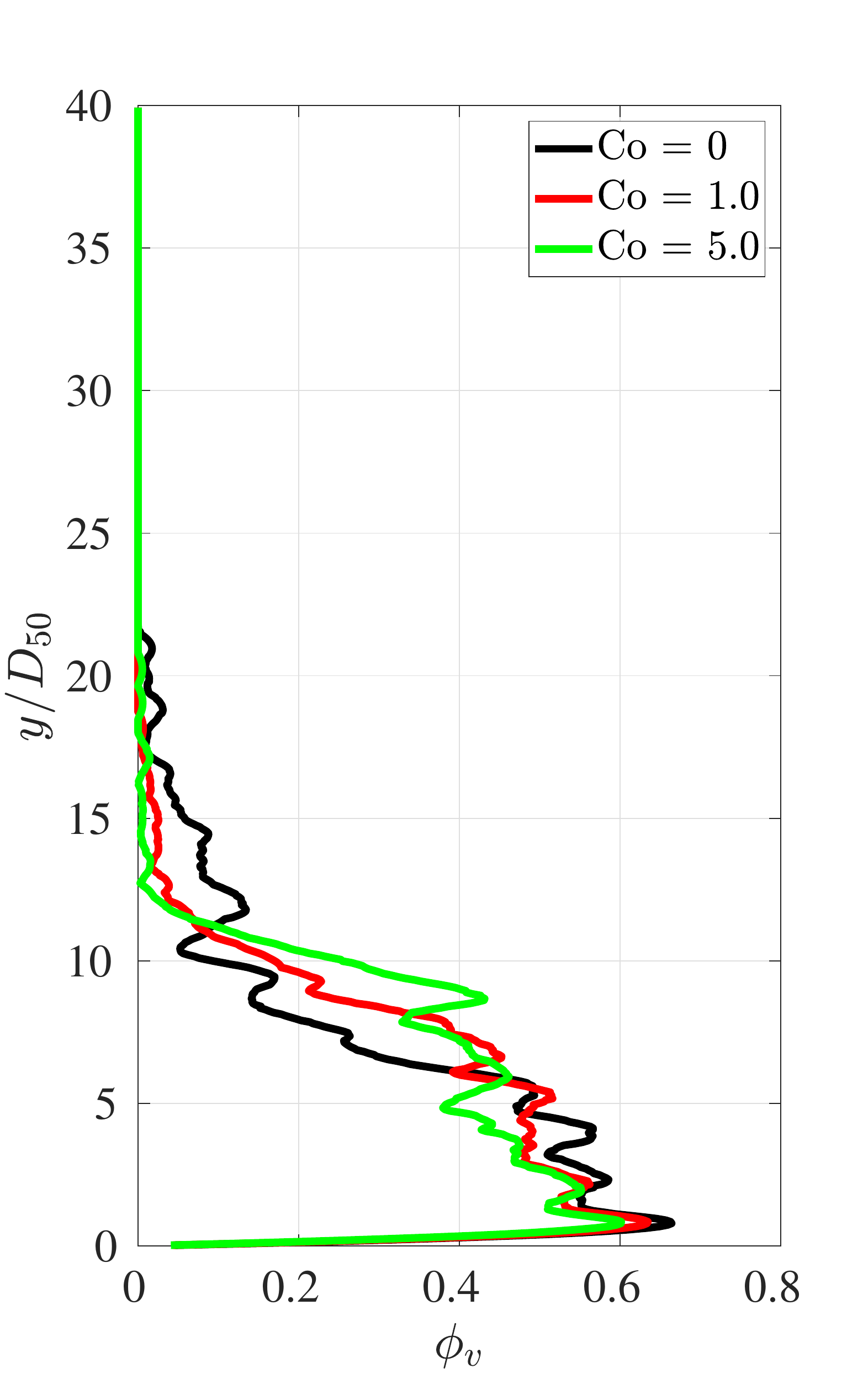}}
  \put(4.5 ,   0.0 ){\includegraphics[width=0.33\textwidth]  {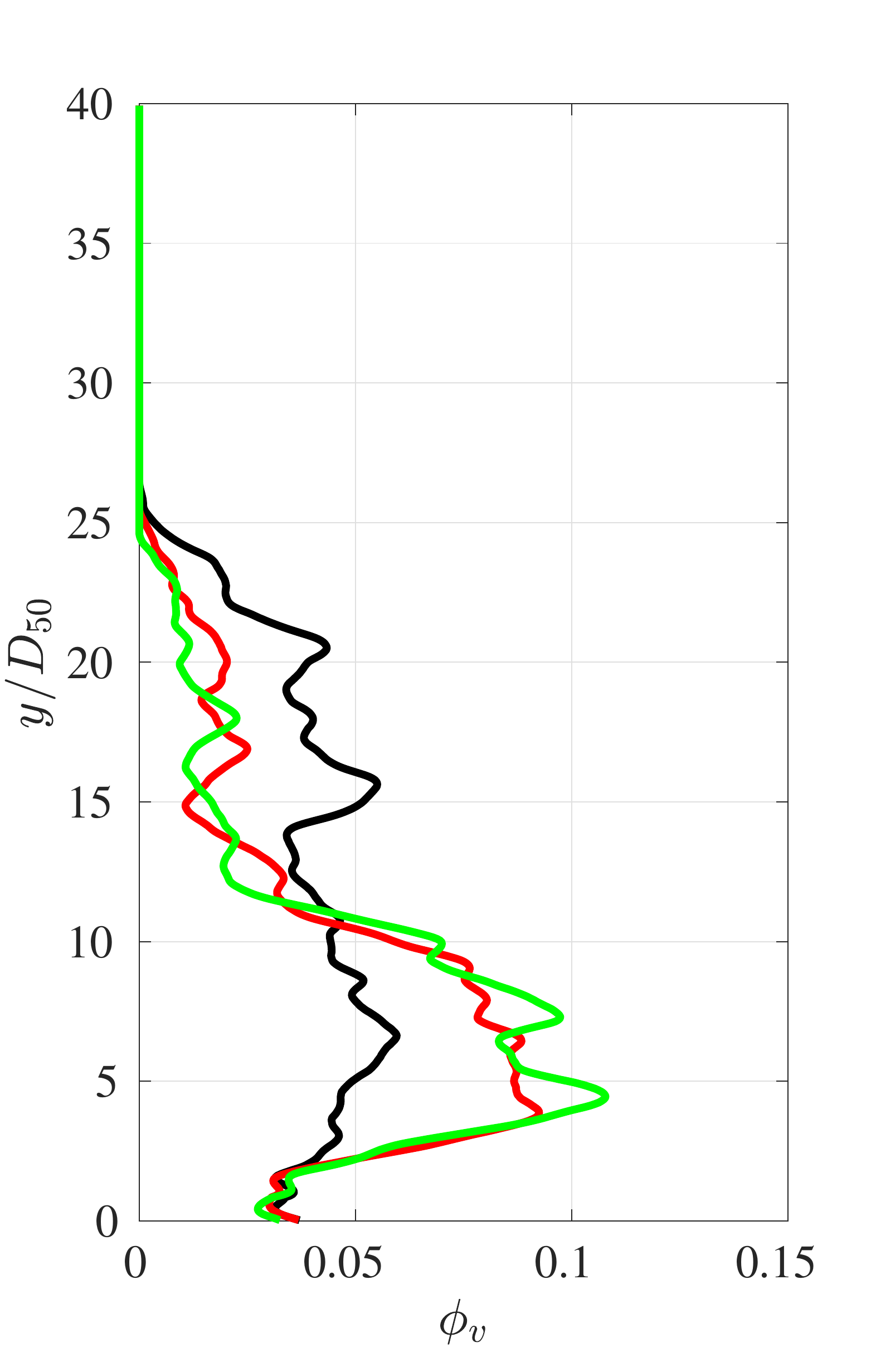}}
  \put(9.00,  -0.05 ){\includegraphics[width=0.328\textwidth]{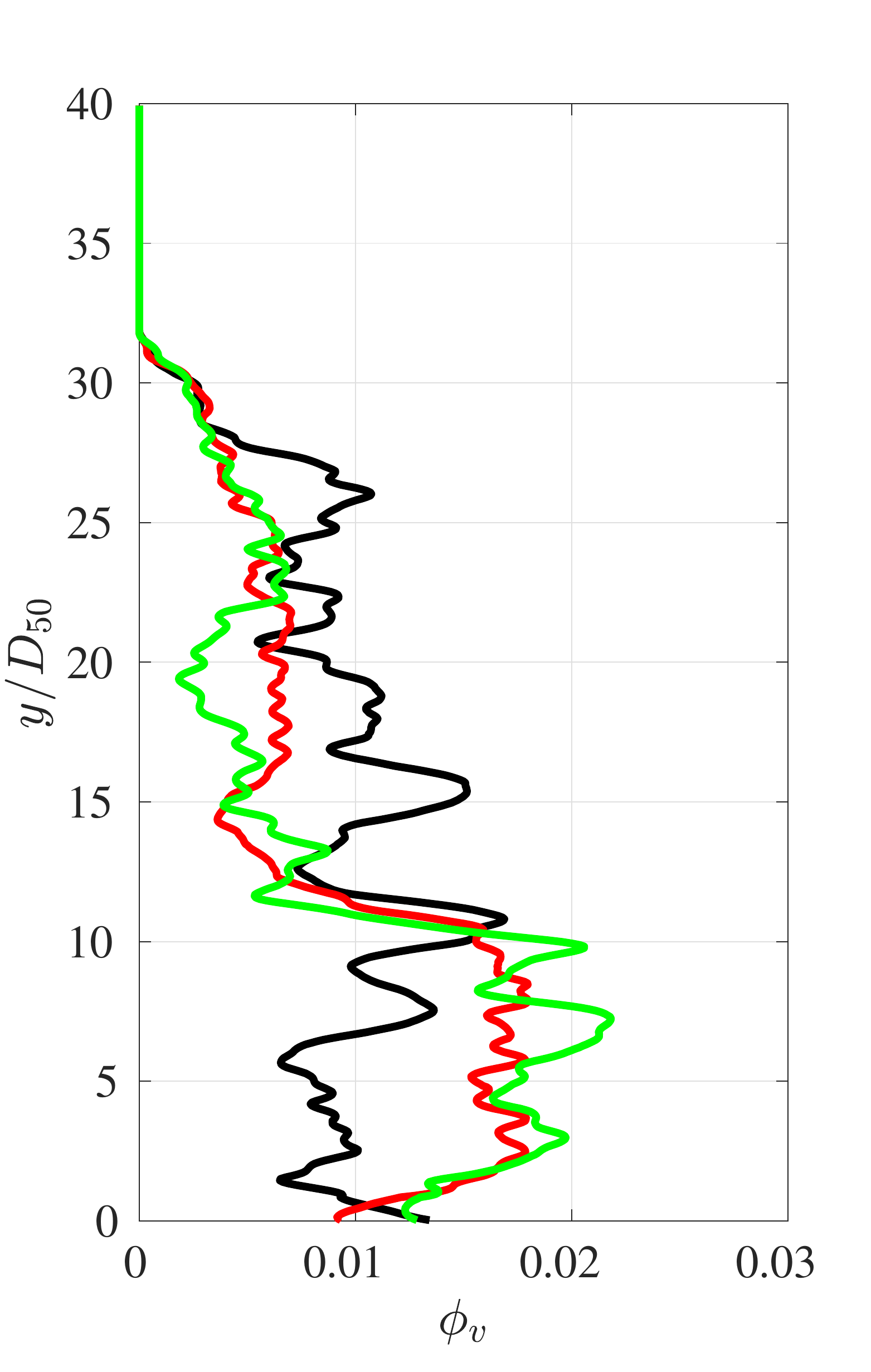}}
    \put(0.0,  6.8)     {a) }
    \put(4.5,  6.8)     {b) }    
    \put(9.0,  6.8)     {c) }       
\end{picture}
        \caption{Particle volume fraction profile for different particle radii \rtwo{at $t=380.1 \tau_s$}: (a) small particles with $D_p \leq D_{33}$, (b) medium sized particles in the range $D_{33} < D_p \leq D_\text{66}$ , and (c) large particles with $D > D_{66}$. Note the different horizontal axis scalings for the individual frames. The results in (a) and (b) were smoothed by a moving average with filter width of $1.5 D_{50}$ for clarity.
        }
    \label{fig:final_distribution}
\end{figure}
%

 \subsection{Hindered settling behavior}\label{sec:settling_behavior}
The impact of cohesive forces on the settling behavior is illustrated by figure \ref{fig:settling}. During the early stages the particle distributions are very similar for all three simulations. Over the course of the simulations, however, the cohesive sediment is seen to settle faster than its noncohesive counterpart. This qualitative observation is confirmed by the concentration profiles of figure \ref{fig:settling_concentration}. At $t=17.6 \tau_s$, when the particle phase has its maximum kinetic energy (cf. \S \ref{sec:budget}), the profiles for all three simulations remain nearly identical, as cohesive forces have not yet had sufficient time to cause a noticeable change. As time progresses, the concentration profiles remain very similar in the dilute region near the top of the tank, where the volume fraction remains below $5\%$ so that particle-particle interactions are negligible \citep{capart2011}, cf. figure \ref{fig:settling_concentration}b. In the lower part of the tank, differences begin to emerge, as cohesive forces result in the formation of flocs with larger settling speeds, so that particles accumulate at the bottom of the tank more quickly. Also, note that the undulations in the profiles are milder for larger cohesive forces. At the final simulation time (figure \ref{fig:settling_concentration}c), cohesive sediment has a lower volume fraction at the very bottom of the tank at ($0 \leq y \leq 2 D_\text{50}$), as compared to cohesionless grains. This reflects the impact of cohesive forces on the consolidation process, as larger cohesive forces yield stable flocs, whereas cohesionless sediment rearranges itself into a denser configuration under the weight of the overlying sediment \citep{been1981}. Above the dense layer of sediment at the very bottom of the tank ($0 \leq y \leq 5 D_\text{50}$), there exists a layer of loosely flocculated particles with a lower volume fraction, which increases in depth over time. This observation holds for both $\text{Co}=1$ and $\text{Co}=5$, and is consistent with experimental observations of freshly sedimented flocs with larger pore spaces above older sediments \citep[e.g.][]{winterwerp2001}. At the final simulation time, we observe a sharp decrease in particle volume fraction near $y \approx 10 D_\text{50}$ for the cohesive sediment, which is more pronounced for the simulation with larger cohesive forces. The cohesionless sediment, on the other hand, shows higher volume fractions at $y > 10 D_\text{50}$, as a result of the unfinished settling process. 

As the particles settle, they replace fluid at the bottom of the tank and generate an upward counterflow. For the current simulations, this counterflow is sufficiently strong to sweep smaller particles upward. It represents one reason for hindered settling and for the separation of the grain sizes for very large water columns \citep{teslaa2015}. This effect is illustrated in figure \ref{fig:final_distribution}, which shows the final volume fraction profiles for the smallest, intermediate and largest third of the particles. The figure demonstrates that small and intermediate cohesive particles settle much more rapidly than their noncohesive counterparts, consistent with the observation of \cite{lick2004}, who found intermediate size particles to be most strongly affected by cohesive forces. These types of grains have their peak concentration in the interval $3 \, D_{50} < y < 10 \, D_{50}$. In contrast, large cohesive grains have a lower volume fraction near the bottom of the tank than large cohesionless grains (figure \ref{fig:final_distribution}c). As a consequence, the effect of size segregation is less pronounced for cohesive grains, as flocculation results in particles of different sizes settling with the same velocity \citep{mehta1989}. 

To validate our simulation results, we compare the effect of cohesive forces on the settling of our polydisperse particles to established empirical relations describing hindered settling of silt. We can compute the settling behavior  of the particle phase in a double-averaged sense \citep{vowinckel2017a}. To this end, we apply an averaging operator to our Eulerian fluid grid that evaluates instantaneous snapshots of the particle velocity distribution $(\phi \, v_p)$, where $\phi$ is the part of a cell occupied by solids and $v_p$ is the settling velocity of the particle taking up this space. Note that we assume rigid-body motion and zero rotation for this analysis. This yields
\begin{equation}\label{eq:horizontal_averaging}
 \langle v_p \rangle(y,t) = \frac{\int_0^{L_z}\int_0^{L_x} \phi(x,y,z,t) \, v_p(x,y,z,t) \, \text{d}x \, \text{d}z \,\text{d}t}{\int_0^{L_z}\int_0^{L_x} \phi(x,y,z,t) \, \text{d}x \,\text{d}z \,\text{d}t} \qquad ,
\end{equation}
where the angular brackets denote horizontal averaging. This data is then evaluated for binned values of $\phi$ using
\begin{equation}\label{eq:double_averaging}
 \overline{\langle v_p \rangle}(\phi) = \frac{1}{N_t N(\phi)}\sum_{N_t}\sum_{N(\phi)} \langle v_p \rangle(y,t) \qquad ,
\end{equation}
where $N(\phi)$ is the number of samples recorded for a given $\phi$, and $N_t$ is the number of the evaluated datasets in time outputted with an interval of $\Delta_t=0.25\tau_s$. Time averaging indicated by the overbar is performed from $t_s = 240 \tau_s$ to $t_e = 380 \tau_s$, which is the end of the simulation as displayed in figure \ref{fig:settling_concentration}c. The averaging time interval was chosen to start well after the initial stir-up phase, as will be discussed in detail below. The sample size hence comprises $N_t = 560$ datasets over a total time of $140 \tau_s$, which is long enough to obtain statistically meaningful data of the well developed settling behavior. Similarly, we evaluate the undisturbed settling velocity  $v_{ra}$ (cf. Appendix \ref{app:char_vel}) for the instantaneous particle distribution $\phi$ to compute $\overline{\langle v_{ra} \rangle}$. As a result, the ratio $\overline{\langle v_p \rangle}/\overline{\langle v_{ra} \rangle}$ is still a function of the volume fraction and can therefore be used to compare with the hindered settling functions available in literature.

\setlength{\unitlength}{1cm}
\begin{figure}
\begin{picture}(7,8.4)
  \put(0.0 ,   4.2 ){\includegraphics[width=\textwidth]{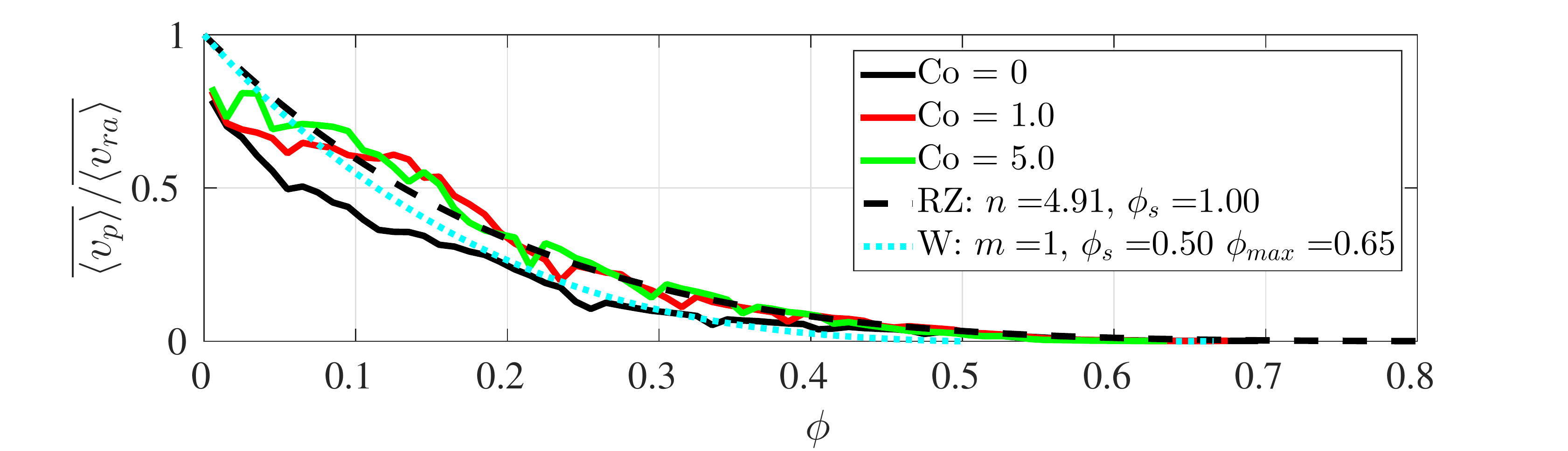}}
  \put(0.0 ,   0   ){\includegraphics[width=\textwidth]{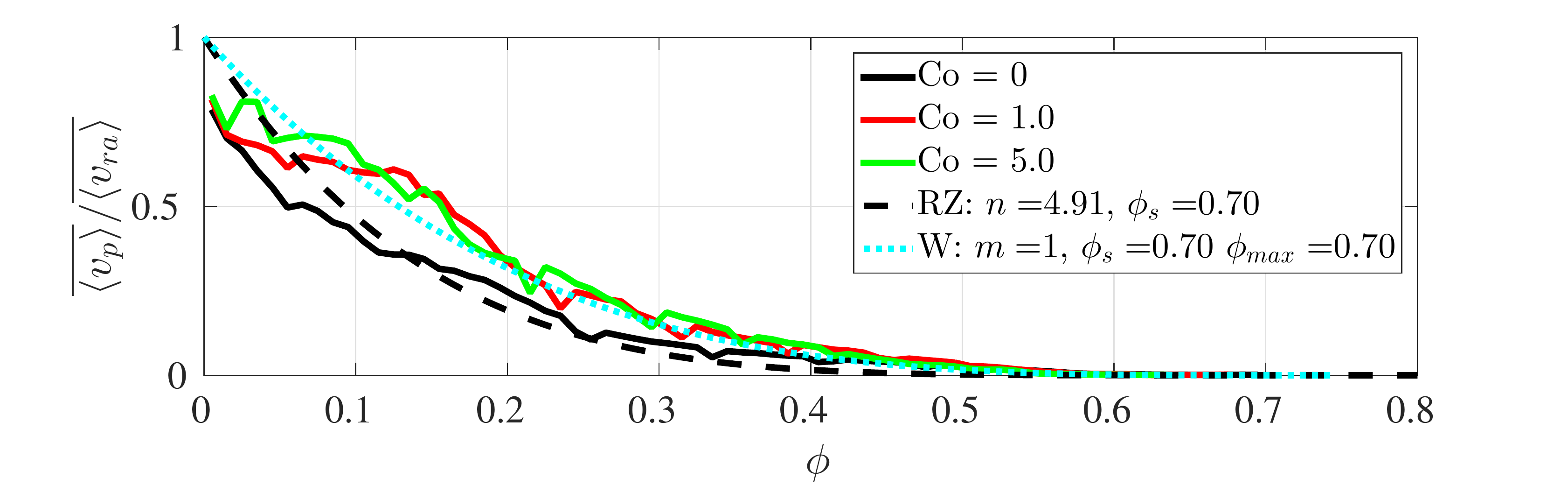}}
    \put(0.0,  8.0)     {a) }
    \put(0.0,  3.8)     {b) }    
\end{picture}
        \caption{Double-averaged settling velocity normalized with the undisturbed settling velocity. Comparison to empirical relationships \eqref{eq:RZ_settling} of  \cite{richardson1954} (RZ) and \eqref{eq:teslaa_settling} of \cite{winterwerp2002} (W). (a) parameterization according to \cite{teslaa2015}, (b) same parameterization except for chosing $\phi_s = \phi_{max} = 0.7$ according to the simulation data of figure \ref{fig:settling_concentration}c.
        }
    \label{fig:hindered_settling}
\end{figure}

One of the first hindered settling functions was proposed by \cite{richardson1954} 
\begin{equation}\label{eq:RZ_settling}
 \frac{\overline{\langle v_p \rangle}}{\overline{\langle v_{ra} \rangle}} = \left(1-\frac{\phi}{\phi_s}\right)^n \qquad ,
\end{equation}
where $\phi_s$ and $n$ are empirical parameters describing the volume fraction of a freshly deposited sediment bed and the particle size and shape, respectively. As argued by \cite{dankers2007}, cohesive sediment such as mud with a significant amount of clay deposits at the bottom in a gel-like structure with a volume fraction that is lower than the maximum possible volume fraction $\phi_{max}$. Hence, these authors define $\phi_s<\phi_{max}$ based on measurements of the settling velocity of mud to separate the effects of hindered settling from the consolidation of the deposited sediment. Due to its simplicity, equation \eqref{eq:RZ_settling} has been very popular in hydraulic engineering. However, as described by \cite{teslaa2015}, this function is known to underestimate the settling velocity for higher concentrations. Instead, these authors have used the hindered settling function of \cite{winterwerp2002}
\begin{equation}\label{eq:teslaa_settling}
 \frac{\overline{\langle v_p \rangle}}{\overline{\langle v_{ra} \rangle}} = \frac{\left(1-\frac{\phi}{\phi_s}\right)^m (1-\phi)}{\left(1-\frac{\phi}{\phi_{max}}\right)^{-\frac{5}{2}\phi_{max}}} \qquad,
\end{equation}
where the numerator represents the effects of the counterflow and the increased buoyancy, respectively, while the denominator reflects the increased viscosity of dense suspensions according to \cite{krieger1959}. Here, $m$ is an empirical exponent, which plays a similar role to the parameter $n$ in \eqref{eq:RZ_settling}. It was shown by \cite{teslaa2015} that equations \eqref{eq:RZ_settling} and \eqref{eq:teslaa_settling} both yield good agreement with experimental results for settling coarse silt. The best agreement for \eqref{eq:RZ_settling} was shown for the parameter values $n=4.91$ and $\phi_s=1$, while \eqref{eq:teslaa_settling} performs best with $m=1$, $\phi_s = 0.5$ and $\phi_{max} = 0.65$. 

We compare our double-averaged simulation results obtained from \eqref{eq:double_averaging} to the hindered settling functions as parameterized by \cite{teslaa2015} in figure \ref{fig:hindered_settling}a \rtwo{to validate our simulations. Our results} agree well with the two empirical hindered settling functions, and they demonstrate the enhanced settling velocity due to the cohesive forces for all concentration values. Unlike the hindered settling functions, the simulated settling velocities do not approach the undisturbed value for very low volume fractions. This can be attributed to the finite size of the computational domain and the limited number of particles. As can be seen in figure \ref{fig:settling_concentration}, these low volume fractions are typically found in the top part of the tank, where the smallest particles are still accelerating with a very low Reynolds number. Nevertheless, the agreement is remarkable, which demonstrates the ability of the current simulation approach to produce physically realistic results.

Surprisingly, the hindered settling function of \cite{richardson1954} as parameterized by \cite{teslaa2015} does not underestimate the settling velocities of higher volume fractions, and the agreement of \eqref{eq:RZ_settling} with our data seems to be even better than for \eqref{eq:teslaa_settling}. This is because the parameter $\phi_s = 1$ was calibrated for best fit to match experimental results. \rtwo{Hence, figure \ref{fig:hindered_settling}a serves as validation of our simulation approach}. However, this parameterization is not in line with the definition of $\phi_s$ as given by \cite{dankers2007}. Moreover, it is immediately obvious that the solid content of a freshly deposited sediment bed cannot reach this value. Hence, to improve the parameterization of \eqref{eq:RZ_settling} and \eqref{eq:teslaa_settling}, we propose to choose $\phi_s = \phi_{max}$ since we do not deal with mud but with coarse silt. The consolidation of the sediment will continue to squeeze out water from the bed until the particle packing jams. This process will maintain a counterflow over very long time scales \citep[e.g.][]{houssais2016}. \rtwo{Based on these observations, we propose to not calibrate $\phi_s$ but to parameterize critical volumetric concentrations by the maximum value of our concentration data as shown in figure \ref{fig:settling_concentration}c. Using this data}, we can immediately parameterize $\phi_s = \phi_{max}=0.7$, \rtwo{which is in line with experimental and computational studies of polydisperse particle packings \citep{desmond2014,sohn1968}. Note that this value is larger than the volume fraction of a randomly closed packing of monodisperse spheres, since we are dealing with polydisperse particles and the operator }\eqref{eq:horizontal_averaging}  \rtwo{fully resolves the volume fraction over intervals smaller than the particle diameter.}
Using this physically based parameterization, we obtain a much better fit of \eqref{eq:teslaa_settling} to our data (dotted line in figure \ref{fig:hindered_settling}b), which illustrates the importance of including the effects of the counterflow, the buoyancy, and the increased viscosity in the formulation of the hindered settling function even during the consolidation phase. On the other hand, \eqref{eq:RZ_settling} underestimates the settling of cohesive sediment, but is very close to the settling behavior of the simulated cohesionless sediment. This was expected, as \eqref{eq:RZ_settling} was derived for cohesionless sediments in the first place.

 \subsection{Energetics of enhanced settling}\label{sec:budget}

We now analyze the energetics of the sedimentation process, with a focus on the conversion of the initial potential energy of the particles into kinetic energy, and on the modulation of this process by hydrodynamic and collision forces. Corresponding studies of the energetics of gravity and turbidity currents have provided useful information for the fluid phase that can facilitate the development of simplified models \citep[e.g.][]{necker2005, konopliv2016}. By integrating the particle equation of motion \eqref{eq:eom_trans} along its trajectory, we obtain for the energy budget of a single particle $p$
       \begin{equation}\label{eq:mech_energy}
       \begin{split}
        \underbrace{\frac{1}{2} m_p \: \textbf({u}_{p}^0)^2}_{=E_{k,p}^0} + \underbrace{V_p(\rho_p -\rho_f) g y_{p}^0}_{=E_{y,p}^0} = &
        \underbrace{\frac{1}{2} m_p\: \textbf{u}_p^2}_{=E_{k,p}(t)} + \underbrace{V_p(\rho_p -\rho_f) g y_p}_{=E_{y,p}(t)}  \\
      & - \underbrace{\int_0^t \textbf{F}_{h,p}\textbf{u}_{p} \text{d} t^*}_{=W_{h,p}(t)} - \underbrace{\int_0^t \textbf{F}_{c,p}\textbf{u}_{p} \text{d} t^*}_{=W_{c,p}(t)}\qquad.
       \end{split}        
       \end{equation}
The left hand side represents the energies of the initial (conservative) budget $E_p^0 = E_{k,p}^0 + E_{y,p}^0$, where $E_{k,p}^0$ and $E_{y,p}^0$ denote the initial kinetic and potential energies, respectively. The right hand side represents the budget  $E_{p}(t)=E_{k,p}(t) + E_{y,p}(t) - W_{h,p}(t) - W_{c,p}(t)$ as a function of time, where $W_{h,p}(t)$ and $W_{c,p}(t)$ indicate the work performed on the particle up to time $t$ by the nonconservative hydrodynamic and collision forces $\textbf{F}_{h,p}$ and $\textbf{F}_{c,p}$, respectively. To compute this work, these non-conservative forces are integrated over the entire particle path. The hydrodynamic forces and collision forces modify the total conservative energy $E_{k,p} + E_{y,p}$.

To assess the energy budget, we store particle information such as position, velocity, hydrodynamic and collision forces every 1,000 timesteps. We can then compute the potential and kinetic energy, along with work performed by the hydrodynamic and collision forces, based on the integration scheme
  \begin{subequations}\label{eq:predictor_corrector}
 \begin{alignat}{1}
  E_{k}^n &= \sum_{N_p}\frac{1}{2} m_p\: (\textbf{u}_p^n)^2	
 	\label{eq:E_pot_poly} \\
  E_{y}^n &= \sum_{N_p} V_p(\rho_p -\rho_f) g y_p^n	
 	\label{eq:E_kin_poly} \\
  W_h^n &= W_h^{n-1} + \frac{\Delta t}{4}
 	\sum_{N_p} \left[ \textbf{F}_{h,p}^n + \textbf{F}_{h,p}^{n-1}\right] 
 	\left[\textbf{u}_p^n + \textbf{u}_p^{n-1} \right]	
 	\label{eq:W_h_poly} \\
  W_c^n &= W_c^{n-1} + \frac{\Delta t}{4}
 	\sum_{N_p} \left[ \textbf{F}_{c,p}^n + \textbf{F}_{c,p}^{n-1}\right] 
 	\left[\textbf{u}_p^n + \textbf{u}_p^{n-1} \right]	
 	\label{eq:W_c_poly}  \qquad ,
 \end{alignat}
 \end{subequations}
where $n$ denotes the index of the output data set. Note that we dropped the subscript $p$ for the quantities on the left hand side of \eqref{eq:predictor_corrector}, as these terms reflect the sum over the entire ensemble of particles. The integration rule for the external work employs linear interpolation between the two consecutive output times $n-1$ and $n$. To monitor the accuracy of our computational analysis, we kept track of the relative error 
\begin{equation}\label{eq:error_E_0}
\epsilon = \frac{\lvert E^0 - E(t)\rvert}{E^0 }
\end{equation}
over time, where $E^0=E^0_y$ and  $E(t)=E_k(t)+E_y(t)-W_h(t)-W_c(t)$. This error remained below 0.0045 for all times, which is sufficiently small to establish confidence in the results. We furthermore note that the error saturates at an almost constant level for larger times and is not expected to grow any further as the particles gradually come to rest at the bottom of the tank towards the end of the simulation.

The results of the energy analysis are shown in figure \ref{fig:energy_budget}. Integrated over all particles, the contributions of the kinetic energy and the work performed by collision forces are orders of magnitude smaller than the contributions of potential energy and the work of the hydrodynamic forces, which indicates that the initially available potential energy $E_p$ is primarily used to overcome the viscous drag force as the particles settle (figure \ref{fig:energy_budget}a and b). During the initial stage $t<80 \tau_s$, the curves for cohesive and noncohesive sediment are nearly indistinguishable. Subsequently, however, as the cohesive sediment forms flocs and settles out more rapidly, its potential energy decays faster and it performs more work against the viscous drag forces.

\setlength{\unitlength}{1cm}
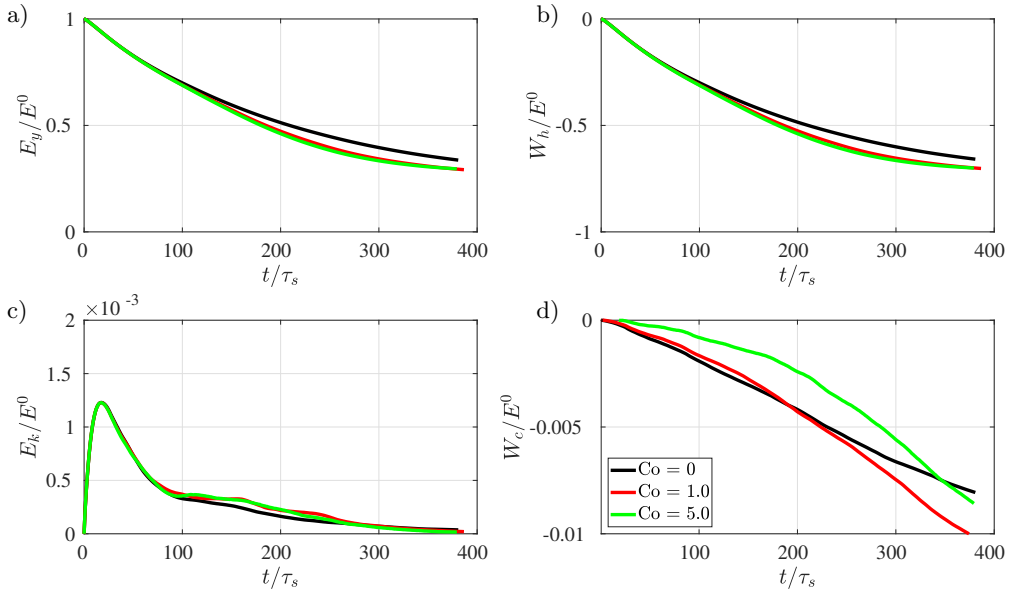
\begin{figure}
\begin{picture}(7,8.1)
  \put(-1.00,  -0.2 ){\includegraphics[width=1.15\textwidth]{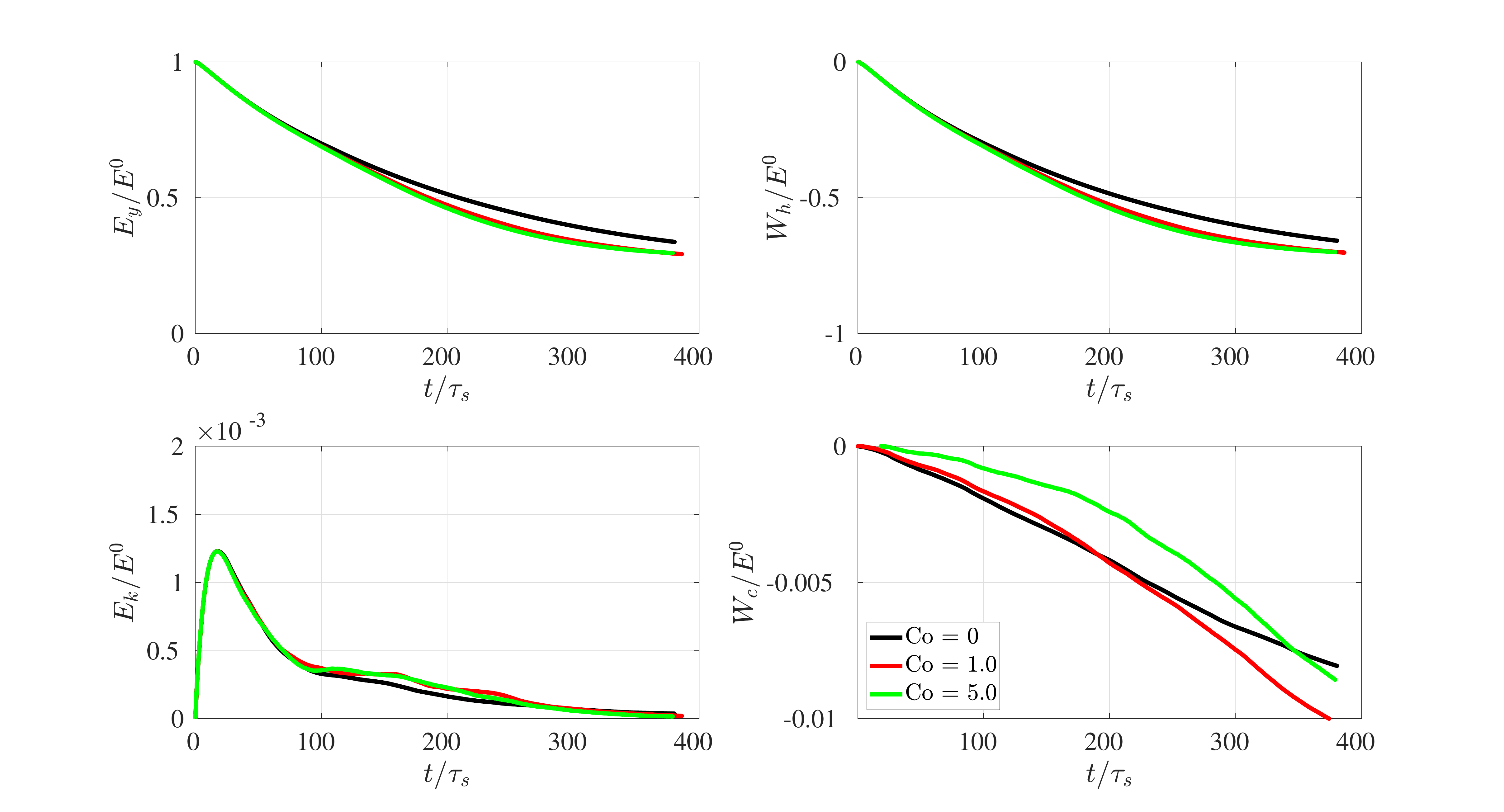}}
    \put( 0.0,  7.5)     {a) }
    \put( 7.0,  7.5)     {b) }
    \put( 0.0,  3.6)     {c) }
    \put( 7.0,  3.6)     {d) }    
\end{picture}
        \caption{Time evolution of the mechanical energy budget of all particles in the flow, normalized by the initial energy $E^0 = E^0_k + E^0_y$. (a) potential energy, (b) work performed by hydrodynamic forces, (c) kinetic energy, (d) work due to collision forces.}
    \label{fig:energy_budget}
\end{figure}
During the initial stage the kinetic energy data in figure \ref{fig:energy_budget}c collapse for all simulations, with a distinct peak at $t = 17.6 \tau_s$ and a subsequent exponential decay. This behavior can be attributed to the fact that particles initially are distributed throughout the entire domain, so that particles close to the bottom wall immediately begin to feel the presence of the confinement. These particles will never accelerate towards their undisturbed settling velocity, and as soon as there are more particles decelerating than accelerating, $E_k$ starts to decay. Hence, the evolution of $E_k$ reflects the behavior of a dissipative dynamical system released from rest, with an initial supply of potential energy, so that its dynamics resemble the temporal evolution of the fluid kinetic energy for a lock-release turbidity current propagating in a channel \citep{necker2005} or spreading radially in a basin \citep{francisco2017}. The difference in settling velocities of the larger and smaller particles during the initial stage results in the strong stirring and mixing of both the fluid and the particles. During the interval $100 \tau_s < t < 270 \tau_s$, the kinetic energy is larger for the cohesive sediment, reflecting its higher settling velocity. As particles begin to deposit, this effect becomes less and less prominent since fewer particles remain in suspension and $E_{k}$ eventually approaches zero for all simulations. During the final simulation stages, the kinetic energy is slightly larger for the cohesionless sediment, since almost all of the cohesive sediment has already settled out.

This behavior is also reflected in the work performed by the particles against the collision forces (figure \ref{fig:energy_budget}d). Even though the forces acting on particle $p$ and $q$ through equation \eqref{eq:particle_forces_cohesive} must be opposite and equal, the total work they perform against the collision forces is nonzero. The cohesive forces modify the amount of work performed by the particles during collisions, as they tend to align the particles and reduce their velocity difference. This observation is more pronounced for $\text{Co}=5$. Since cohesive sediment settles out more quickly than cohesionless sediment, the late stages of the cohesive simulations see more collisions of flocculated particles with deposited particles, so that the work performed against the collision forces is larger for cohesive sediments.


\setlength{\unitlength}{1cm}
\begin{figure}
\begin{picture}(7,8)  
  \put(-1.00,  -0.2 ){\includegraphics[width=1.15\textwidth]{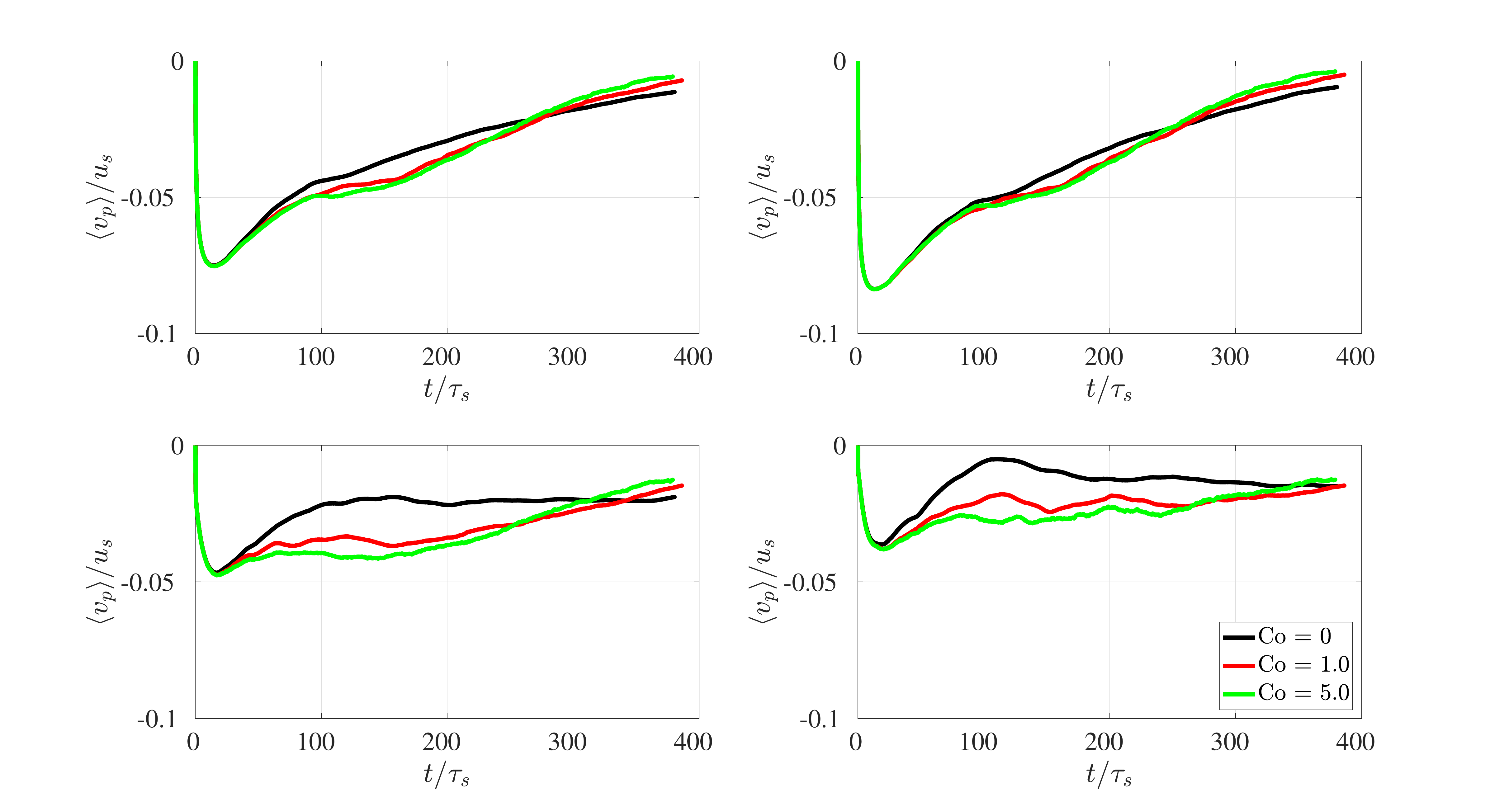}}

    \put( 0,  7.6)     {a) }
    \put( 7.0,7.6)     {b) }    
    \put( 0  ,3.7)     {c) }        
    \put( 7  ,3.7)     {d) }            
\end{picture}
        \caption{Average velocity of the particle center of mass $\langle v_p \rangle$ as function of time. (a) all particles, (b) $D_p > D_{66}$, (c) $D_{33} < D_p \leq D_{66}$, and (d) $D_p \leq D_{33}$.            
        }
    \label{fig:mean_vertical_velocity}
\end{figure}
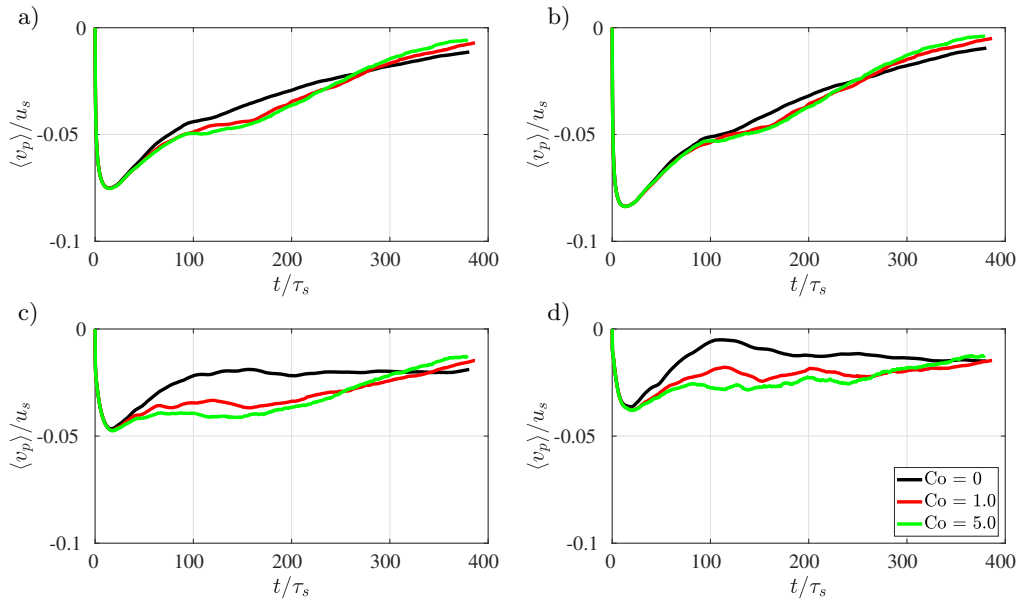

The above demonstrates that cohesive forces modify the processes by which potential energy $E_y$ is converted into kinetic energy $E_k$. As a result, the effective settling rate of the particle ensemble is altered, as reflected by the vertical velocity component of the center of mass
\begin{equation}
 \langle v_p \rangle = \frac{1}{\sum_{p=1}^{N_p}M_p}\sum_{p=1}^{N_p}M_p v_p \qquad ,
\end{equation}
cf. figure \ref{fig:mean_vertical_velocity}a. Figures \ref{fig:mean_vertical_velocity}b-d display the ensemble-averaged velocity $\langle v_p \rangle$, conditioned by particle size in the same way as in figure \ref{fig:final_distribution}. These data confirm that the enhanced kinetic energy of the cohesive sediment during the time interval $100 \tau_s \leq t \leq 270 \tau_s$ can mainly be attributed to faster settling velocities, rather than enhanced horizontal velocity fluctuations. After peaking at $t = 17.6 \tau_s$, the settling process slows down most noticeably for cohesionless sediment. Consistent with our earlier observations, the settling of medium and small cohesive grains is accelerated most strongly by cohesive forces (figure \ref{fig:mean_vertical_velocity}c and d). We note that for small cohesionless grains the ensemble-averaged settling velocity decays almost to zero at $t = 110 \tau_s$, as a significant fraction of them are swept upwards by the counterflow. Subsequently these smaller particles settle towards the bottom, reaching a constant settling velocity over time. By contrast, small cohesive grains attach to larger ones and consequently settle more rapidly. In addition, the settling velocity of the smaller particles follows the decelerating behavior of the larger ones.

The above observations confirm the enhanced settling of cohesive sediment. We can quantify this effect by computing the relative increase in the settling velocity
\begin{equation}
 \Delta \langle v_p \rangle = \frac{\langle v_p \rangle(\text{Co}) - \langle v_p \rangle(\text{Co}=0)}{\langle v_p \rangle(\text{Co}=0)} \qquad ,
\end{equation}
cf. figure \ref{fig:enhanced_settling}. After the acceleration phase up to $t= 75 \tau_s$, the cohesive particles with $\text{Co}=1$ and $\text{Co}=5$ settle up to 24\% and 29\% faster than cohesionless sediment. Beyond $t > 250 \tau_s$, more and more of the cohesive particles have reached the sediment bed, so that the relative settling velocity increase $\Delta \langle v_p \rangle$ loses its meaning.

 \setlength{\unitlength}{1cm}
\begin{figure}
\begin{picture}(7,3.5)
  \put(0.0 ,   0  ){\includegraphics[width=0.46\textwidth]{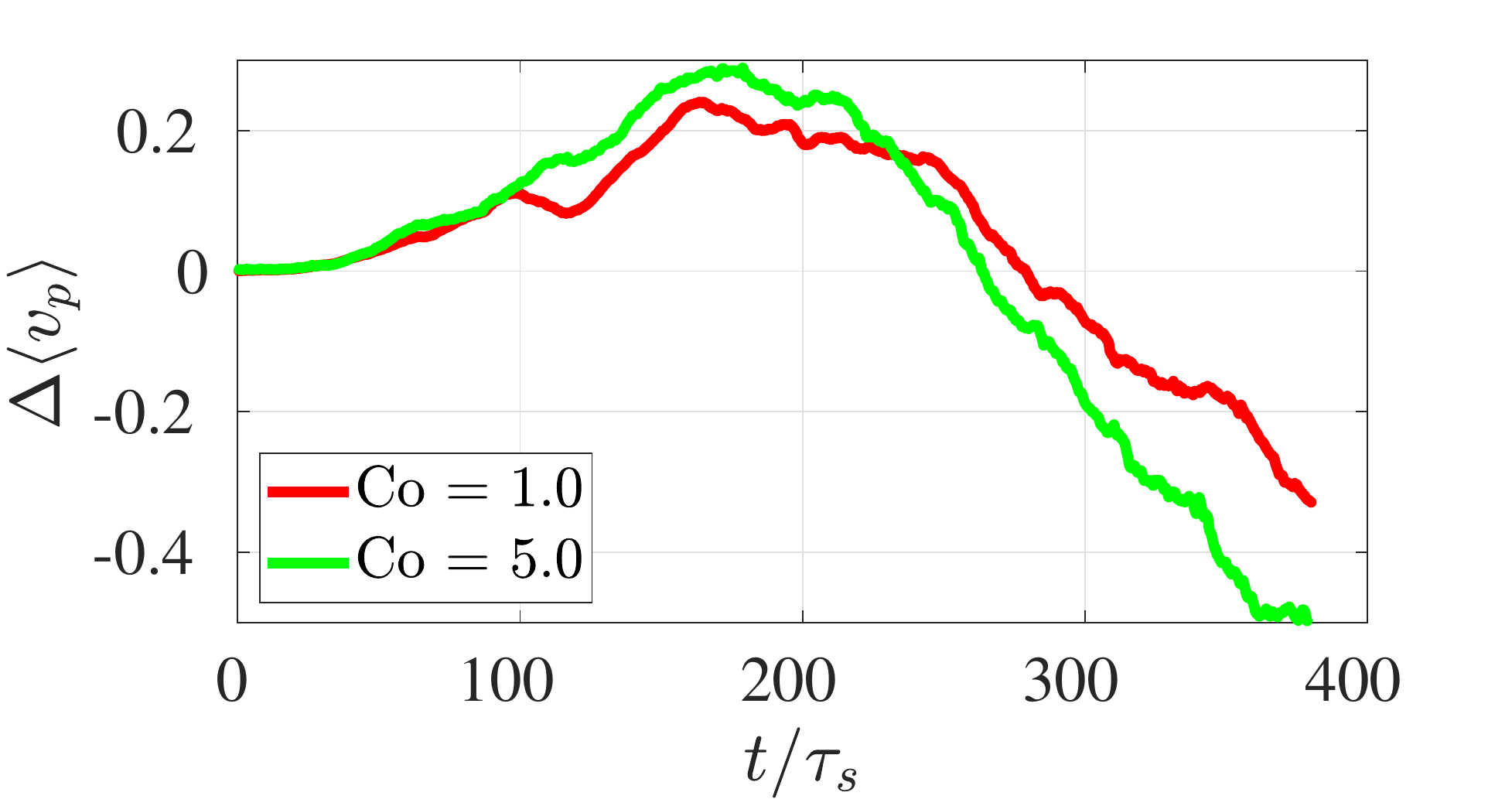}}   
  \put(5.8 ,   0  ){\includegraphics[width=0.58\textwidth]{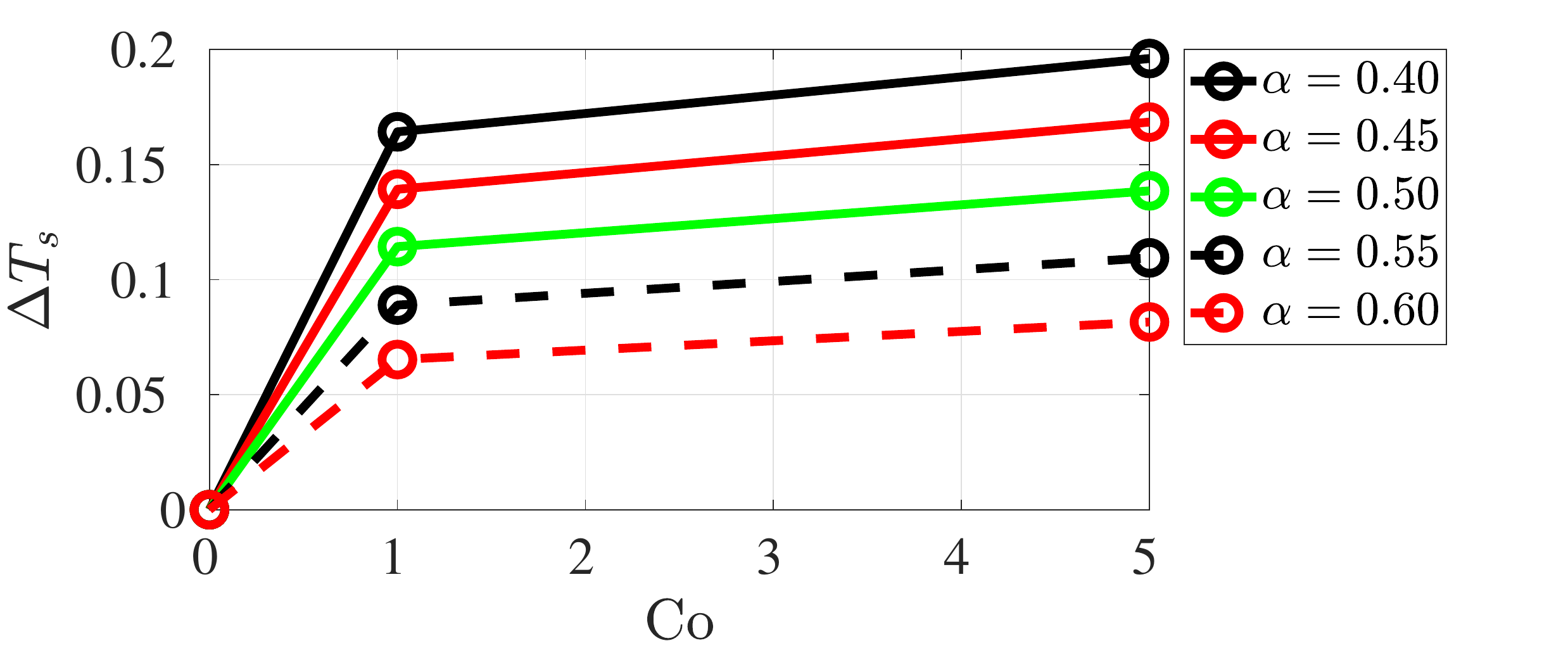}}     
     \put( 0.0,  3.)     {a) }
     \put( 5.8,  3.)     {b) }     
\end{picture}
  \caption{(a) Relative settling velocity increase as a function of time, (b) relative enhancement of the decay of potential energy for different values of $\alpha = E_y / E^0$.}
    \label{fig:enhanced_settling}
\end{figure}    

An alternative way of quantifying the enhanced settling behavior of cohesive sediment relies on the time $T_\alpha$ it takes for the initial potential energy to decay to a relative value of ${\alpha = E_y / E^0}$. We define the relative settling time reduction as 
\begin{equation}
 \Delta T_s = \frac{T_\alpha(\text{Co}=0) - T_\alpha(\text{Co})}{T_\alpha(\text{Co}=0)} \qquad ,
\end{equation}
and show corresponding computational results in figure \ref{fig:enhanced_settling}b. The relative settling time reduction is larger for lower values of $\alpha$, as cohesive sediment continues to settle out faster than cohesionless particles for later times. The speedup is most pronounced as we compare cohesionless sediment with the cohesive case $\text{Co}=1$. A further increase of the cohesive forces to $\text{Co}=5$ results in only slightly more rapid settling.

We conclude that the energy budget analysis represents a suitable tool for clarifying the mechanisms by which cohesive forces accelerate the settling of particles, as observed in \S \ref{sec:settling_behavior} as well as in experiments \citep{mehta1989}.

\subsection{Preferred particle interaction configurations}\label{sec:bonding}
We proceed to explore if the observations of \S \ref{sec:binary_interaction} for interacting particle pairs can help explain the origins of enhanced settling for large particle ensembles. There we had found that, while cohesionless grains tend to undergo the Drafting-Kissing-Tumbling (DKT) process, cohesive particle pairs bond to each other in certain preferred geometrical configurations that depend on the ratio $R_p/R_q$ of the particle radii. With a ratio of $R_p/R_q \neq 1$, i.e. a polydisperse size distribution, particles were seen to align obliquely or even vertically. 

In the following, we define particle $p$ as having a larger $y$-coordinate than particle $q$, so that $y_p > y_q$. To test how much of the behavior observed for isolated cohesive particle pairs can still be found in the context of many settling particles, we analyze the probability density function (PDF) of the angles of interacting particles. Due to the symmetry of the problem, we can consider the angle of the contact point with respect to the vertical coordinate of particle $p$ as
\begin{equation}
 \text{cos}\, \theta = \frac{y_p - y_q}{||\textbf{r}_{pq}||} \ ,
\end{equation}
and we can define vertical, oblique and horizontal contacts as having angles $0^\circ < \theta \le 22.5^\circ$, $22.5^\circ < \theta \le 67.5^\circ$, and $67.5^\circ < \theta \le 90^\circ$, respectively. Furthermore, we distinguish between direct contact ($\zeta_n<0$) and cohesive bonding ($0\leq \zeta_n \leq \lambda$).

\setlength{\unitlength}{1cm}
\begin{figure}
\begin{picture}(7,14)
  \put(0.0 ,   9.4  ){\includegraphics[width=0.5\textwidth]{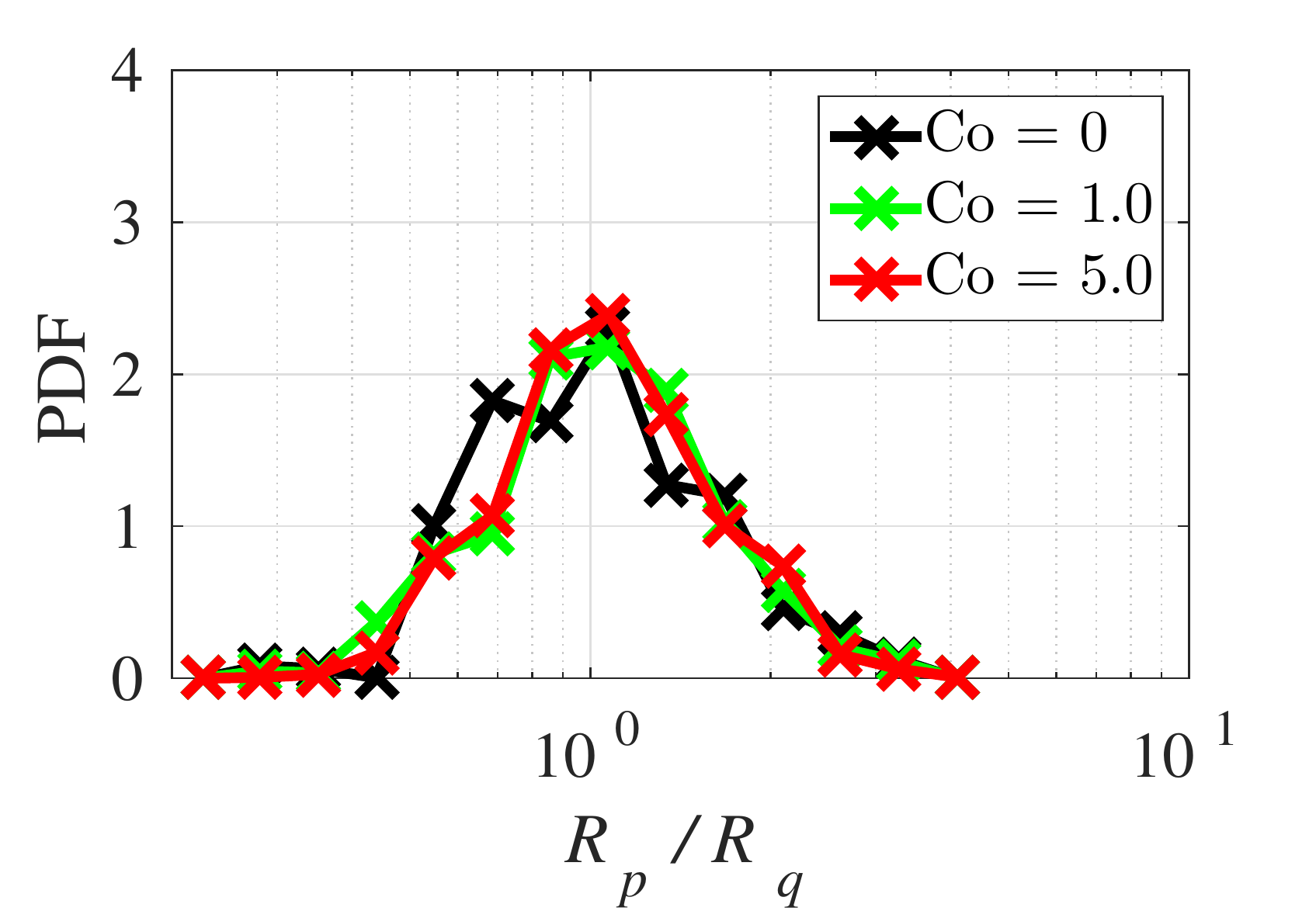}}
  \put(7.0 ,   9.4  ){\includegraphics[width=0.5\textwidth]{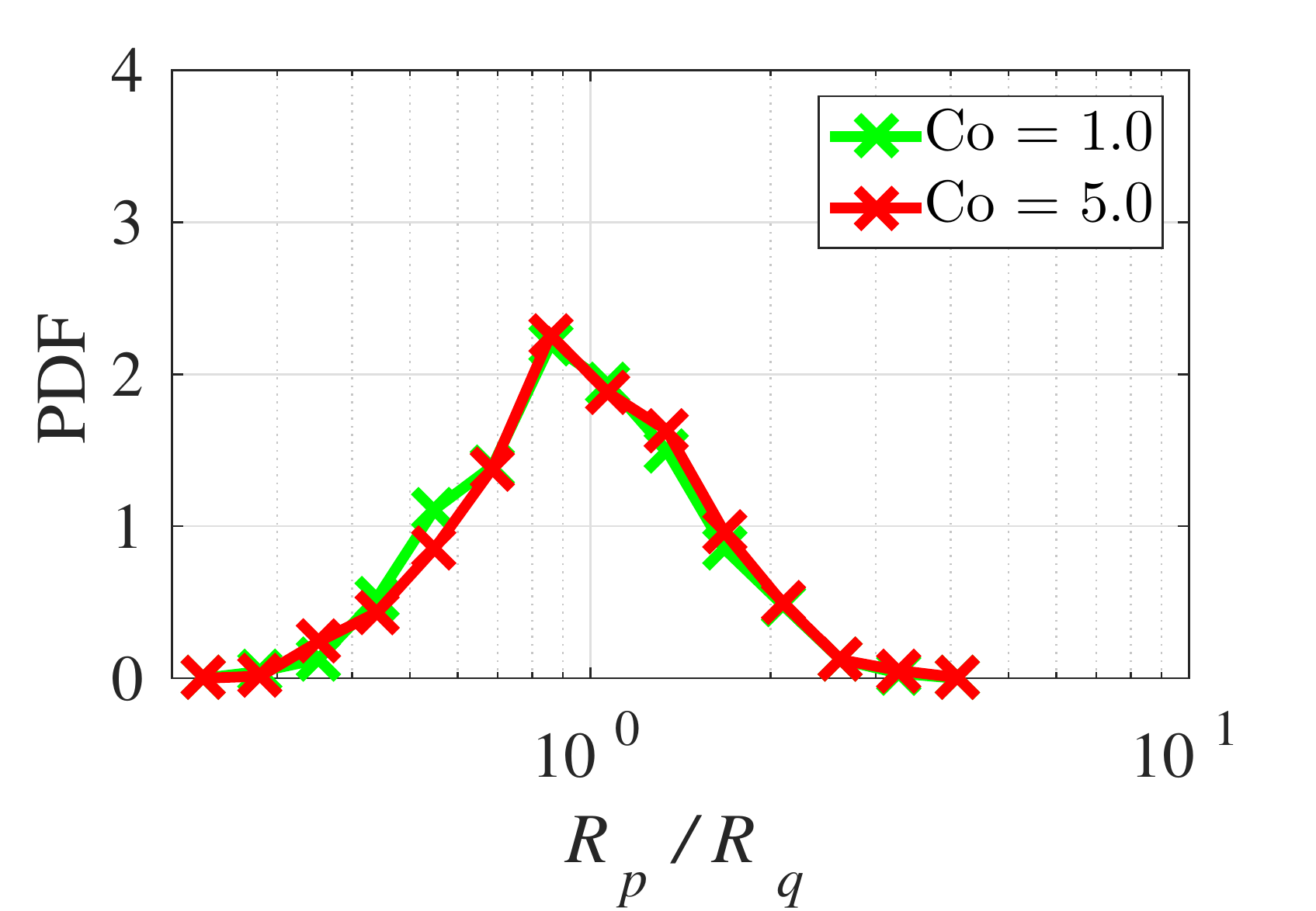}}
  \put(0.0 ,   4.7  ){\includegraphics[width=0.5\textwidth]{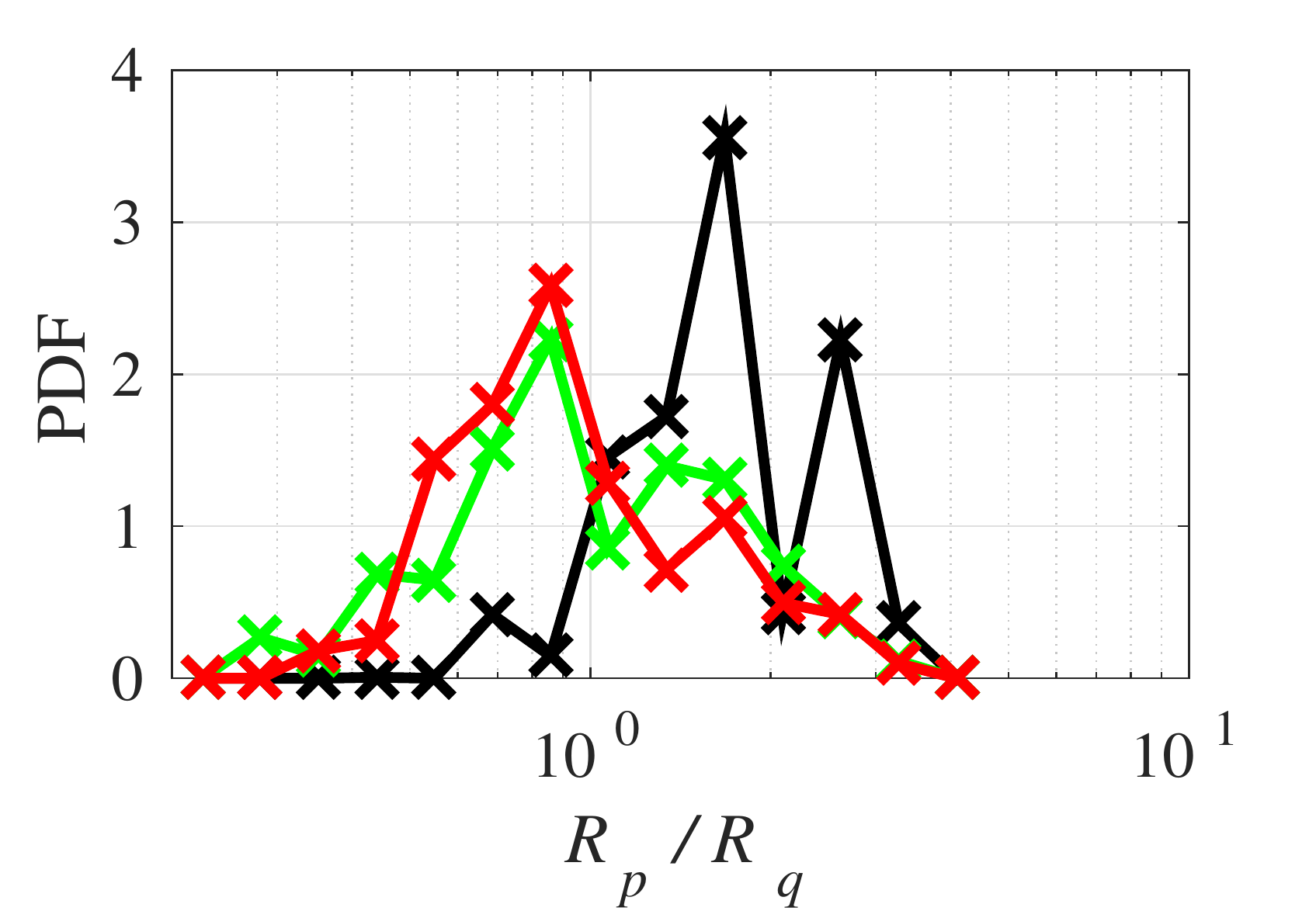}}
  \put(7.0 ,   4.7  ){\includegraphics[width=0.5\textwidth]{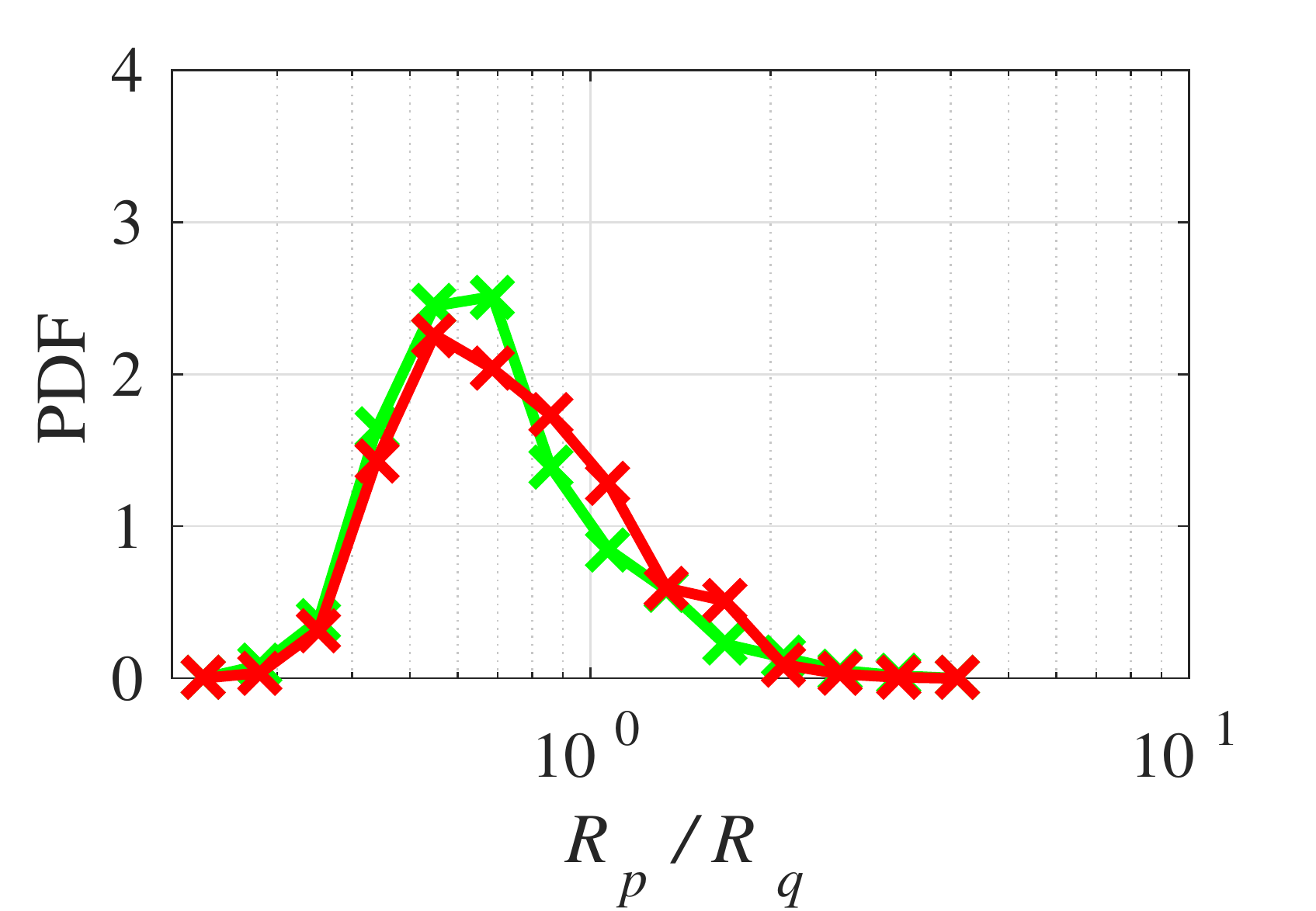}}
  \put(0.0 ,   0    ){\includegraphics[width=0.5\textwidth]{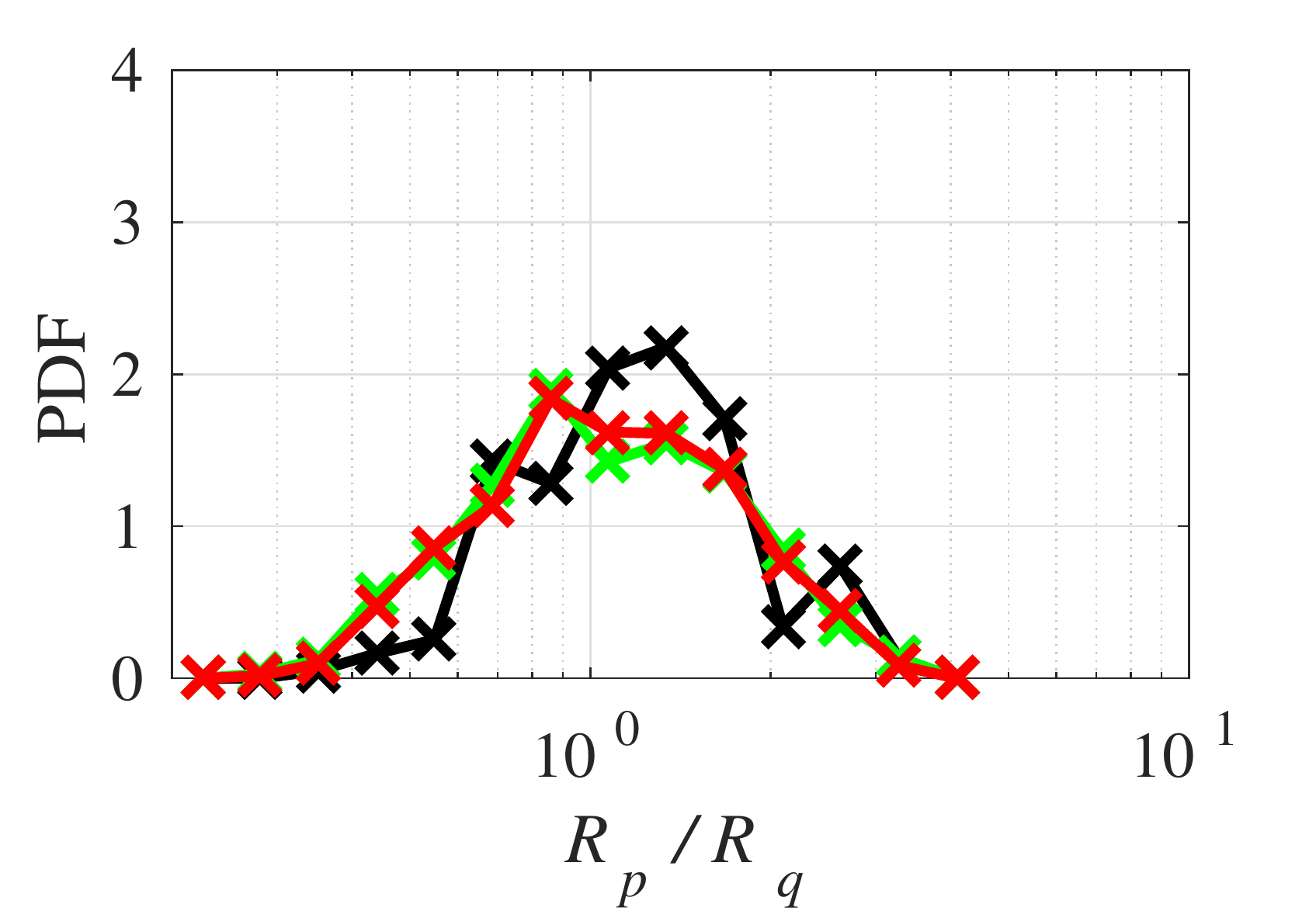}}  
  \put(7.0 ,   0    ){\includegraphics[width=0.5\textwidth]{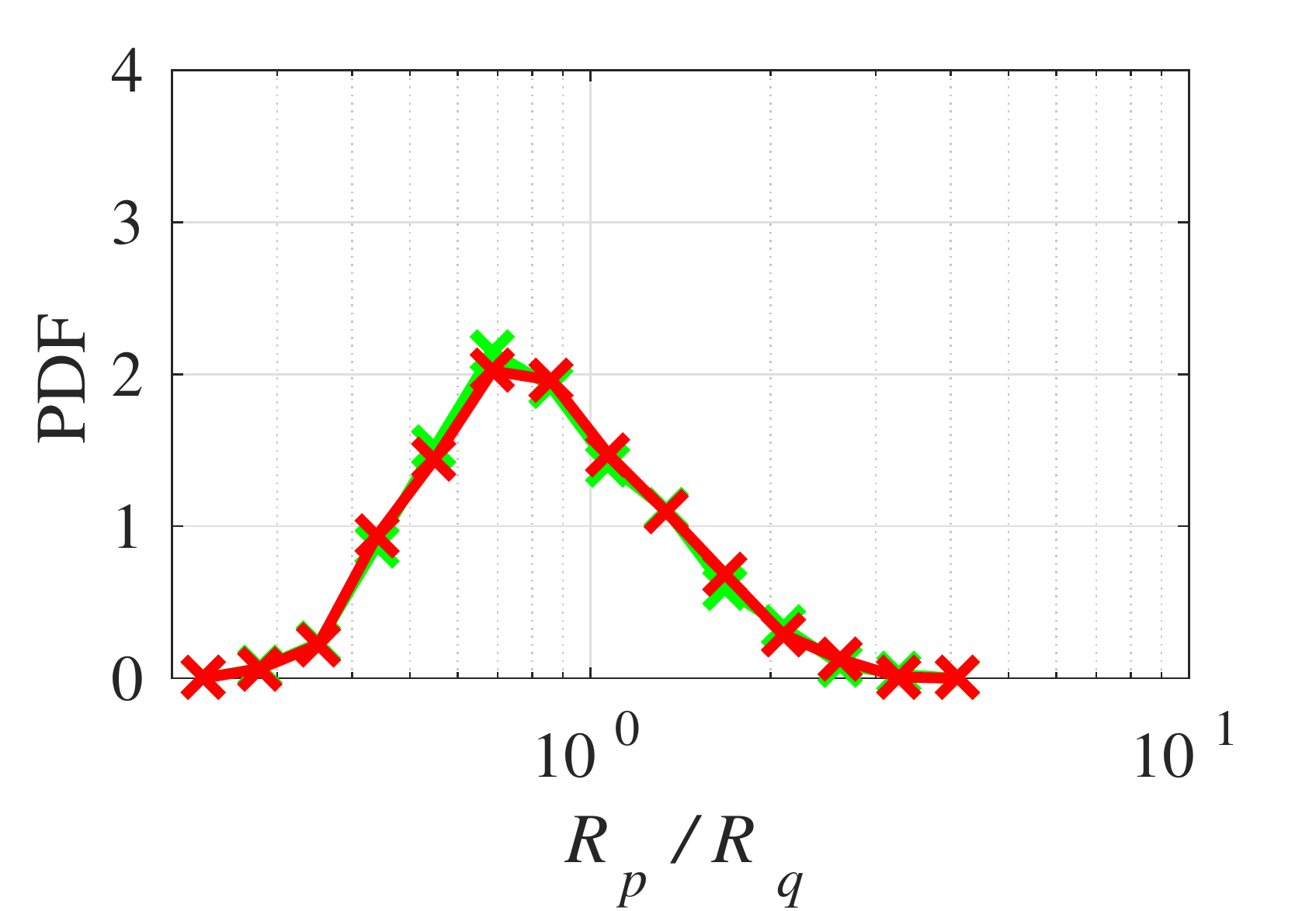}}    
    \put( 1,  13.3)     {\fcolorbox{black}{white}{horizontal contact}}
    \put( 8,  13.3)     {\fcolorbox{black}{white}{horizontal bonding}}        
    \put( 1,  8.6)     {\fcolorbox{black}{white}{vertical contact}}
    \put( 8,  8.6)     {\fcolorbox{black}{white}{vertical bonding}}  
    \put( 1,  3.9)     {\fcolorbox{black}{white}{oblique contact}}
    \put( 8,  3.9)     {\fcolorbox{black}{white}{oblique bonding}}      
    \put( 0,  13.7)     {a) }
    \put( 7.0,13.7)     {b) }    
    \put( 0,   9.0)     {c) }
    \put( 7.0, 9.0)     {d) }    
    \put( 0  , 4.4)     {e) }        
    \put( 7  , 4.4)     {f) }            
\end{picture}
        \caption{Probability density function of the radii ratio with respect to their contact angles. (a) horizontal direct contact, (b) horizontal cohesive bonding, (c) vertical direct contact, (d) vertical cohesive bonding, (e) oblique direct contact, and (f) oblique cohesive bonding. Particles bonding horizontally are preferentially of similar size, while particles in vertical or oblique configurations tend to have different sizes. Interacting cohesive grains most often have the smaller particle trailing the larger one, whereas for cohesionless grains the larger particle tends to catch up with the smaller one from above.
        }
    \label{fig:PDF_bonding}
\end{figure}
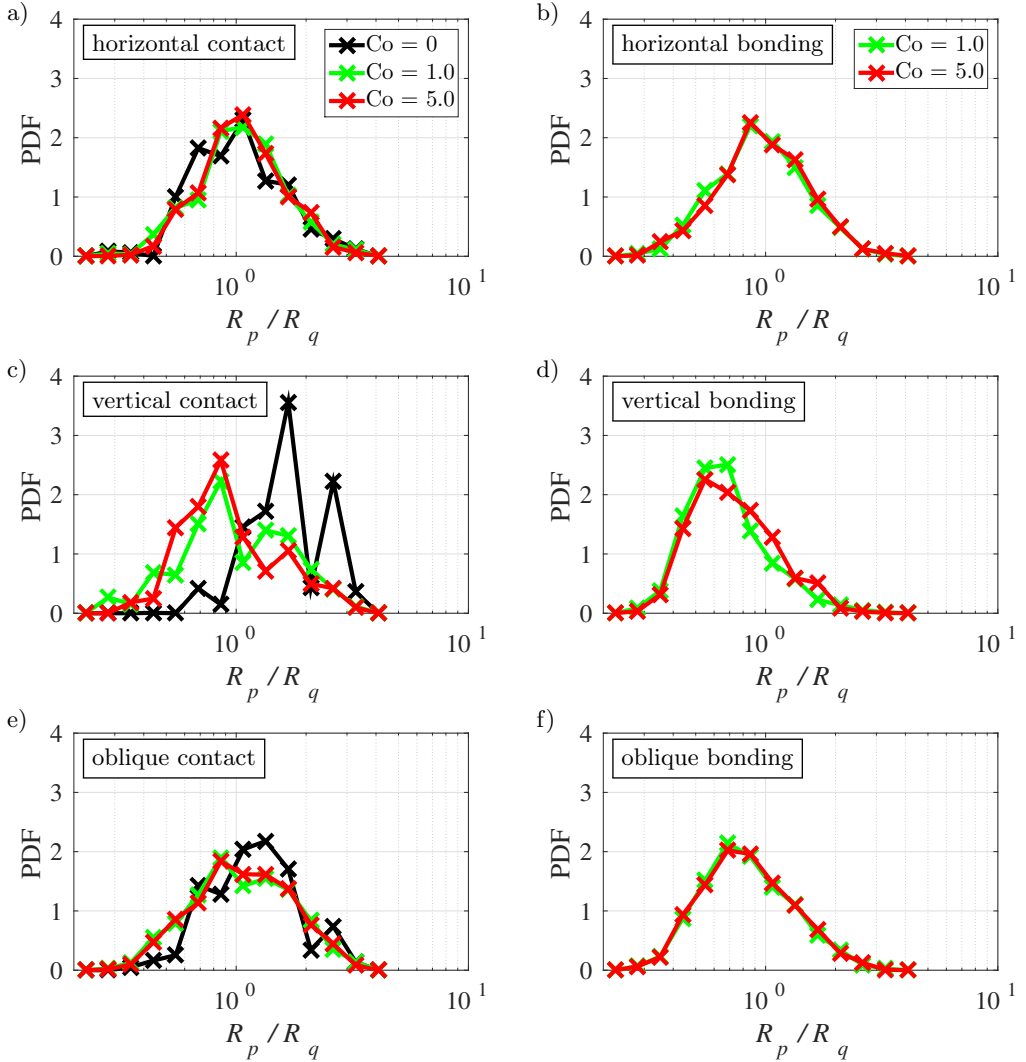

 \begin{table}
   \begin{center}
 \def~{\hphantom{0}}
   \begin{tabular}{l| c | c | c| c| c| c}
                         & \multicolumn{3}{ c|}{Direct contact} & \multicolumn{3}{ c}{Cohesive bonding} \\
                         & horizontal & vertical & oblique      & horizontal & vertical & oblique       \\
       \hline                                                                  
       $\text{Co} = 0$   & 1.02        & 1.54    & 1.19          &  -         & -        &  -            \\
       $\text{Co} = 1.0$ & 1.07        & 0.94    & 1.04          & 0.93       &0.64      & 0.77          \\       
       $\text{Co} = 5.0$ & 1.03        & 0.89    & 1.05          & 0.95       &0.71      & 0.80          \\       
   \end{tabular}
   \caption{Median values of particle radii ratios $R_p/R_q$ at different contact angles and direct contact and cohesive bonding.   }
   \label{tab:median_contact_angle}
   \end{center}
 \end{table}

Figure \ref{fig:PDF_bonding} shows PDFs of the particle radii ratio $R_p/R_q$ for horizontal, vertical and oblique particle interactions, respectively. These PDFs were calculated from all contact points throughout the simulation between particle pairs with a vertical velocity greater than $5 \%$ of the characteristic velocity $u_s = \sqrt{g' D_{50}}$. This conditioning by velocity is necessary to avoid counting particles that have already settled out. The median values of the PDFs displayed in figure \ref{fig:PDF_bonding} are summarized in table \ref{tab:median_contact_angle}. Figures \ref{fig:PDF_bonding}a and b confirm the observation of \S \ref{sec:DKT} that particles interacting horizontally preferentially are of similar size, as the median value of the PDF is very close to unity. This holds for direct contact as well as for cohesive bonding, and for all three simulations. For vertical particle interactions, figures \ref{fig:PDF_bonding}c and d show that cohesive particles interact very differently from cohesionless grains. While cohesive particle pairs tend to arrange themselves with the smaller particle trailing the larger one, cohesionless grains behave oppositely. This reflects the fact that for cohesionless particle pairs of different size the larger particle tends to settle more rapidly, so that it catches up with the smaller one from behind. For cohesive particle pairs, on the other hand, DKT causes the smaller particle to arrange itself in the wake of the larger one, so that the PDF has a distinct maximum for particle radii ratios below unity. For oblique contact angles (figure \ref{fig:PDF_bonding}e and f), the direct contact PDFs have maxima slightly below (above) unity for cohesive (cohesionless) sediment. For oblique cohesive bonding, on the other hand, the PDFs have median values clearly below unity, although somewhat larger than for vertical contact angles. In summary, we find that many of the features observed for isolated particle pairs in \S \ref{sec:DKT} remain relevant within a larger ensemble of sedimenting polydisperse particles.

%% file: 99_conclusions.tex
The present paper develops a physical and computational model for performing fully coupled, grain-resolving DNS simulations of cohesive sediment. This model distributes the cohesive forces over a thin shell surrounding each particle, while preserving the overall energy of the true physical van-der-Waals forces. It thus allows for the spatial and temporal resolution of the cohesive forces during particle-particle interactions, along with direct contact and lubrication forces, thereby enabling us to conduct a detailed analysis of their influence on the overall dynamics of the sediment. The influence of the cohesive forces is captured by a single dimensionless parameter in the form of a cohesion number, which represents the ratio of cohesive and gravitational forces acting on a particle. 

The cohesive force model is tested and validated for binary particle interactions in the well-known Drafting-Kissing-Tumbling configuration. In contrast to noncohesive particles, cohesive sediment grains can remain attached to each other during the tumbling phase following the initial collision, which forms the basis for the formation of flocs. The DKT simulations demonstrate that cohesive particle pairs settle in a preferred orientation, which depends on the ratio of the particle radii. When the particles are very different in size, they tend to align themselves in the vertical direction, with the smaller particle being drafted in the wake of the larger one.

The preferred orientation of cohesive particle pairs is seen to remain influential within much larger simulations of 1,261 polydisperse particles released from rest, for different values of the cohesion number. These simulations reproduce several earlier experimental observations by other authors, such as the accelerated settling of sand and silt particles due to particle bonding, the stratification of cohesive sediment deposits, and the consolidation process of the deposit. We find that cohesive forces accelerate the overall settling process primarily because smaller grains attach to larger ones and settle in their wakes, which speeds up their downward motion, consistently with our DKT simulations. For the cohesion number values simulated in the present study, we observe that settling can be accelerated by up to 29\%. Based on the simulation results, we revisit hindered settling functions proposed by earlier authors, and propose \rtwo{physically based parameterization that does not involve arbitrary calibration} to account for cohesive interparticle forces. A detailed investigation of the energy budget provides quantitative information on the work performed by the hydrodynamic and collision forces. While the work of the collision forces is much smaller than that of hydrodynamic forces, it nevertheless substantially modifies the processes that convert potential into kinetic energy and vice versa. 

In summary, the grain-resolving DNS simulations of cohesive sediment dynamics identify three characteristic phases for the polydisperse settling process: (i) an initial stir-up phase of increasing particle kinetic energy, during which flocculation remains limited; (ii) a phase of increased flocculation and enhanced settling as the particle kinetic energy decays, and (iii) the dewatering phase, during which the freshly settled sediment flocs consolidate. 

The model developed in the present paper lends itself well for further computational investigations into the physics of cohesive sediment. Among the interesting issues to be addressed are the interaction of cohesive sediment with a turbulent flow field, as well as the erosion of a cohesive sediment bed by an imposed flow. Efforts in these directions are currently underway.

%% file: App_comparison.tex
As outlined in \S \ref{sec:cohesion_physics}, the DLVO theory involves both repulsive and attractive forces. A model incorporating both effects has been proposed by \cite{pednekar2017}
\begin{equation}\label{eq:DLVO}
 F_\text{DLVO} =\underbrace{ A_R \, R_\text{eff} \exp{\left(-\kappa \zeta_n\right)}}_{F_\text{rep}} - \underbrace{\frac{A_H R_\text{eff}}{12(\zeta_n^2 + \zeta_0^2)}}_{F_\text{att}} \qquad ,
\end{equation}
where $A_R$ is a repulsive force scale due to the particles' surface potential for $\zeta_n = 0$, $\kappa$ is the Debye length, $A_H$ is the Hamaker constant, and $\zeta_0$ is the surface roughness preventing $F_\text{att}$ from diverging to infinity for vanishing gap size. These four parameters need to be adjusted according to the physical system. Moreover, \eqref{eq:DLVO} does not incorporate an explicit scaling with the median grain size $D_{50}$. In the following, we will compare \eqref{eq:DLVO} to different conditions occurring in different environmental systems to obtain reasonable parameter ranges for our cohesive force model \eqref{eq:cohesive_forces_dimensionless}, which conveniently only involves one tunable parameter, i.e. the Cohesive number.

\begin{figure}
\centering
\includegraphics[width=0.5\textwidth]{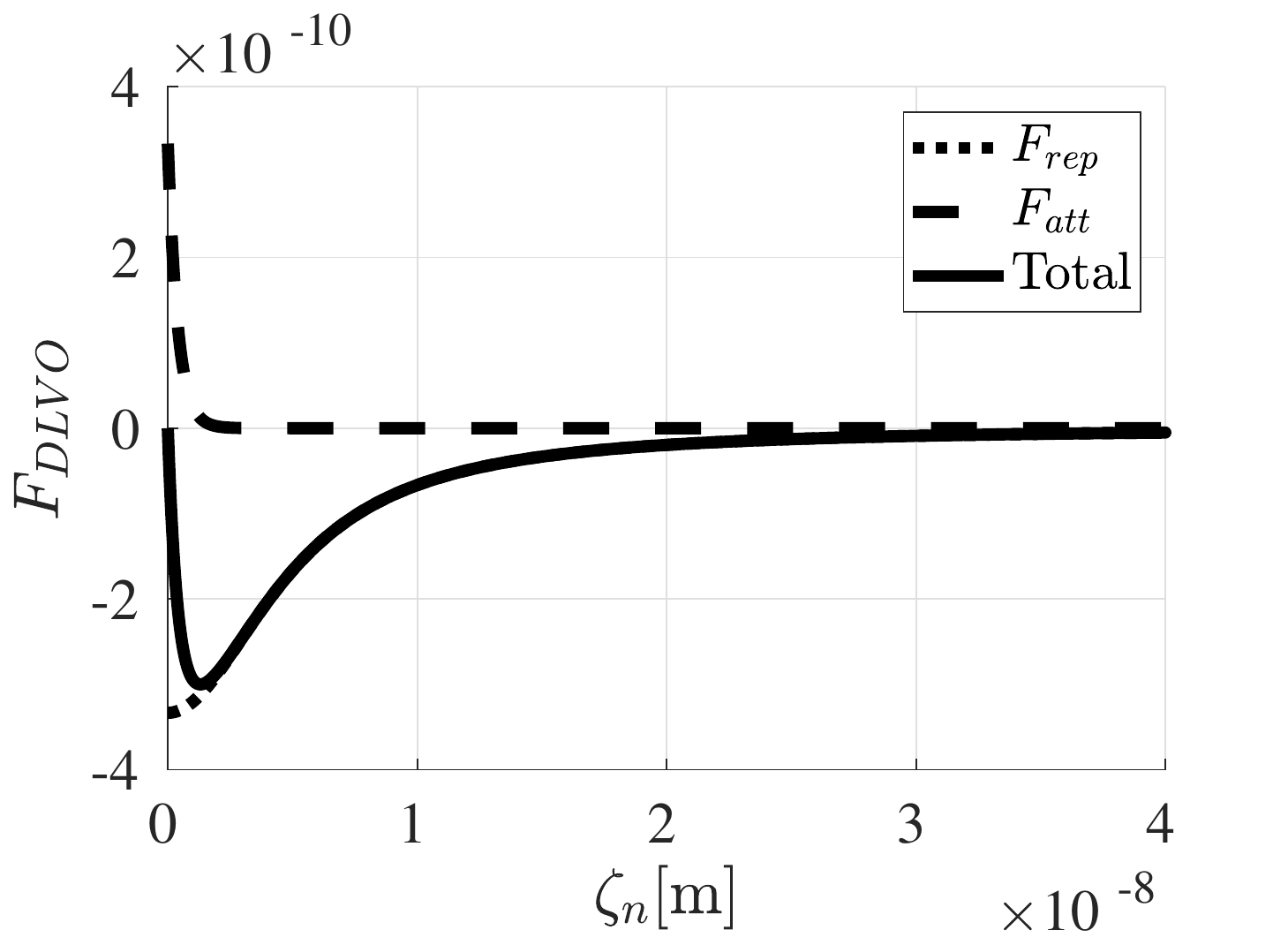}

\caption{DLVO curve for $A_H = 1\cdot10^{-20}$, $D_p = 20 \mu$m,  $\zeta_0 = R_p \cdot 5\cdot 10^{-4}$, and $C_\text{salt} = 35$ppt.  }
    \label{fig:baseline}
\end{figure}

As a baseline application, we choose the following parameters: (i) $A_H = 1\cdot10^{-20}$ J, which reflects silica materials in water according to \cite{bergstrom1997}, (ii) $R_\text{eff} = \frac{R_p}{2} = 5 \mu$m for monodisperse silt particles of grain size $D_p = 20 \mu$m, (iii) $\zeta_0 = R_p \cdot 5\cdot 10^{-4}$ which is below the surface roughness $\zeta_\text{min} = R_p \cdot 3\cdot 10^{-3}$ of the lubrication model \eqref{eq:lubrication}, but rougher than the values of $R_p \cdot 1\cdot 10^{-4}$ reported by \cite{gondret2002} for glass spheres; (iv) we determine $\kappa$ using the approximation for the  monovalent salt sodium chloride given by \cite{berg2010} as $\kappa^{-1} = \frac{0.304\cdot 10^{-9}\text{m}^{-1}}{|z| \sqrt{C_\text{salt}}}$ in meters, where $z$ is the valency of the salt and $C_\text{salt}$ is the salt concentration in mol/liter. Here, we choose the salinity of sea water with 35ppt. These parameters yield $C_\text{salt} = 0.6$mol/liter and $\kappa = 0.393$nm; (v) Since a key feature of our model is to have vanishing forces for particle contact, i.e. $\zeta_n = 0$, we set $A_R = A_H/(12 \zeta_0^2)$. The DLVO curve for this case is displayed in figure \ref{fig:baseline}. In this scenario, the total force $F_\text{DLVO}$ follows the attractive forces with a distinct minimum at $\zeta_n \approx 1$nm. The minimum force is $| \min(F_\text{DLVO})|= 3\cdot 10^{-10}$N, while the weight becomes $F_g = \pi \, g (\rho_p - \rho_f) D_p^3/6 = 6.78 \cdot 10^{-11}$N, where we set gravitational acceleration, particle density, and fluid density to be $g = 9.81 \text{m/s}^2$, $\rho_p = 2650 \text{kg/m}^3$, and $\rho_f = 1000 \text{kg/m}^3$, respectively. This yields a cohesive number of $\text{Co} =  | \min(F_\text{DLVO})|/F_g = 4.43$, which is within the parameter ranges addressed in \S \ref{sec:binary_interaction} and \S \ref{sec:settling}.

\begin{table}
   \begin{center}
 \def~{\hphantom{0}}
   \begin{tabular}{r l | r || r  l | r  }
       $A_H$ [J]              &  System                                      & Co    & $D_p [\mu$m] & System      &  Co         \\
       \hline                                                                  
       $1.8\cdot 10^{-18}$    &  Ionic crystals in water \citep{visser1972}  & 796.60 &  2         & Clay        &  13,345   \\
       $1.0\cdot 10^{-20}$    &  Quartz in water \citep{bergstrom1997}       & 4.43   &  6.3       & Fine silt   &  333.23        \\  
       $6.3\cdot 10^{-23}$    &  Silicate in water \citep{lick2004}          & 0.03   &  20.0      & Medium silt &  4.43   \\       
      \multicolumn{3}{c ||}{}                                                        &  63.0      & Coarse silt &  0.05   \\         
       \hline
       \hline
       $\zeta_0 / R_p$        &  System                             & Co      & $C_\text{salt}$ [ppt] & System      & Co         \\
       \hline                                                                 
       $1\cdot 10^{-4}$       &  according to \citep{gondret2002}    & 64.09   &  0.1                  & Fresh water &  0.51   \\
       $5\cdot 10^{-4}$       &                                     & 4.43    &  2.5                  & Oligohaline &  3.04        \\
       $1\cdot 10^{-3}$       &                                     & 1.18    &  12.0                 & Mesohaline  &  4.02   \\     
       $3\cdot 10^{-3}$       &  calibrated by \citep{biegert2017}   & 0.14    &  35.0                 & Ocean water &  4.43   \\  
   \end{tabular}
   \caption{Sensitivity of the four material parameters $A_H$, $D_p$, $\zeta_0$ and $C_\text{salt}$ entering \eqref{eq:DLVO}.}
   \label{tab:sensitivity}
   \end{center}
 \end{table}

\setlength{\unitlength}{1cm}
\begin{figure}
\begin{picture}(7,10)
  \put(0.0 ,   5  ){\includegraphics[width=0.5\textwidth]{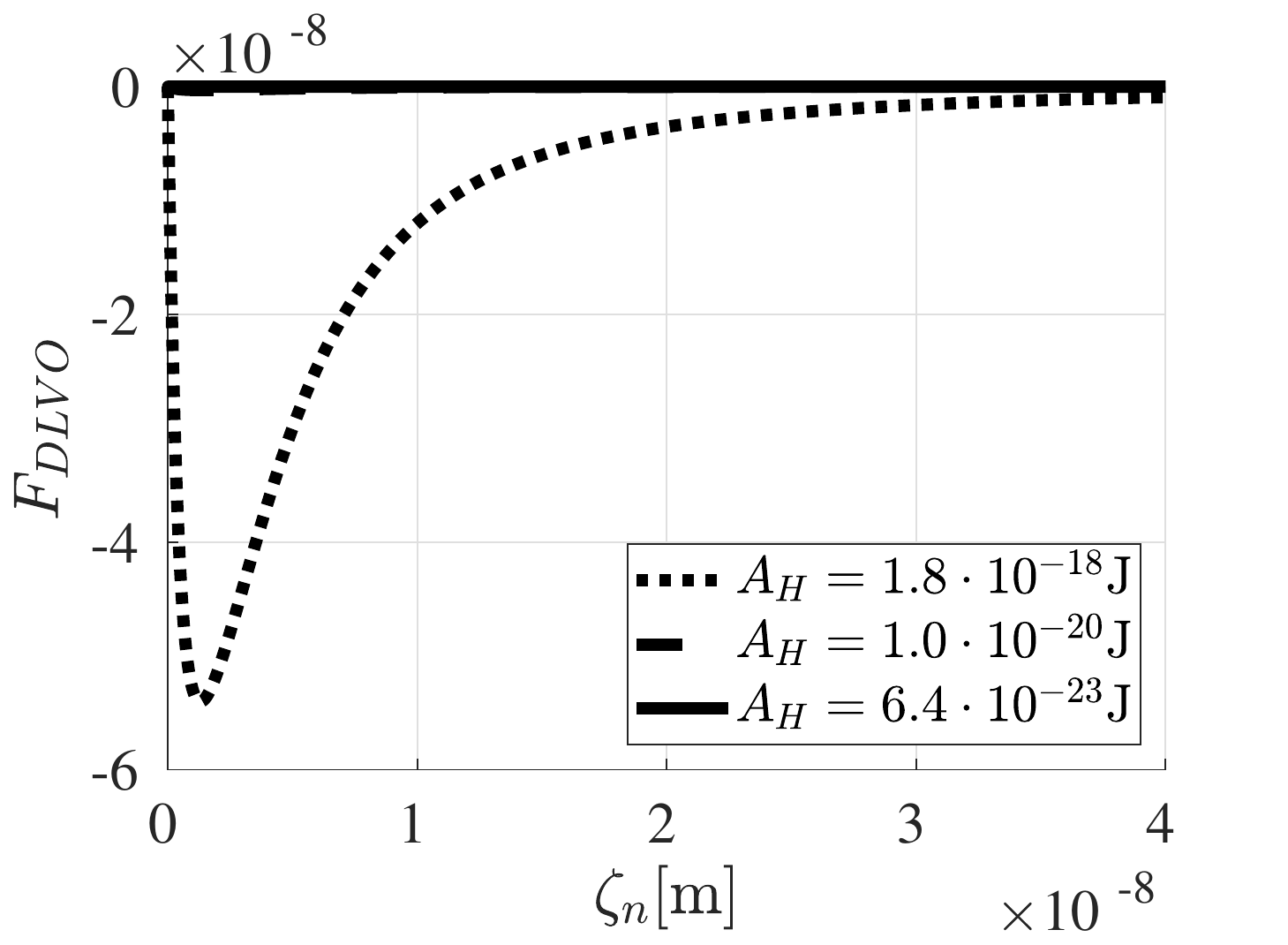}}
  \put(7.0 ,   5  ){\includegraphics[width=0.5\textwidth]{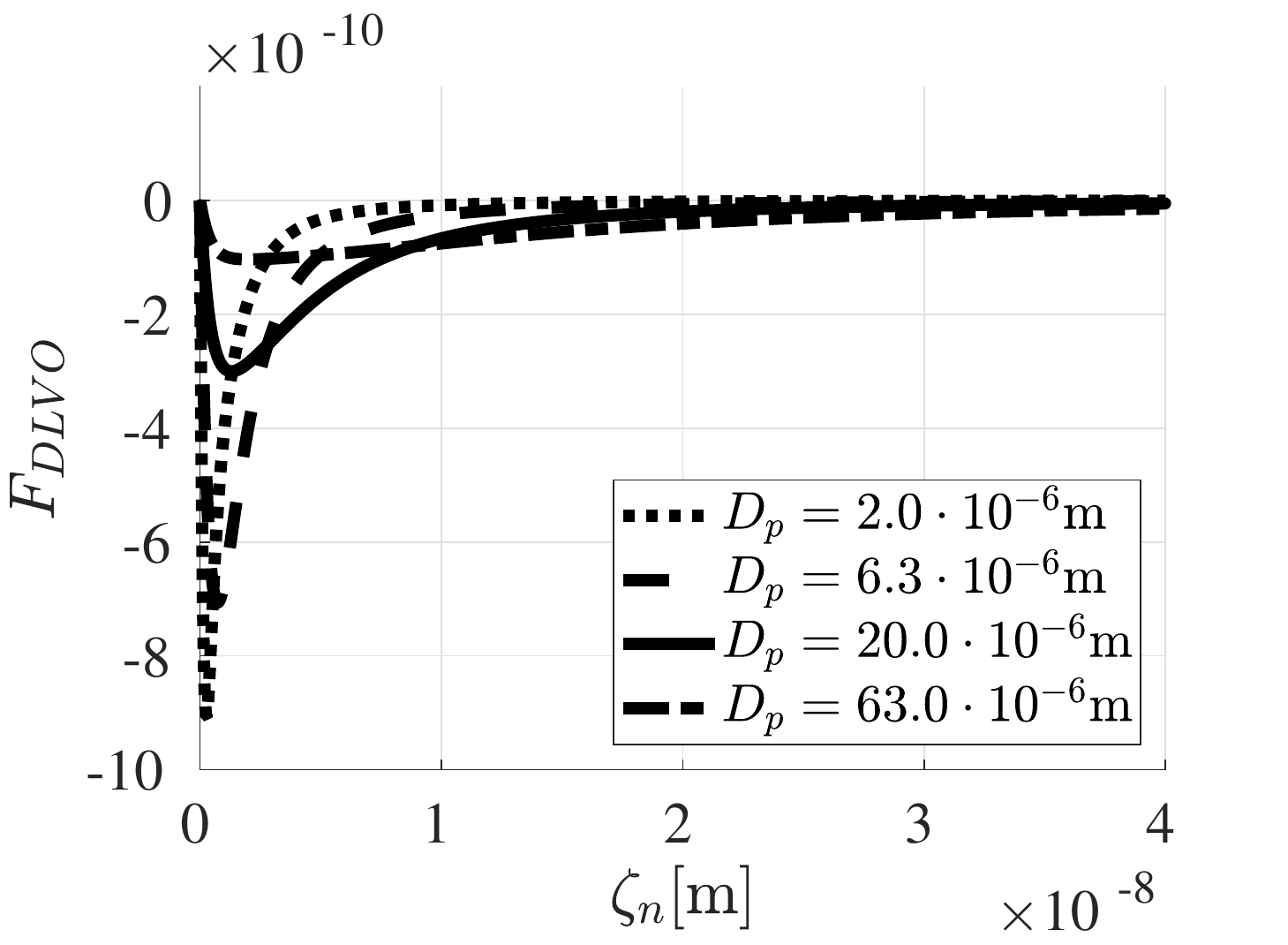}}
  \put(0.0 ,   0  ){\includegraphics[width=0.5\textwidth]{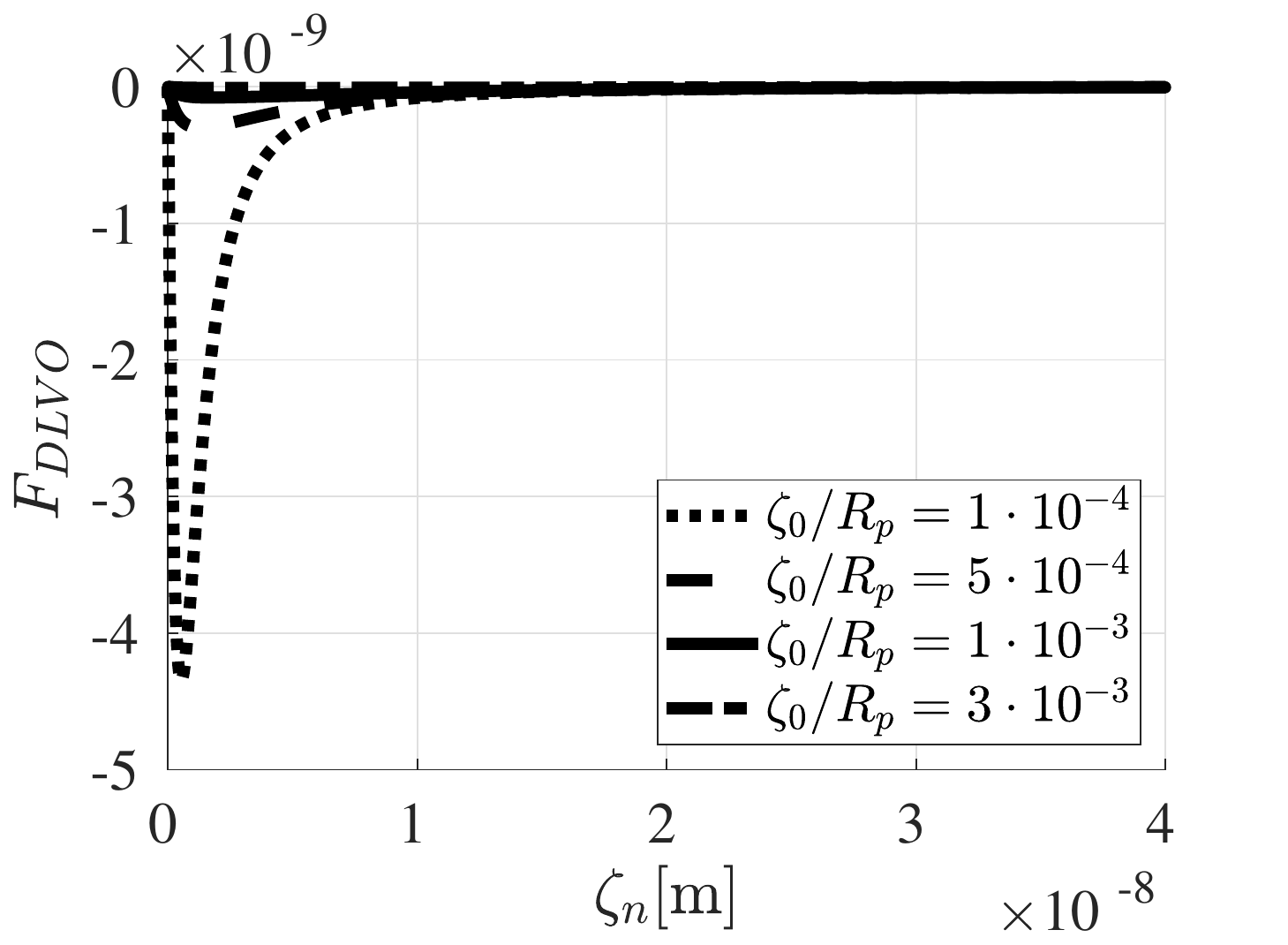}}  
  \put(7.0 ,   0  ){\includegraphics[width=0.5\textwidth]{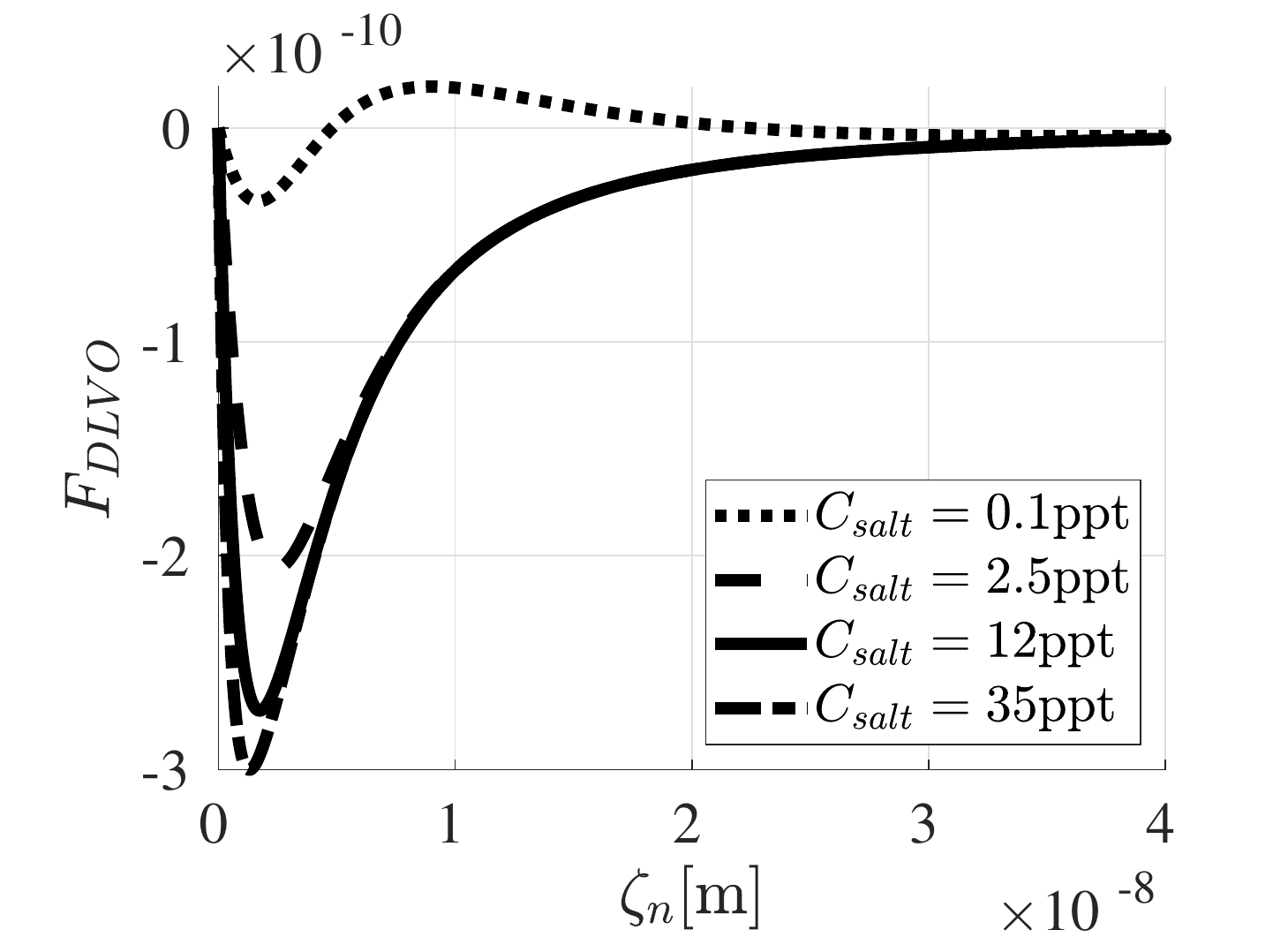}}    

    \put( 0,  9.8)     {a) }
    \put( 7.0,9.8)     {b) }    
    \put( 0  ,4.8)     {c) }        
    \put( 7  ,4.8)     {d) }
\end{picture}
        \caption{Sensitivity of $F_\text{DLVO}$ with respect to the material parameters entering \eqref{eq:DLVO}. (a) $A_H$, (b) $D_p$, (c) $\zeta_0$, and (d) $C_\text{salt}$.            
        }
    \label{fig:sensitivity}
\end{figure}
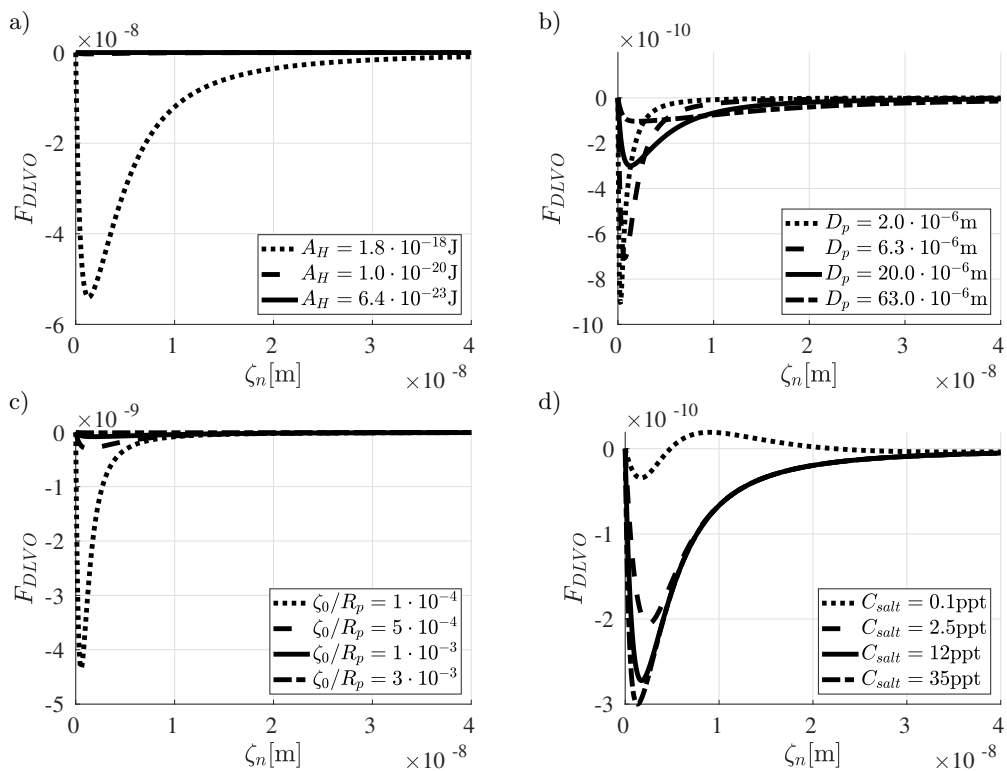

To get a better understanding of the applicability of our modeling approach, we conducted a sensitivity analysis of \eqref{eq:DLVO} reflecting parameter ranges of our interest. The results of $F_\text{DLVO}$ are illustrated in figure \ref{fig:sensitivity} and the sensitivity in terms of $\text{Co}$ is summarized in table \ref{tab:sensitivity}. Out of the four parameters investigated, $D_p$ has the strongest influence. This is caused by the dependency of $\zeta_\text{min}$ and, hence, $A_R$ as well as $F_g$ on this parameter. As a consequence, we conclude that our model \eqref{eq:cohesive_forces_dimensionless} might not be applicable for very small grains as Co becomes very large. Note that at this grain size, particles are also considered to experience Brownian motion \citep{metcalfe2012}, which is not incorporated in our simulation approach, either. The Hamaker constant $A_H$ has the potential to change the Cohesive number, but we retain the linear dependency of Co on $A_H$. The big change in Co is mainly because the numbers suggested in the literature vary by five orders of magnitude. The results also retain their quadratic dependency on the surface roughness $\zeta_0$. Changes in the Debye length as a function of salt concentration do not strongly influence our system. However, it must be noted that if the Debye length exceeds the surface roughness, we no longer obtain a distinct minimum for $F_\text{DLVO}$. We can therefore conclude that this behavior imposes another constraint on the model \eqref{eq:cohesive_forces_dimensionless}. This, however, is a reasonable assumption, since we are dealing with natural sediments of macroscopic silica grains.

%% file: App_particle_eom_dimesionless.tex
To obtain the non-dimensional form of equation \eqref{eq:eom_trans}, we scale all variables by corresponding characteristic quantities
\begin{subequations}\label{eq:nondim_fluid}
\begin{equation}\label{eq:nondim_p}
  p = \rho_f u_s^2 \tilde{p}                                         \qquad ,
\end{equation}
\begin{equation}\label{eq:nondim_fibm}
  \textbf{f}_\textit{IBM} = g' \tilde{\textbf{f}}_\textit{IBM}                                         \qquad ,
\end{equation}
\begin{equation}\label{eq:nondim_u}
  \textbf{u} = u_s \tilde{\textbf{u}} = \sqrt{g'D_{50}} \tilde{\textbf{u}}              \qquad ,
\end{equation}
\begin{equation}\label{eq:nondim_time}
  t = \frac{D_{50}}{u_s}\tilde{t} = \sqrt{\frac{D_{50}}{g'}}\tilde{t}                            \qquad ,
\end{equation}
\begin{equation}\label{eq:nondim_L}
  L = D_{50} \tilde{L}                                         \qquad .
\end{equation}
\end{subequations}
Here $\textbf{u}$ represents any velocity vector and $L$ any length appearing in equations \eqref{eq:navier_stokes} and \eqref{eq:eom_trans}. The tilde symbol indicates dimensionless variables. In this way we obtain the dimensionless momentum conservation and contintuity equations
\begin{equation} \label{eq:navier_stokes_dimless}
\frac{\partial{\tilde{\textbf{u}}}}{\partial{\tilde{t}}}+\tilde{\nabla}\cdot(\tilde{\textbf{u}}\tilde{\textbf{u}}) = - \:\tilde{\nabla} \tilde{p} + \frac{1}{\Rey} \tilde{\nabla}^2 \tilde{\textbf{u}} + \tilde{\textbf{f}}_\textit{IBM}  \hspace{0.5cm},
\end{equation}
\begin{equation}\label{eq:continuity_dimless}
\tilde{\nabla}\cdot \tilde{\textbf{u}}=0 \qquad ,     \hspace{0.5cm}
\end{equation}
where $\Rey = D_{50} u_s / \nu_f$ denotes the Reynolds number.

In a similar fashion, we introduce characteristic scales for the hydrodynamic force, the surface roughness, and the stiffness and damping coefficients
\begin{subequations}\label{eq:nondim_particle}
\begin{equation}\label{eq:nondim_m}
  m_{p} = m_{50} \tilde{m}_{p}  = \rho_f V_{50} \tilde{m}_{p}       \qquad ,
\end{equation}
\begin{equation}\label{eq:nondim_V}
  V_{p} = V_{50} \tilde{V}_{p}          \qquad ,
\end{equation}
\begin{equation}\label{eq:nondim_Fhn}
  \textbf{F}_{p,h} = \rho_f g'V_{50} \tilde{\textbf{F}}_{p,h}          \qquad ,
\end{equation}
\begin{equation}\label{eq:nondim_zeta_app}
  \max(\zeta_n,\zeta_\text{min}) = D_{50} \tilde{\zeta}_n           \qquad ,
\end{equation}
\begin{equation}\label{eq:nondim_kn}
  k_n = \rho_f g' \sqrt{V_{50}} \tilde{k}_n          \qquad ,
\end{equation}
\begin{equation}\label{eq:nondim_dn}
  d_n = \rho_f  \sqrt{\frac{g'}{D_{50}}} V_{50} \tilde{d}_n              \qquad ,
\end{equation}
\begin{equation}\label{eq:nondim_kt}
  k_t =  \rho_f \frac{g'}{D_{50}} V_{50} \tilde{k}_t            \qquad ,
\end{equation}
\begin{equation}\label{eq:nondim_dt}
  d_t = \rho_f  \sqrt{\frac{g'}{D_{50}}} V_{50}  \tilde{d}_t             \qquad ,
\end{equation}
\begin{equation}\label{eq:nondim_lambda_app}
  \lambda_n = D_{50} \tilde{\lambda}_n             \qquad .
\end{equation}
\end{subequations}
Introducing these into equation \eqref{eq:eom_trans} yields as the characteristic scale of the forces acting on the particles $F_i = m_{50} u_{s}^2/D_{50} =   m_{50} g'$. We thus obtain
\begin{equation}\label{eq:eom_trans_non_dim_1}
      \begin{split}
    \tilde{m}_p \frac{\text{d}\tilde{\textbf{u}}_p}{\text{d} \tilde{t}} =& 
    \tilde{\textbf{F}}_{p,h}
    + \tilde{V}_{p} \textbf{e}_g
    -36 \frac{\nu_f }{ D_{50} u_s} \frac{\tilde{R}_\text{eff}^2 \tilde{\textbf{g}}_n}{\tilde{\zeta}_n}
    -\left( \tilde{k}_n |\tilde{\zeta}_n|^{3/2}\textbf{n} + \tilde{d}_n \tilde{\textbf{g}}_n \right)  \\
   &-(\tilde{k}_t |\tilde{\boldsymbol{\zeta}}_t| + \tilde{d}_t \tilde{\textbf{g}}_{t,cp})
    + \frac{\text{max}(|| \textbf{F}_{\text{coh},50} ||)}{m_{50} g'} \, \frac{8 \, \tilde{R}_\text{eff}}{\tilde{\lambda}^2}(\tilde{\zeta}_n^2 - \tilde{\zeta}_n \tilde{h}) \textbf{n} \ .
     \end{split}
\end{equation}
By defining the Cohesive number as  $\text{Co} = \text{max}(|| \tilde{\textbf{F}}_\text{coh} ||)/(m_{50} g')$ this results in
\begin{equation}\label{eq:eom_trans_non_dim}
      \begin{split}
    \tilde{m}_p \frac{\text{d}\tilde{\textbf{u}}_p}{\text{d} \tilde{t}} =& 
     \tilde{\textbf{F}}_{p,h}
    + \tilde{V}_{p} \textbf{e}_g
    - \frac{36 }{\Rey} \frac{\tilde{R}_\text{eff}^2 \tilde{\textbf{g}}_n}{\tilde{\zeta}_n}
    - \left( \tilde{k}_n |\tilde{\zeta}_n|^{3/2}\textbf{n} + \tilde{d}_n \tilde{\textbf{g}}_n \right) \\ 
    &-  (\tilde{k}_t |\tilde{\boldsymbol{\zeta}}_t| + \tilde{d}_t \tilde{\textbf{g}}_{t,cp})
    + \text{Co} \, \, \frac{8 \, \tilde{R}_\text{eff}}{\tilde{\lambda}^2}(\tilde{\zeta}_n^2 - \tilde{\zeta}_n \tilde{h}) \textbf{n} \ .
     \end{split}
\end{equation}
Here, $\textbf{e}_g$ denotes the unit vector pointing into the direction of $\textbf{g}$. Equation \eqref{eq:eom_trans_non_dim} demonstrates that, once the $k$- and $d$-values have been specified by the collision model, the particle behavior is governed by two dimensionless similarity paramaters in the form of the Reynolds and Cohesive numbers. 

%% file: App_char_velocity.tex
For steady-state motion of a single particle in an otherwise quiescent fluid, we obtain a balance between the hydrodynamic force $\textbf{F}_{h,p}$ and the buoyant weight $\textbf{F}_{g,p}$
\begin{equation}\label{eq:steady_state}
 0 = \textbf{F}_{h,p} + \textbf{F}_{g,p} \qquad ,
\end{equation}
where the buoyant weight of a sphere is given by
\begin{equation}\label{eq:weight}
 \textbf{F}_{g,p} = \frac{4}{3} \pi R_p^3 ( \rho_p-\rho_f )\: \textbf{g} \qquad .
\end{equation}
The classical settling velocity value based on Stokes' drag law \cite[e.g.][]{biegert2017peps} is limited to Reynolds numbers $\Rey \ll 1$. For higher Reynolds numbers, Lord Rayleigh formulated the drag equation 
\begin{equation}\label{eq:drag_equation}
\textbf{F}_{h,p} = \frac{1}{2} \rho_f \textbf{u}_p^2 C_d  \underbrace{A_p}_{\frac{1}{4} \pi D_p^2} \ ,
\end{equation}
where the drag coefficient $C_d$ depends on the Reynolds number, and $A_p$ denotes the projected area of the sphere. Substituting \eqref{eq:drag_equation} and \eqref{eq:weight} into \eqref{eq:steady_state} yields the settling velocity
\begin{equation}\label{eq:settling_velocity}
 v_{ra} = \sqrt{\frac{4}{3} \frac{1}{C_d}\frac{\rho_p - \rho_f}{\rho_f}g \, D_p} =  \sqrt{\frac{4}{3} \frac{1}{C_d} g' \, D_p}\qquad. 
\end{equation}
Choosing $C_d = 4 / 3$ simplifies \eqref{eq:settling_velocity} to $u_s= \sqrt{g' \, D_p}$, which is the characteristic velocity used in the present study. We can use the empirical correlation of \cite{clift2005}
\begin{equation}\label{eq:drag}
 C_d = \frac{24}{\Rey}[1+0.1935 \, \Rey^{0.6305}]
\end{equation}
to obtain the corresponding Reynolds number as $\Rey = v_{ra} D_p / \nu_f$. The definition of the Reynolds number together with equations \eqref{eq:settling_velocity} and \eqref{eq:drag} also define the set of equations to iteratively determine $v_{ra}$.